\documentclass[12pt,a4paper,twoside]{article}
\usepackage[top=3cm, bottom=3cm, left=2cm, right=2cm]{geometry}
\usepackage{setspace}
\usepackage{microtype}
\usepackage[utf8x]{inputenc}
\usepackage[T1]{fontenc}
\usepackage{amsmath}
\usepackage{amsfonts}
\usepackage{amssymb}
\usepackage{amsthm}
\usepackage{comment}
\usepackage{caption}
\usepackage{booktabs} % for professional tables
\usepackage{makecell}
\usepackage{cellspace} 
\usepackage{multirow}
\usepackage{xcolor}
\usepackage[pagebackref=true]{hyperref}
\hypersetup{
	colorlinks=true,
	urlcolor=blue,
    citecolor=blue,
	linktoc=all,
	linkcolor=blue}
	
\renewcommand*{\backrefalt}[4]{%
	\ifcase #1 \footnotesize{(Not cited.)}%
	\or        \footnotesize{(Cited on page~#2.)}%
	\else      \footnotesize{(Cited on pages~#2.)}%
	\fi}
\usepackage{graphicx}
\usepackage{float}
\usepackage{natbib}
\usepackage[english]{babel}
\usepackage{amssymb, mathtools}
\usepackage{xfrac} % for \sfrac
\usepackage{comment}
\usepackage{xspace}
\graphicspath{{./figures/}}
\usepackage{amsmath}
\usepackage{mathrsfs}
\usepackage{multirow}
\usepackage{algorithm}
\usepackage{algorithmic}
\usepackage{subfig}
\usepackage[all]{nowidow}

\usepackage{fancyhdr}

\usepackage{xr}
\makeatletter
\newcommand*{\addFileDependency}[1]{% argument=file name and extension
  \typeout{(#1)}
  \@addtofilelist{#1}
  \IfFileExists{#1}{}{\typeout{No file #1.}}
}

\makeatother

\newcommand*{\myexternaldocument}[1]{%
    \externaldocument{#1}%
    \addFileDependency{#1.tex}%
    \addFileDependency{#1.aux}%
}
\myexternaldocument{a_neutral-suppl}
\newcommand{\update}[1]{#1}

\usepackage{nth}
\usepackage{enumerate} 

\usepackage[acronym]{glossaries}
\makeglossaries
\newacronym{ssd}{SSD}{Species Sensitivity Distribution}
\newacronym{cec}{CEC}{Critical Effect Concentration}
\newacronym{ecx}{$\mathrm{EC}_x$}{Effect Concentration at $x\%$}
\newacronym{lc50}{$\mathrm{LC}_{50}$}{Lethal Concentration $50\%$}
\newacronym{ec50}{$\mathrm{EC}_{50}$}{Effect Concentration $50\%$}
\newacronym{hc5}{$\mathrm{HC}_{5}$}{Hazardous Concentration for $5\%$ of the Species}
\newacronym{iid}{\emph{iid}}{identically and independantly distributed}
\newacronym{cpo}{CPO}{conditional predictive ordinate}
\newacronym{loo}{LOO}{Leave-One-Out}
\newacronym{bnp}{BNP}{Bayesian nonparametric}
\newacronym{rivm}{RIVM}{National Institute for Public Health and the Environment}
\newacronym{kde}{KDE}{Kernel Density Estimate}

\def\VI{\mathcal{VI}\xspace}
\DeclareMathOperator*{\argmin}{arg\,min}
\newacronym{dpm}{DPM}{Dirichlet process mixtures}
\newacronym{nrmi}{NRMI}{normalized measures with independent increments}

\newcommand{\R}{\mathbb{R}}

\def\d{\mathrm{d}}
\def\simind{\stackrel{\mathrm{ind}}{\sim}}
\def\simiid{\stackrel{\mathrm{i.i.d.}}{\sim}}

\newcommand{\Exp}{\mathbb{E}}

\newcommand{\ddr}{\mathrm{d}}

\newcommand{\bX}{\boldsymbol{X}}

\newcommand{\NRMIname}{\ensuremath{\text{NRMI}}\xspace}

\newcommand{\MISE}{\text{MISE}\xspace}

\usepackage{xcolor}
\usepackage{tikz}

\definecolor{norange}{RGB}{250,164,58}
\definecolor{nblue}{RGB}{93,165,218}
\definecolor{ngreen}{RGB}{96,189,104}
\definecolor{nred}{RGB}{241,90,96}
\definecolor{dblue}{RGB}{90,155,212}
\definecolor{dgreen}{RGB}{122,195,106}

\DeclareRobustCommand{\orange}{\begin{tikzpicture}
\draw[draw=norange,very thick,dashed](0,0.5)--(0.3,0.5);
\end{tikzpicture}}
\DeclareRobustCommand{\nblue}{\begin{tikzpicture}
\draw[draw=nblue,very thick,dashed](0,0.5)--(0.3,0.5);
\end{tikzpicture}}
\DeclareRobustCommand{\ngreen}{\begin{tikzpicture}
\draw[draw=ngreen,very thick,dashed](0,0.5)--(0.3,0.5);
\end{tikzpicture}}

\DeclareRobustCommand{\orangeTri}{\begin{tikzpicture}
\draw[draw=norange,very thick,loosely dashed](0,0.5)--(0.4,0.5);
\fill[norange] (0.12,0.44)-- (0.32,0.44) -- (0.2,0.6) -- cycle;
\end{tikzpicture}}
\DeclareRobustCommand{\nbluePoint}{\begin{tikzpicture}
\draw[draw=nblue,very thick,loosely dashed](0,0)--(0.4,0);
\fill[nblue](0.21,0) circle(0.07);
\end{tikzpicture}}
\DeclareRobustCommand{\ngreenSquare}{\begin{tikzpicture}
\draw[draw=ngreen,very thick,loosely dashed](0,0.5)--(0.4,0.5);
\fill[ngreen] (0.13,0.43) rectangle (0.28,0.57);
\end{tikzpicture}}

\DeclareRobustCommand{\red}{\begin{tikzpicture}
\draw[draw=nred,very thick](0,0.5)--(0.3,0.5);
\end{tikzpicture}}
\DeclareRobustCommand{\blue}{\begin{tikzpicture}
\draw[draw=dblue,very thick](0,0.5)--(0.3,0.5);
\end{tikzpicture}}
\DeclareRobustCommand{\green}{\begin{tikzpicture}
\draw[draw=dgreen,very thick](0,0.5)--(0.3,0.5);
\end{tikzpicture}}

\DeclareRobustCommand{\orangeTri}{\begin{tikzpicture}
\draw[draw=norange,very thick,loosely dashed](0,0.5)--(0.4,0.5);
\fill[norange] (0.12,0.44)-- (0.32,0.44) -- (0.2,0.6) -- cycle;
\end{tikzpicture}}
\DeclareRobustCommand{\nbluePoint}{\begin{tikzpicture}
\draw[draw=nblue,very thick,loosely dashed](0,0)--(0.4,0);
\fill[nblue](0.21,0) circle(0.07);
\end{tikzpicture}}
\DeclareRobustCommand{\ngreenSquare}{\begin{tikzpicture}
\draw[draw=ngreen,very thick,loosely dashed](0,0.5)--(0.4,0.5);
\fill[ngreen] (0.13,0.43) rectangle (0.28,0.57);
\end{tikzpicture}}

\usepackage[symbol]{footmisc}

\makeindex

\title{Species Sensitivity Distribution revisited:\\
a Bayesian nonparametric approach}
\date{}
\begin{document}
\maketitle
\vspace{-5em}
\begin{center}
Louise Alamichel$^{1}$\footnote{Corresponding author: \href{mailto:louise.alamichel@unibocconi.it}{louise.alamichel@unibocconi.it}}, Julyan Arbel$^2$, Guillaume Kon Kam King$^3$, Igor Pr\"unster$^1$

\bigskip
{\it
$^{1}$Bocconi Institute for Data Science \& Analytics, Bocconi University, 20136 Milano, Italy
}\\
{\it
$^{2}$Univ. Grenoble Alpes, CNRS, Inria, Grenoble INP, LJK, 38000 Grenoble, France
}\\
{\it
$^{3}$Université Paris-Saclay, INRAE, MaIAGE, 78350 Jouy-en-Josas, France
}
\end{center}
\bigskip

\begin{abstract}
    We present a novel approach to ecological  risk assessment by %reexamining
    recasting the Species Sensitivity Distribution (SSD) method within a Bayesian nonparametric (BNP) framework. Widely mandated by environmental regulatory bodies globally, SSD has faced criticism due to its historical reliance on parametric assumptions when modeling species variability. By adopting nonparametric mixture models, we address this limitation, establishing a %more 
    statistically robust foundation for SSD.
    Our BNP approach offers several advantages, including its efficacy in handling small datasets or censored data, which are common in ecological risk assessment, and its ability to provide principled uncertainty quantification alongside simultaneous density estimation and clustering. We utilize a specific nonparametric prior as the mixing measure, chosen for its robust clustering properties, a crucial consideration given the lack of strong prior beliefs about the number of components.
    Through simulation studies and analysis of real datasets, we demonstrate the superiority of our BNP-SSD over classical SSD methods.
    We also provide a BNP-SSD Shiny application, making our methodology available to the Ecotoxicology community.
    Moreover, we exploit the inherent clustering structure of the mixture model to explore patterns in species sensitivity. Our findings underscore the effectiveness of the proposed approach in improving ecological risk assessment methodologies.
\end{abstract}


\newpage
\section{Introduction}
\label{sec:intro}
\update{\textbf{Ecotoxicological context.}}  
Assessing the response of a community of species to environmental stress is critical for ecological risk assessment. Methods developed for this purpose vary greatly in levels of complexity and realism. \gls{ssd} represents an intermediate tier method, more refined than rudimentary assessment factors \citep{Posthuma2010} but practical enough to be used routinely by environmental managers and regulators in most developed countries (Australia and New Zealand, \citealp{armcanz2000australian}, Canada, \citealp{canada_guidelines}, China, \citealp{liu2014setting}, EU, \citealp{ECHA2008}, South Africa, \citealp{Roux1996}, USA, \citealp{USEPASSD}). The \gls{ssd} approach is intended to provide, for a given contaminant, a description of the tolerance of all species possibly exposed using information collected on a sample of species. This information consists of \glspl{cec}, a concentration specific to a species that marks a limit over which the species suffers a critical level of effect. Such levels of effect are, for instance, the concentration at which $50\%$ of the tested organisms died (\gls{lc50}), or the concentration which inhibited growth or reproduction by $50\%$ compared to the control (\gls{ec50}). Each \gls{cec} is the summary of costly bioassay experiments for a single species, so data is usually in short supply. The European Chemical Agency (ECHA) sets the minimal data requirement to a sample size of 10, preferably 15 \citep{ECHA2008}.\smallskip

\noindent\update{\textbf{Limitations of existing parametric approaches.}}  
To describe the tolerance of all species to be protected, the distribution of the \glspl{cec} is estimated from the sample of tested species. In practice, a parametric distributional assumption is often adopted \citep{Forbes2002}: the \glspl{cec} are assumed to follow a log-normal \citep{Wagner1991,aldenberg2002normal}, log-logistic \citep{Aldenberg1993,Kooijman1987a}, triangular \citep{VanStraalen2002,Zhao2016} or BurrIII \citep{Shao2000} distribution. Once the response of the community is characterized, a safe concentration is defined, typically the \gls{hc5}, the \nth{5} percentile of the distribution. Instead of the raw estimate, practitioners often use the lower extreme of a confidence interval, further corrected by a safety factor. 
\update{The difficulty of justifying the choice of a single parametric distribution has sparked a lot of research into distributional comparisons \citep{Xu2015,He2014, Jagoe1997, VanStraalen2002,Xing2014,Zhao2016}. The general consensus emerging from these works is that no single distribution provides a uniformly superior fit across datasets, and that the most appropriate model is strongly dataset dependent \citep{Forbes2002}. Moreover, because goodness-of-fit tests tend to have low statistical power when sample sizes are small, practical considerations often dominate theoretical ones. In this context, the log-normal distribution has emerged as the conventional choice, primarily due to its mathematical simplicity, interpretability, and the convenience it offers for estimating confidence intervals of the \gls{hc5}.}\smallskip

\noindent\update{\textbf{Alternatives and the need for flexibility.}}  
Another research direction has aimed at avoiding parametric assumptions altogether, using distribution-free approaches such as the empirical distribution function \citep{SuterIi1999,Jones1999}, rank-based methods \citep{VanderHoeven2001,Chen2004}, bootstrap resampling \citep{Jagoe1997,Grist2009a,Wang2008} or nonparametric kernel density estimation \citep{Wang2015}. However, these methods require large sample sizes, in contradiction to the scarcity of ecotoxicological data and to the ethical aim of reducing animal testing. \update{This limitation highlights a persistent trade-off between model generality and data availability: while flexible, nonparametric methods are attractive in principle, their reliability diminishes sharply when data are sparse or noisy.} A further line of work considers that the distribution of \glspl{cec} might not be homogeneous but instead composed of several subgroups, each with its own distributional shape \citep{Zajdlik2009,Kefford2012a,PeterCraig2013}. This is ecologically realistic: taxonomy, habitat, or mode of action of contaminants may induce multimodality, especially for compounds targeting specific \update{taxonomic} groups \update{(e.g. certain pesticides). Yet current official recommendations when multimodality is suspected are to retain only the most sensitive group \citep{ECHA2008,Zajdlik2009}, a pragmatic but information-wasting strategy that oversimplifies ecological variability. These considerations reinforce the need for methods that can adapt flexibly to complex, multimodal, and data-poor settings, capturing heterogeneous sensitivity patterns without imposing restrictive distributional assumptions.}\smallskip

\noindent\update{\textbf{Rationale for Bayesian nonparametrics.}
Given the uncertainty on determinants of species sensitivity, there is little prior knowledge on group structure, which calls for a nonparametric approach. In such ecological settings, the relationships among species and their sensitivities are complex and often cannot be captured by fixed parametric forms. \gls{bnp} models are particularly appealing because they allow the data to reveal the underlying structure without imposing restrictive assumptions on the number or shape of sensitivity groups. Existing frequentist nonparametric methods rely on asymptotic guarantees and thus may perform poorly with small datasets. \gls{bnp} mixture models offer an effective solution for both large and small samples, because the complexity of the mixture adapts to the available data. {\gls{bnp} models allow the number of mixture components to be learned from the data rather than fixed in advance, avoiding arbitrary assumptions that are hard to justify in ecological applications.} \update{They also naturally quantify uncertainty on key regulatory quantities such as the \gls{hc5}, providing full posterior distributions instead of point estimates.} Compared to kernel density methods, they reduce overfitting risks and offer principled uncertainty quantification \citep{barrios2013modeling}. Importantly, the low information in small datasets can be complemented by informative priors, leveraging external knowledge from other species or contaminants \citep{Awkerman2008,PeterCraig2013,craig2012species}. In this way, \gls{bnp} mixture models reconcile ecological realism (through flexible multimodal structures) with regulatory needs (through reliable uncertainty quantification), making them particularly well suited for \gls{ssd} applications.}\smallskip

\noindent\textbf{Objectives and outline.}
In Section~\ref{sec:methods} we present a Bayesian nonparametric approach to \gls{ssd} based on a nonparametric mixture model. We show that our \gls{bnp}-\gls{ssd} approach can include censored data, which are common in ecotoxicology  \citep{KonKamKing2014}, and that it provides a rigorous description of the uncertainty on the variable of interest, the \gls{hc5}. We showcase the value of our approach by comparing the \gls{bnp}-\gls{ssd} with the most standard normal-\gls{ssd} approach \citep{Aldenberg2000a} and with a nonparametric \gls{ssd} method based on \gls{kde} \citep{Wang2015}. The comparison is performed on simulated data in Section~\ref{sec:simulation}, to demonstrate the higher accuracy of the \gls{bnp}-\gls{ssd}, and we study real censored and noncensored datasets in Section~\ref{sec:contaminant-wise} highlighting the robustness of our method. Additionally, we perform an exploratory analysis\footnote{The code is available on GitHub at \url{https://github.com/alamichL/BNP_SSD/}.} to describe what biological insight we can gain with \gls{bnp}-\gls{ssd} by studying patterns of species or contaminants induced by the clustering in Section~\ref{sec:res_clus}. Finally, we conclude and describe further research directions in Section~\ref{sec:discussion}.
\section{Methods\label{sec:methods}}

Due to the wide spectrum of variation of \gls{ssd} concentrations and to their positivity, it is common practice to work on log-transformed concentrations. We consider a sample of $n$ log-concentrations that we denote by $\bX = (X_1,\ldots,X_n)$, that typically represents the \gls{cec} for a collection of $n$ species tested with a given contaminant. Moreover, the data are standardized: observations are centered and rescaled to have a variance of one. After the inference, all estimations are transformed back to the original scales.

We carry out density estimation for the \gls{ssd} based on a sample $\bX$ using Bayesian nonparametric mixtures. The method of mixtures of probability density kernels with a nonparametric prior as mixing measure is due to \citet{lo1984class}, where \gls{dpm}\index{Dirichlet process mixture} is introduced.
Generalizations of the \gls{dpm} correspond to allowing the mixing distribution to be any discrete nonparametric prior. A large class of such prior distributions is obtained by normalizing random measures known as \textit{completely random measures} \citep{kingman1967completely}. The normalization step gives rise to so-called \gls{nrmi} as defined by \citet{rlp2003}. See \citet{lijoi2010models,Jordan2010hierarchical,barrios2013modeling} for reviews. More details on specific choices of the  \gls{nrmi} prior and their inferential impact are provided in the sequel.
An \gls{nrmi} mixture model is defined as
\begin{align}
	\label{eq:mix1}
	X_i  \mid \tilde P & \simiid \tilde f(x)=\int k(x \mid \theta) \tilde P(\d\theta), \\
	\tilde P           & \sim \NRMIname\nonumber
\end{align}
where $k$ is a probability density kernel parametrized by some $\theta\in\Theta$ and $\tilde P$ is a random probability on $\Theta$
whose distribution is an \gls{nrmi}.
Alternatively, the mixture model can also be formulated hierarchically as
\begin{align}\label{eq:DPM}
	X_i  \mid  \theta_i    & \simind k( x  \mid
	\theta_i),\quad i=1,\ldots,n,
	\nonumber
	\\
	\theta_i \mid \tilde P & \simiid \tilde P,\quad i=1,\ldots,n,
	\\
	\tilde P               & \sim \NRMIname.
	\nonumber
\end{align}

Specifically, we consider location-scale mixtures and denote by $\theta_i = (\mu_i,\sigma_i)$ the vectors of individual means and standard deviations, $\theta_i \in \R\times\R_+$.
As discussed in the Introduction, log-concentrations are commonly fitted with a normal distribution, or with mixtures of such distributions. Our aim is to move from these parametric models to the nonparametric specification in~\eqref{eq:DPM}, and in order to allow for comparisons with competing approaches, we stick to the normal specification for $k$ on the log-concentrations $\bX$, $k(x\mid \mu,\sigma) = \mathcal{N}(x\mid \mu,\sigma^2)$.

Within this framework, density estimation is carried out by evaluating the posterior mean
\begin{equation}
	\hat{f}(x \mid  \bX)=\Exp \bigl(\tilde f(x)  \mid  X_1,
	\ldots,X_n \bigr) \label{eq:predictive}
\end{equation}
for any $x$ in $\R$.
\update{For posterior sampling, we use \texttt{BNPdensity}\footnote{Available at \url{https://CRAN.R-project.org/package=BNPdensity}.}, an R package which performs \gls{bnp} density estimation and clustering under a general specification of NRMI prior based on generalized gamma processes \cite[see][]{barrios2013modeling,arbel2021BNPdensity}. \texttt{BNPdensity} leverages the popular Ferguson and Klass algorithm \citep{ferguson1972representation}, extended with a Metropolis--Hastings-within-Gibbs scheme.}

We compare the proposed \gls{bnp}-\gls{ssd} to two competitors. First, the most commonly used model for the SSD, the normal distribution \citep{Aldenberg2000a}, with estimated density $\hat{f}_{\mathcal{N}}(x) = \mathcal{N}(x  \mid  \hat \mu, \hat{\sigma}^2)$ where $\hat \mu$ and $\hat{\sigma}$ are the data empirical mean and standard deviation. Second, the frequentist nonparametric kernel density method recently applied to the SSD by \citet{Wang2015}, with estimated density $\hat{f}_{\mathrm{KDE}}(x) = \frac{1}{n} \sum_{i=1}^n \mathcal{N}(x  \mid  X_i, h_n^2)$ where  $h_n = 1.06 \hat{\sigma}n^{-\frac{1}{5}}$ is the asymptotically optimal default bandwidth recommendation of \cite{silverman1986density}, also used by \citet{Wang2015}.

\subsection{Censored data\label{sec:censored-data}}

\cite{KonKamKing2014} explained how to use censored data with the normal \gls{ssd}, and showcased the drawbacks of the common approach, which consists of transforming or discarding censored data in \gls{ssd}. It is similarly possible to incorporate censored data into the \gls{bnp}-\gls{ssd}.

Indeed, only the first line of the hierarchical model defined in~\eqref{eq:DPM} needs to be suitably adapted, while the other levels in the hierarchy remain unchanged.
More specifically, denote by $F$ the cumulative density function of the kernel $k$, then: $k(x\mid\theta)$ has to be replaced by $F(x\mid\theta)$ for a left-censored observation, by $1-F(x\mid\theta)$ for a right-censored observation, and by $F(x_r\mid\theta)-F(x_l\mid\theta)$ for an interval-censored observation $[x_l,x_r]$.
This approach can be used for the standard normal \gls{ssd} and any likelihood-based inference, but there is no widely available tool to perform \gls{kde} on all types of censored data: the {R} package \texttt{ICE} %\footnote{Available at \url{https://CRAN.R-project.org/package=ICE}.}, for instance,
does not handle left/right censored data (not maintained anymore on CRAN); R packages \texttt{muhaz} or \texttt{Kernelheaping} can deal with right or interval-censored data respectively, but there does not seem to be an available implementation for all three types of censored data.

In this paper, we illustrate the differences among the various approaches on a censored dataset and, for comparison purposes, we study censored and non-censored versions of the dataset. To obtain a non-censored dataset from a censored dataset, we follow the customary approach, which consists of discarding left and right censored data and replacing interval-censored data with the central value of the interval.

\subsection{Prior specification}

The class of \gls{nrmi}  priors is very broad {and we refer the reader to \cite{lijoi2010models} for an extensive review. Here we focus on a specific member of the class known as the normalized stable process \citep{kingman1975random}, which as argued in \cite{barrios2013modeling} represents a natural default choice.} It is specified in terms of a stability parameter $\gamma \in (0,1)$ and a base measure $P_0$, which corresponds to the expectation of the random probability measure.
The stability parameter $\gamma$ controls the variability of the prior distribution on the number of clusters: {heuristically,} a small $\gamma$ corresponds to an informative prior, while a large $\gamma$ indicates a vague prior.
Large values of $\gamma$ can require expensive computations to preserve the quality of the posterior sampling. We chose $\gamma = 0.4$ as a compromise between the flexibility of the model and the computational requirements. See \cite{lijoi2007,barrios2013modeling} for details.
\update{A sensitivity analysis of the model regarding the choice of $\gamma$ is available in Figure~\ref{fig:sens_gam}.}

As mentioned previously, we shall consider location-scale mixtures, meaning that the \gls{nrmi} prior should be defined on $\R\times\R_+$, the space of locations and scales. Thus the base measure $P_0$ is defined on $\R\times\R_+$, and we denote by  $f_0$ its density with respect to the Lebesgue measure on $\R\times\R_+$.
We assume that the locations $\mu$ and scales $\sigma$ are a priori independent.
Thus, we use the notation $f_0(\mu,\sigma)=f_{0}^1(\mu) f_{0}^2(\sigma)$ with possible hyperparameters for $f_{0}^1$ and $f_{0}^2$. The possibility to specify independent priors for $\mu$ and $\sigma$ is a beneficial feature of \glspl{nrmi}, which does not require any conjugacy structure in the prior. Therefore, the prior specification can be derived in a straightforward way.

The specific choice of distribution on the location parameters of the clusters $\mu$ is a normal distribution  $f_0^1(\mu) = \mathcal{N}(\mu\mid \varphi_1,\varphi_2^{-1})$ where mean $\varphi_1$ and precision $\varphi_2$ are assigned a normal-gamma conjugate hyperprior.
This corresponds to a vague prior for the location of the clusters, which can just as well be at the center of the dataset or at the borders.
Regarding the scale parameter $\sigma$ of the clusters, given the standardization of the data during the pre-processing where the variance is set to one, $\sigma$ should be smaller than one, approaching one only in cases of unimodality. Moreover, {$\sigma$ should also be a priori bounded from below} since numerous extremely small clusters do not make biological sense regarding species sensitivity distributions.
Therefore, we choose a uniform prior  $f_0^2(\sigma)=\text{Unif}_{[0.1, 1.5]}(\sigma)$, leaving room around the upper bound of one to allow for {potential} unimodality.
We studied the sensitivity with respect to this prior specification by varying its extreme points and also with respect to a left-truncated normal prior of mean $0.5$, variance $1$, and lower bound $0.1$. Note that the latter has approximately $3/4$ of its mass within the support of the $\text{Unif}_{[0.1, 1.5]}$ distribution.
\update{The sensitivity analyses (available in Figure~\ref{fig:sens_sig}) showed little variation to moderate changes in the parameters of these two prior distributions.}

\subsection{Estimation of the \gls{hc5}}

The main quantity of interest for ecological risk assessment is the  \gls{hc5}, which corresponds to the \nth{5} percentile of the \gls{ssd} distribution.
In our \gls{bnp} framework, we rely on the posterior expectation.
\update{The posterior distribution induced on the \gls{hc5} is obtained by considering the (random) cumulative distribution function $\tilde F(x) = \int F_k(x|\theta) \tilde P (d\theta)$, where $F_k$ denotes the cumulative distribution function of kernel $k$.
The inverse of the cumulative distribution function is the quantile function, so $\text{HC}_5 = \tilde F^{-1}(0.05)$. 
This inversion is performed at each iteration, that is for each realization of $\tilde F$, of the MCMC chain, giving us access to the posterior distribution of the \gls{hc5}.
The $95\%$ credible bands are formed by the $2.5\%$ and $97.5\%$ quantiles of the \gls{hc5} posterior distribution.
In practice, our method resorts to numerical inversion of the realizations of the cumulative distribution function, based on a truncation approximation of the mixing measure $\tilde P$.
The truncation of $\tilde P$ results in a finite mixture. Hence, its cumulative distribution can be readily evaluated and it is amenable to numerical inversion.
The truncation method is a dynamic approximation designed to preserve the moments of the infinite-dimensional mixing measure \citep{arbel2017moment}.}

Similarly, the \nth{5} percentile of the \gls{kde} can be obtained by numerical inversion of the cumulative distribution function, and the confidence intervals using nonparametric bootstrap. The \nth{5} percentile of the normal \gls{ssd} and its confidence intervals were obtained following the classical method by \cite{Aldenberg2000a}.

\subsection{Robustness comparison using Leave-One-Out cross validation\label{sec:CPO}}

We compare the robustness of the three \gls{ssd} methods using a predictive performance measure, \gls{loo} cross-validation.
We compute the \gls{loo} for each method as:
\begin{equation}
	\text{LOO}_i = \hat f(X_i \mid  \bX_{-i}),
\end{equation}
where $\hat f(\,\cdot \mid  \bX_{-i})$ is the {density estimate based on $\bX$ with $X_i$ left out for each of the three methods}.
\gls{loo} cross-validation evaluates the robustness of a method by assessing how well it predicts each observation when that observation is excluded from the dataset.
Overfitting methods tend to produce predictions that vary significantly with the inclusion or exclusion of specific points, leading to poor \gls{loo} performance.
In contrast, robust methods should exhibit stable predictions regardless of the removal of individual points, resulting in better \gls{loo} performance.

For the \gls{bnp} models, \glspl{loo} are referred to as \glspl{cpo} statistics.
They are commonly used in applications, see for instance \citet{gelfand1996model}.

A CPO statistic is defined for each data point $X_i$ as
\begin{equation*}
	\operatorname{CPO}_i=\hat f(X_i\mid \bX_{-i}) = \int k(X_i\mid \theta) \pi(\ddr \theta\mid \bX_{-i}),
\end{equation*}
%where $\bX_{-i}$ denotes the  whole sample $\bX$ but $X_i$, 
where $\pi(\ddr \theta \mid \bX_{-i})$ is the posterior distribution associated to $\bX_{-i}$ and $\hat f(\,\cdot \mid  \bX_{-i})$  is the (cross-validated) posterior predictive distribution of~\eqref{eq:predictive}.
CPOs can be easily approximated by Monte Carlo as
\[
	\widehat{\operatorname{CPO}_i}= \Biggl(\frac{1}{T}\sum
	_{t=1}^T \frac{1}{k(X_i\mid \theta^{(t)})}\Biggr)^{-1},
\]
where $\{\theta^{(t)}, t=1,2,\ldots,T\}$ is an MCMC sample obtained through the \texttt{BNPdensity} R package.
For the two frequentist models, the  \glspl{loo} can be computed by fitting them directly on the leave-one-out data $\bX_{-i}$ for each $i$.

\subsection{Clustering estimation}

%\begin{itemize}
%\item It's a long standing question how to give an optimal estimate of the clustering from an MCMC sample.
% \item Output of MCMC is a posterior distribution on the space of exchangeable partitions, which is very large, so it is under-explored
% \item Typically, a partition is visited at most once or twice by the chain and the optimal partition (largest posterior probability) might not even have been visited.
% \item \cite{Wade2015} proposes a approach for determining the optimal partition from decision-theoretic arguments based on a Variation of Information loss function
% \item Using a greedy approach for exploring the space of partitions, it is able to determine an optimal partition even if it has not been visited by the MCMC chain.
% \item Seems a good approach, does not present the problem of creating unnecessary tiny clusters as other clustering estimation methods do.
%\end{itemize}

The question of how to estimate data clustering based on an MCMC posterior sample is a long-standing problem in Bayesian statistics \citep[see][]{dahl2006model,lau2007bayesian}.
% Defining an unsupervised clustering of the data is akin to partitioning it. 
Estimating a clustering structure is computationally expensive, owing to the extremely rapid growth in the cardinality of the partition space with the sample size $n$, known as the Bell number of order $n$.
Enumeration of all partitions is infeasible in practice; thus, one typically resorts to approximations.
Many ad-hoc procedures have been devised in the literature. However, as noted by \citet{dahl2006model}, it seems counter-intuitive to apply an ad-hoc clustering method on top of a model that itself produces clusterings.

We adopt instead a fully Bayesian route by undertaking clustering on decision-theoretic grounds. We consider a loss function $\mathcal{L}$ and propose a Bayesian point estimator $\hat c$ for a clustering structure obtained as an argument that minimizes the posterior expected loss given the data $\bX$ %\citep{wade2015bayesian}
\begin{align}\label{eq:clustering_point_estimator}
    \hat c \in \argmin_{c^\prime}\sum_{c}\mathcal{L}(c^\prime,c)\pi(c\mid \bX),
\end{align}
where \update{the summation over $c$ is over all possible clusterings, and where $\pi(c\mid \bX)$ denotes the posterior distribution of clustering $c$}.

The maximum a posteriori (MAP), often considered in the literature, is an example of such a Bayesian estimator, based on the very crude $0-1$ loss function $\mathcal{L}_{0-1}$.
However when $n$ is large, a posterior sample generally hardly visits twice the same clustering, thus making the empirical MAP of the MCMC output very sensitive to the initialization of the chain and of very limited validity in practice.

Manifestly, many other loss functions can be considered and expected to perform better than $\mathcal{L}_{0-1}$.
One particular choice of a loss function stands out from these in best estimating the number of groups in a clustering.
It is known as the variation of information, denoted by $\VI$, which is a loss function firmly established in information theory \citep{meila2007comparing,wade2015bayesian}.
The variation of information between two clusterings is defined as the sum of their information (their Shannon entropies) minus twice the information they share.
Simulations indicate that the variation of information is a sensible choice: when other losses such as the Binder loss \citep{binder1978bayesian} typically tend to overestimate the number of clusters, the variation of information instead seems to {effectively} %consistently
recover it (see for instance the simulated examples, and more specifically Figures 6--8, of \citealp{wade2015bayesian}).
%One specificity of the $\VI$ loss which underlies its success is that it penalizes singletons and small clusters much more than the Binder loss, for instance. 
%More formal elaborations on posterior consistency properties are ongoing work by the present authors.

A merit of the approach presented in \citet{wade2015bayesian} is that it rests on a greedy search algorithm to determine the minimum loss clustering of~\eqref{eq:clustering_point_estimator}. Starting from the MCMC output, this greedy approach explores the space of partitions not restricted to those visited by the MCMC chain to find the optimum. \update{Different efficient implementations were proposed, e.g. in \cite{Rastelli2018} and \cite{dahl2022search}. In this work, we chose to use the algorithm proposed by \cite{dahl2022search}. Results in \cite{dahl2022search} indicate that this algorithm achieves superior performance and greater computational efficiency compared with other implementations.}
\section{Simulation study\label{sec:simulation}}

{In order to compare the performance of \gls{bnp}-\gls{ssd}, the normal-\gls{ssd} \citep{Aldenberg2000a} and the nonparametric \gls{kde}-\gls{ssd} \citep{Wang2015}, we perform a simulation study with synthetic datasets corresponding to different scenarios.}
%Here, we consider synthetic datasets to check the performance of our model and to compare with existing methods.

\subsection{Simulated data}

We consider three distinct simulated data scenarios corresponding to various situations: 
\begin{itemize}
 \item[(a)] a standard normal distribution: this is a situation where the normal assumption made for \gls{ssd} is justified;
 \item[(b)] a $t$-distribution with three degrees of freedom and noncentrality parameter equal to $-2$: this is a situation where some species are relatively more sensitive, creating a heavier tail on the left of the distribution;
 \item[(c)] a bimodal distribution corresponding to a mixture of normals $\sfrac{1}{3}\,\mathcal{N}(-2,1)+\sfrac{2}{3}\,\mathcal{N}(5,1)$: this is a situation where a group of species is much more sensitive than all the others, typical of some pesticides which disproportionately affect the target species.
\end{itemize}
These three scenarios represent the diversity of empirical distributions found in real data such as the \gls{rivm} database \citep{de2001observed}. 

For all settings, we sampled independently $S=40$ datasets of sizes 10, 20, 50, 100. These sizes are representative of the dataset sizes in the field, as described in the Introduction. 
{Figure~\ref{Illustration} depicts the data generating densities and the different estimates obtained from the three different approaches with datasets of size 20.} 

\def\colorlegend{Orange (\orange) for the BNP model, blue (\nblue) for the normal model, and green (\ngreen) for the KDE model.}

\begin{figure}
    \centering
    \includegraphics[width = \textwidth]{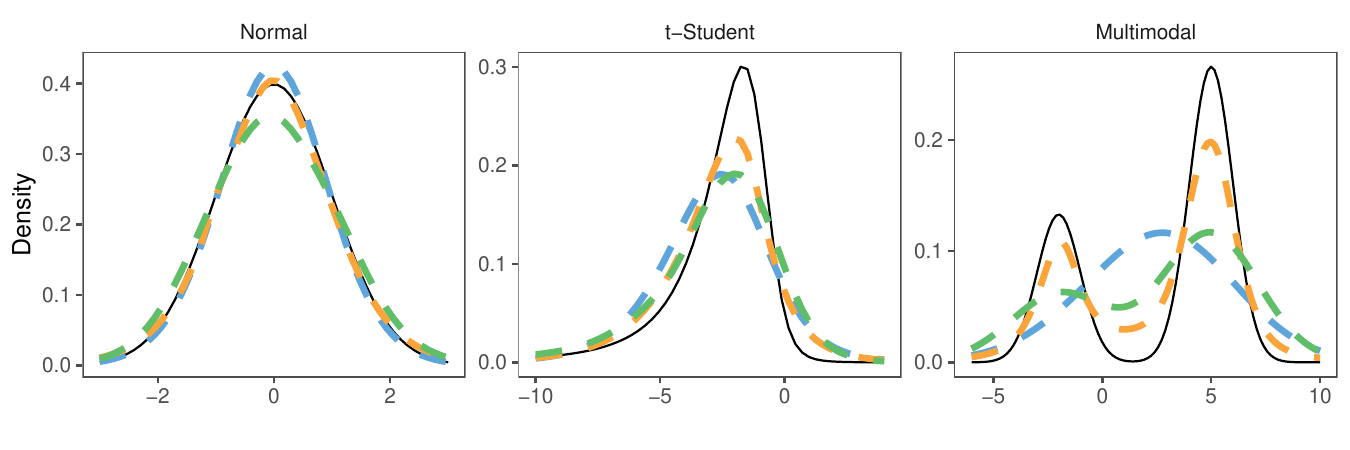} 
    \caption{Three simulation scenarios: data generating density (solid line) and density estimates for each model based on datasets of size $n=20$ (dashed lines).
    \colorlegend
    \label{Illustration}}
\end{figure}

% \includegraphics[width=14cm, height=7cm]{illustration_simulated_data} 
% \captionof{figure}{Three simulation scenarios: data generating density (solid line) and density estimates for each model based on datasets of size $n=20$ (dashed lines).\label{Illustration}
% }

\subsection{Performance comparison of the three approaches}

% For comparison, we estimate our model, the normal model \citep{Aldenberg2000a} and the nonparametric \gls{kde} \citep{Wang2015}(Figure~\ref{Illustration}).

\newcommand{\average}[1]{\big\langle #1\big\rangle_S}  

For each sampled dataset $i\in\{1,\ldots, S\}$ and each model considered, we estimate the data generating density $f$ by a density denoted $\hat f_i$, and compute the associated \nth{5} percentile $\hat q_i$, {which is used} as an estimate of the true \gls{hc5} denoted by $q_0$. 
We denote by $(\hat l_i, \hat u_i)$ a $95\%$ {confidence/credible} interval for $\hat q_i$, and by $\hat \ell_i = \hat u_i - \hat l_i$ its length. To account for sampling variation, we compute averaged summaries (using the notation $\average{\,\cdot\,}$ to denote averaging over the $S$ independent samples).

We compute two performance indicators, the mean absolute error $\text{MAE}= \average{|\hat q_i-q_0|}$, and the mean integrated squared error $\MISE= \average{\int(\hat f_i -  f)^2}$. 
Moreover, we compute the mean {confidence/credible} interval length $\text{MCIL} = \average{\hat \ell_i}$ as a measure of uncertainty: as the \gls{bnp} model captures model uncertainty, we expect it to give a more conservative estimate of uncertainty than the other models. However, we would not want to be conservative to the point that the estimates are useless for the practical purpose of estimating an \gls{hc5}.

The density estimates give a first intuition of the superiority of the \gls{bnp}-\gls{ssd} over the other two models in recovering the true density (Figure~\ref{Illustration}). The results from the simulation study are presented in Figure~\ref{graph_simul_data}, which we describe from top to bottom and left to right.

On the normal simulated data, the well-specified normal model obviously performs best. 
However, the mean absolute error on the \gls{hc5} of the \gls{bnp} is very similar to that of the normal. For small sample sizes, the MISE of the \gls{bnp} is almost the same as that of the normal. 
This illustrates the fact that the BNP model complexity scales with the amount of data and that in data-poor contexts, it essentially reduces to a normal model. 
For small dataset sizes, the mean CI length is larger for the \gls{bnp} model than for the normal, reflecting the model uncertainty {built into the \gls{bnp} model, which is a sign of its potential to flexibly adapt in case of deviations as more data become available}. For larger sizes, the model uncertainty decreases and the \gls{bnp} and normal model coincide.
 
For the t-Student simulated data, the \gls{bnp} model outperforms the other two. The normal and \gls{bnp} have a smaller mean absolute error than the \gls{kde}, the \gls{bnp} and \gls{kde} have a smaller MISE and the mean CI length for the normal model is misleadingly small.

Finally, for the multimodal simulated data, the \gls{bnp} model is clearly superior to the other two models in terms of mean absolute error and MISE. 
The mean CI length is relatively larger than the other two models for small dataset sizes and smaller for larger dataset sizes.

\def\colorshapelegend{Orange (\orangeTri) for the BNP model, blue (\nbluePoint) for the normal model, and green (\ngreenSquare) for the KDE model.} 

\begin{figure}
    \centering
     \includegraphics[width = \textwidth]{
     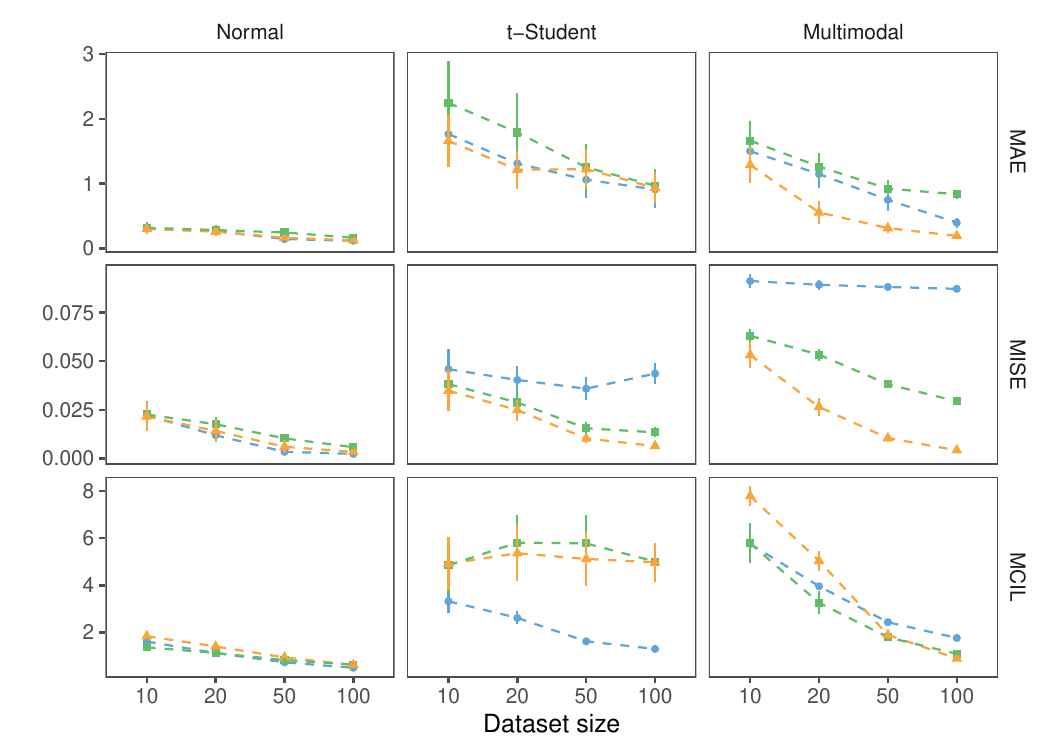}
    \caption{
    Normal, t-Student, and normal mixture simulation scenarios (from left to right); mean absolute error (MAE), mean integrated squared error (MISE), and mean {confidence/credible} interval length (MCIL) as a function of the dataset size (from top to bottom). \update{Uncertainty estimated from the $S=40$ simulations is reported via error bars.}
    \colorshapelegend}
    \label{graph_simul_data}
\end{figure}

% \begin{center}
%  \includegraphics[width=14cm, height=11cm]{graph_comparison_simulated_data} 
% \captionof{figure}{Normal, t-student and normal mixture simulation scenarios (from left to right); mean absolute error (MAE), mean integrated squared error (MISE) and mean \tre{confidence/credible} interval length (MCIL) as a function of the dataset size (from top to bottom). Red denotes the \gls{bnp} model, green the \gls{kde} model and blue the normal model. \tm{this last sentence is a bit redundant give the labels' box; if we keep it we should add it also to Fig. 1} 
% \textcolor{cyan}{changer les noms en ordonnée à droite: MAE et MCIL. \tm{i would also use t-Student or Student's t instead of Student only (applies also to fig 1. i changed it in the text}}\label{graph_simul_data}}
% \end{center}

% \begin{table}
% \begin{center}
% \begin{tabular}{l | ccccc}
% & $\bar a$ &  $\bar b$ &  $\bar \ell$ &  $\overline{\sd}$ & \RMISE \\
% \hline
% Normal & 0.327 & 0.152	 & 1.277 & 0.366 & 0.979 \\
% Logistic & 0.327 & 0.152	 & 1.277 & 0.366 & 0.979 \\
% \gls{kde} & 0.327 & 0.152	 & 1.277 & 0.366 & 0.979 \\
% \gls{bnp} & 0.327 & 0.152	 & 1.277 & 0.366 & 0.979 \\
% \end{tabular}
% \caption{\label{tab:simulation}
% Simulation study results.}
% \end{center}
% \end{table}

\section{Analysis of contaminant-wise clustering\label{sec:contaminant-wise}}

\subsection{Real data description}
\label{sec:data}
We illustrate the advantages of the proposed Bayesian nonparametric approach by means of a selection of contaminants extracted from a large database collected by the \gls{rivm} and first presented in \citet{de2001observed}.

% \subsection{Preprocessing of the data}

We study the dataset already curated by \cite{Hickey2012} with the same restrictions concerning data quality and homogeneity.
We consider both the censored and the non-censored versions of the dataset, the non-censored version being obtained by following the traditional approach of discarding left and right-censored data and taking the central value of interval-censored data.
% \textcolor{magenta}{[Say more about the data: how many contaminants, how many species. Any idea of an informative plot to describe it? This is because later on, in the clustering section, we speak about the data without having described them much.]}
The dataset is an aquatic ecotoxicity research database, with 1,557 species and 3,448 distinct chemicals. Specifically, the dataset records the following covariates: species, chemical, and concentration.

To deal with the presence of multiple \glspl{cec} values for one species, we used the classical approach to replace these values by the geometric mean of the values as a surrogate \citep{ECHA2008} for non-censored data, and followed \cite{KonKamKing2014} in the case of censored data.
% We reduced the concentrations by an affine transformation, so that the processed data are centred and with unit variance. 
% The contaminants were . 
% This database was curated, studied and published by \citet{Hickey2012} and .

\subsection{Density, quantiles and \gls{hc5} estimation}
\label{sec:est_data}
For illustration purposes, we present three categories of contaminants: contaminants with large datasets, consisting of more than 50 values, medium datasets, with around 25 values, and small datasets, with a little over 10 values.
\update{For each of these categories, we select datasets, associated with a contaminant, exhibiting approximately unimodal, skewed, and bimodal behavior, as in the simulation study.
This selection was performed for uncensored datasets.}
The three models (\gls{bnp}, \gls{kde}, and normal) were fitted on each dataset and we studied the estimate of the \gls{hc5} and its credible interval, the LOO error, and the shape of the estimated density compared to the histogram.
The censored version of the same datasets was also studied with the \gls{bnp} and normal model, while there does not seem to exist any implementation of the \gls{kde} model for censored data (see Section~\ref{sec:censored-data}).
The results are displayed in Appendix~\ref{sec:app_results}, see Figure~\ref{fig:S1} to Figure~\ref{fig:S6}.

The \gls{bnp} model is both more flexible than the \gls{kde} model, as apparent from the density estimates for the bimodal datasets,  and comparably, or even more, robust.
The length of the confidence/credible intervals {does not exhibit substantial differences among the three methods. This represents strong evidence in favor of the claim that being less restrictive (in terms of distributional assumptions) than the normal model does not result in over-conservative estimates of the \gls{hc5}. This is of great importance since over-conservative estimates would seriously compromise} a wide adoption of the \gls{bnp}-\gls{ssd} approach.
More precisely, in the case of the roughly normal datasets, the \gls{bnp} method results in an estimate for the \gls{hc5} comparable to that of the normal model. When the datasets strongly deviate from the normal model, the \gls{hc5} estimates from the normal and \gls{bnp} model differ substantially and strongly support the use of the more flexible \gls{bnp} model over the normal.

\subsection{Contaminant-wise clustering}

We have so far demonstrated the advantages of the \gls{bnp} method over the existing approaches to \gls{ssd} for density estimation and for the determination of the \gls{hc5}.
There is an additional benefit connected to the \gls{bnp}-\gls{ssd}: the mixture model induces a clustering of the species, which conveys interesting information from the biological point of view.
Indeed, one of the long-standing questions around \gls{ssd}, and ecotoxicology in general, is to understand what drives the sensitivity of species to a contaminant.
\cite{PeterCraig2013} assumes that taxonomy is a driving factor and effectively imposes a clustering based on taxonomic units. \cite{de2001observed} investigates the influence of habitat by comparing freshwater and saltwater species, while \cite{Kefford2012a} study the variations in sensitivity in different regions of the world.

All these approaches start from a possible clustering structure and test for a significant difference among cluster units.
{The \gls{bnp}-\gls{ssd} takes the opposite path by endogenizing the clustering in a probabilistically principled way. Indeed, it allows} the clustering structure to emerge from the data, and, by using meta-data about the species, this structure can be examined a posteriori to verify whether it matches {certain scientific hypotheses} %our intuitions 
about the driving forces behind species sensitivity.

\begin{figure}
    \centering
    \includegraphics[width=\textwidth]{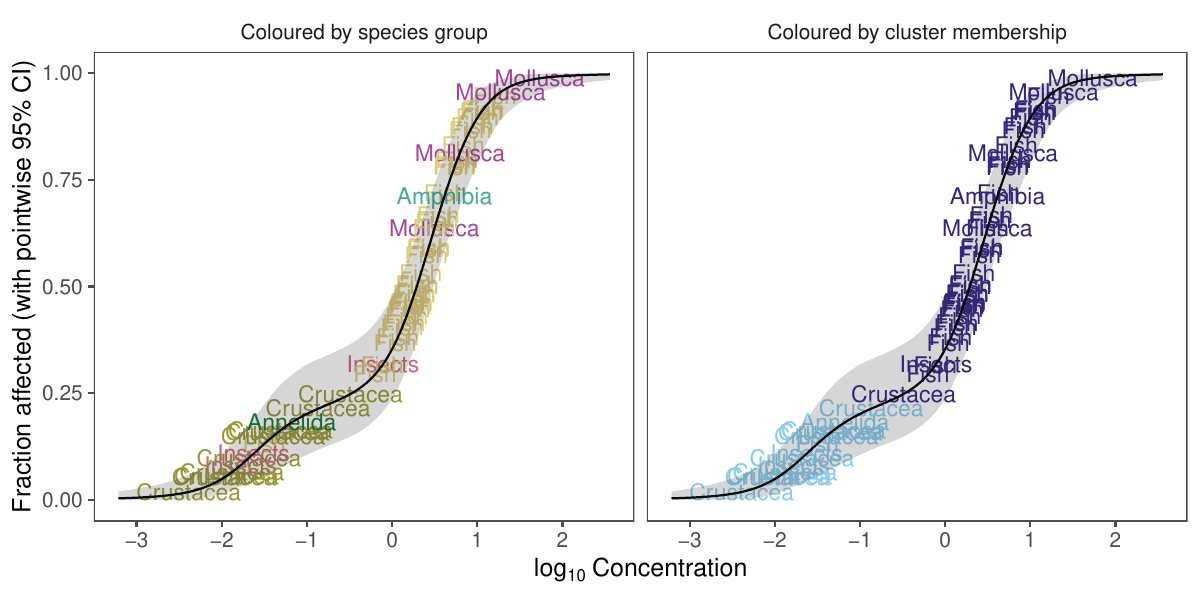}
    \caption{\gls{ssd} for Carbaryl (CAS: 63-25-2) with the quasi-taxonomic group of each species overlaid on the estimate of the cumulative distribution function (solid line). Left: Species coloured by quasi-taxonomic group. Right: Species coloured by cluster membership in the \gls{bnp} model. 
    \update{Light grey denotes $ 95\%$-pointwise credible bands computed from the posterior distribution of the BNP model.}
    \label{Illustration_clustering}}
\end{figure}

The clustering induced by the \gls{bnp} model may or may not coincide with particular information about the species, which can challenge or support existing theories about the determinants of species sensitivity.
\update{Figure~\ref{Illustration_clustering} compares the cumulative distribution functions, along with credible bands, as well as the estimated clustering structure for Carbaryl and a quasi-taxonomic grouping expected to be relevant for species sensitivity in \cite{de2001observed}}. %\update{(Figure~\ref{fig:Illustration_clustering_with_credible_bands} with credible bands around the cumulative distribution function is available in the Supplement)}. %The clustering was obtained from the MCMC sample using a greedy algorithm \cite{Wade2015}
Two clusters emerge, one predominantly composed of crustaceans and containing all crustaceans but one, and another predominantly composed of fishes, containing all fishes but one, and all the mollusks.
The three insects are scattered over the two clusters; the only annelid is grouped with the crustaceans, while the only amphibian is grouped with the fishes.
Thus, for Carbaryl, the estimated clustering structure seems strongly associated with the quasi-taxonomic grouping and supports the theory that species sensitivity is dependent on taxonomy, with fish forming a cluster relatively resilient to Carbaryl while crustaceans form a more sensitive cluster.
However, this parallel between taxonomy and sensitivity is not observed for every contaminant; indeed, it is possible to identify contaminants for which the estimated clustering does not match the quasi-taxonomic grouping. A general tendency observed over many contaminants is that fishes tend to group in a single cluster. This insight could be used to argue for reducing the number of fish species to be tested, as their contribution to the complete \gls{ssd} could be emulated by giving a higher weight to a few representative species, as done in \cite{Garnier-laplace2006}.
% This observation is specific to Fenthion, it is possible to find contaminants for which the estimated clustering does not match the quasi taxonomic grouping. \textcolor{red}{Check that again after new fit}.

\subsection{BNP-SSD Shiny application}
The method described above can be used directly with the \texttt{BNPdensity} package, but this requires a certain level of fluency in the R language.
Thus, we developed a Shiny application, named BNP-SSD, tailored to SSD problems and based on the functions of the package \texttt{BNPdensity}, available at \url{https://alamichl.shinyapps.io/BNP_SSD/}.
This application is inspired by the application \texttt{shinyssdtools} of \cite{dalgarno2021shinyssdtools}.

In this application, the BNP model described in Section~\ref{sec:methods} is fitted to censored or uncensored data.
Before fitting the model, the concentration data are cleaned, dealing with the possible presence of multiple \glspl{cec} values for one species, transformed using a $\log$-scale, and centered-scaled.
For greater flexibility, some options are left to the user, such as the number of iterations of the MCMC algorithm.
Once the model has been fitted, the estimated density is plotted, along with some goodness-of-fit graphs.
In another panel, an estimate of \gls{hc5} is made using the posterior distribution over quantiles. It is also possible to estimate a percentile of the distribution other than the $5$th percentile.
The credible bands of this estimate are also computed.
Finally, in the last panel, the induced optimal clustering is computed and plotted.

\section{Cross-contaminant clustering}
\label{sec:res_clus}
%%%% Explain the post-processing procedure

It would be highly interesting to establish whether similar patterns occur commonly for contaminants by studying the clustering structure for all contaminants in the dataset. A complete clustering analysis would require a hierarchical model with a contaminant effect, which is beyond the scope of the present paper and will be the object of future work.
Here we approach the issue in a simple yet insightful way by fitting the model independently for all contaminants.
Consequently, we present a post-processing of the clustering structure estimated for each contaminant.
The general idea is, over all contaminants, to assess how often each pair of species is grouped.
This defines sensitivity communities of species, which we compare to the quasi-taxonomic grouping.

We restrict ourselves to contaminants tested by at least eight species, which is a little below the minimum threshold recommended for fitting a Species Sensitivity Distribution \citep{ECHA2008}.
\update{Our original dataset, described in section \ref{sec:data}, contains more than 3,000 contaminants. Of these contaminants, after excluding those that have been tested on fewer than eight species, only $179$ remain.}
We first fit the model on all such contaminants.
We then combine the information from the clustering for each contaminant to understand if some common patterns may be observed.
To extract information from the clustering structure for each contaminant, we transform each estimated clustering into an association matrix.
Stacking all the association matrices on top of each other forms a three-dimensional array, also called a tensor, each slice corresponding to a contaminant. %This tensor is denoted by $\mathbf{X}$. 
One difficulty is that contaminants are tested on different sets of species, with potentially little overlap. This results in a large proportion of missing values (pairs of contaminants-species that have not been tested) that need to be dealt with.

\subsection{Non-negative tensor factorization}
%%%% Explain NTF
We perform non-negative three-way tensor factorization \citep{cichocki2009nonnegative}, which is a tensor generalization of principal component analysis.
It is a dimension-reduction technique that decomposes the association tensor into a sum of $R$ rank-one tensors.
Developed in Chemometrics, this technique has also been employed in Biostatistics, Signal Processing, Linguistics, and Machine Learning \citep{Gauvin2014}.
The technique also allows the imputation of missing values.

We use a Parallel Factors Analysis (PARAFAC), also referred to as Canonical Decomposition (CANDECOMP) factorization; Section~\ref{app:ntf} in the Supplement provides details on this technique and background on tensor properties.
Denoting by $\mathbf{Y}$ the tensor of the data described previously, we have that $\mathbf{Y}\in \mathbb{R}^{n_S\times n_S\times n_C}$ is a symmetric tensor in the first two dimensions, where $n_S$ and $n_C$, respectively, denote the number of species and contaminants.
The general PARAFAC factorization for some tensor $\hat{\mathbf{Y}}\in\mathbb{R}^{I\times J\times K}$ is denoted by
$$\hat{\mathbf{Y}} = [\![A,B,C]\!] =\sum_{r=1}^R a_r \circ b_r \circ c_r,$$
where $A=[a_1,\ldots,a_R] \in \mathbb{R}^{I\times R}$, $B=[b_1,\ldots,b_R] \in \mathbb{R}^{J\times R}$ and $C=[c_1,\ldots,c_R] \in \mathbb{R}^{K\times R}$ are three components or factors matrices, and $\circ$ stands for the vector outer product.
For the considered data, the symmetry of the tensor $\mathbf{Y}$ in the first two dimensions implies that the PARAFAC factorization can be simplified as $\mathbf{Y} = [\![A,A,C]\!]$, where $A=[a_1,\ldots,a_R] \in \mathbb{R}^{n_S\times R}$ and $C=[c_1,\ldots,c_R] \in \mathbb{R}^{n_C\times R}$.
Note that the factorization is only an approximation and incurs in some additive error $\mathbf{E}$ in the form of $\mathbf{Y} = [\![A,A,C]\!]  + \mathbf{E}$.
To give physical meaning to the different components found, we use the non-negative PARAFAC factorization \citep{xu2013block} from the \texttt{multiway} R package \citep{leeuw_multiway_2011}. This adds non-negativity constraints on the component matrices  $a_{ir}\in\mathbb{R}_+$ and $c_{jr}\in\mathbb{R}_+$ for all $i\in\{1,\ldots,n_S\}$, $j\in\{1,\ldots,n_C\}$ and $r\in\{1,\ldots,R\}$.
% \citep{harshman1970foundations,carroll1970analysis}
%\texttt{TensorDecomposition}\footnote{https://github.com/yunjhongwu/TensorDecompositions.jl} package of the Julia \citep{BEKS14} language.

%%% Explain selection methods
A popular heuristic to determine the number of components $R$ is the core-consistency diagnostic \citep{Bro2003}. This diagnostic requires the imputation of the missing values for efficient computation.
As the number of missing values in our type of data is large, we used instead a cross-validation method.
The cross-validation consists of removing a chosen proportion of the tensor non-missing values, performing the decomposition for different ranks, and then evaluating the reconstruction error on the removed values, a type of K-fold cross-validation. We measure the reconstruction error using the Frobenius distance between this tensor and the original one on the non-missing values (see Figure~\ref{CrossV}).

The decomposition can be performed once the rank of the decomposition is chosen.
The result of the three-way decomposition consists of three factor matrices, two of which with dimension $n_S \times R$, and one with dimension $n_C \times R$.
The first two encode the degree of membership of each species to each component of the decomposition and are equal by construction.
The third matrix encodes, for each contaminant, its degree of membership to each component.
To facilitate the interpretation of the results, we threshold the membership degrees and decide whether each species and contaminant belongs to a component or not. To do this, we use K-means clustering on the degree of membership vectors to adaptively threshold the membership degrees. Species and contaminants may be allocated to 0 or several components (see Figure~\ref{Kmeans} and Figure~\ref{Kmeans_spec}).

\update{This algorithmic approach is chosen for its ability to handle missing data. In tensor factorization, various contributions \citep[e.g.][]{rai_scalable_2014} propose a model-based approach that avoids the problem of choosing the rank of the factorization. However, in practice, when the proportion of missing data is very high, as in our case, these methods fail.}

\subsection{Results}
%%% The results

\begin{figure}[ht]
    \begin{minipage}[b]{0.49\linewidth}
        \centering \includegraphics[page=1, width=\textwidth]{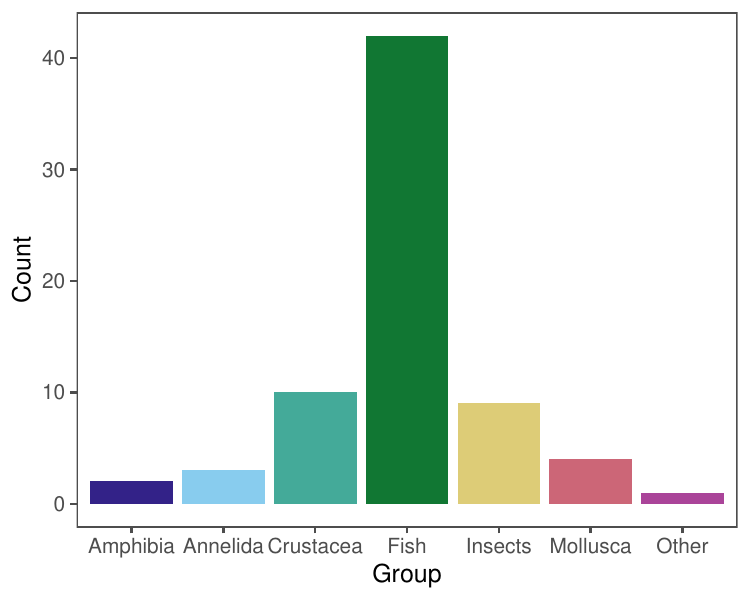}
    \end{minipage}\hfill
    \begin{minipage}[b]{0.49\linewidth}
        \centering
        \includegraphics[page=1, width=\textwidth]{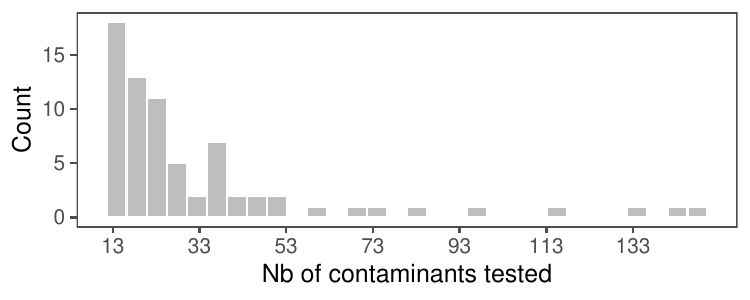}
        \includegraphics[page=3, width=\textwidth]{figures/contaminant_data.pdf}
    \end{minipage}
    \caption{Left: Quasi-taxonomic composition of the species in the part of the data considered. The groups are defined according to the classification in \cite{de2001observed}.
        Right: (Top) Number of contaminants tested for each species in the data considered (at least 13 contaminants by species). (Bottom) Number of species tested for each contaminant.}
    \label{fig:data}
\end{figure}
\begin{figure}[h]
    \centering
    \includegraphics[width = \textwidth]{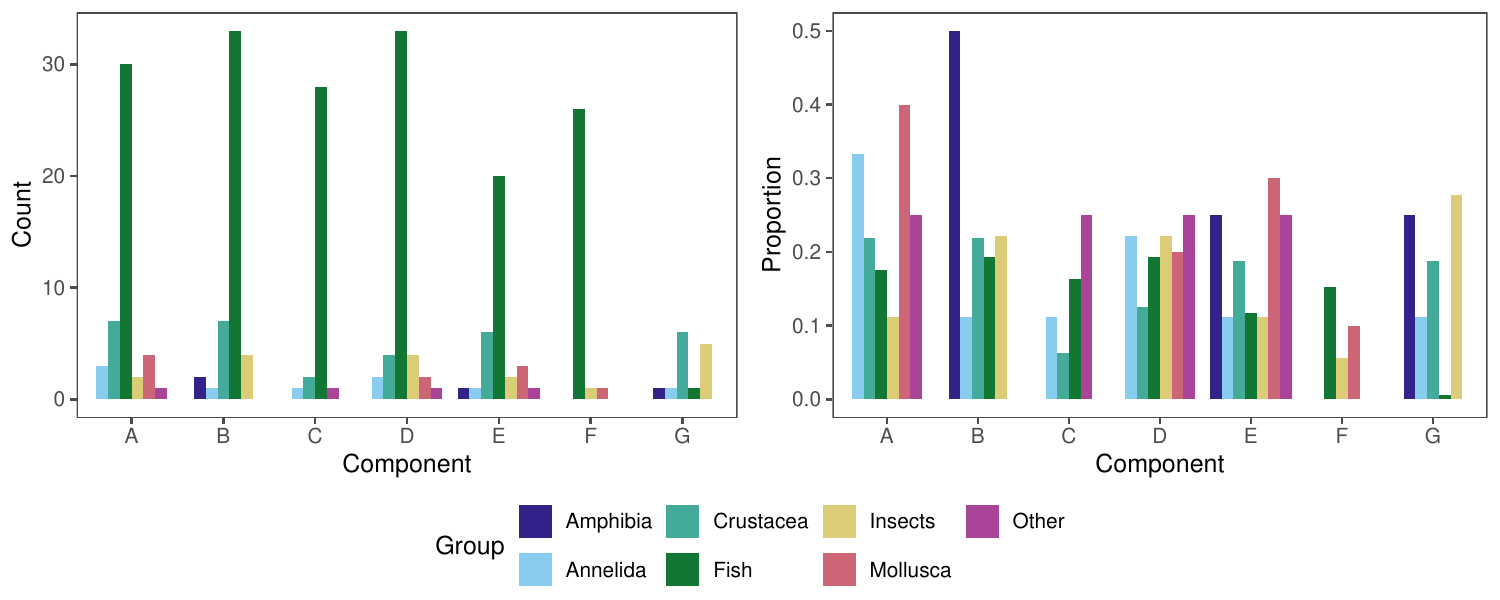}
    \caption{Quasi taxonomic composition of the components. The groups are defined according to the classification in \cite{de2001observed}.
        Left: Number of species in each component. Right: Proportion of species compared to the distribution of species in the whole data in each component.}
    \label{fig:comp_species}
\end{figure}
\begin{figure}[ht]
    \centering
    \includegraphics[width = 0.99\textwidth]{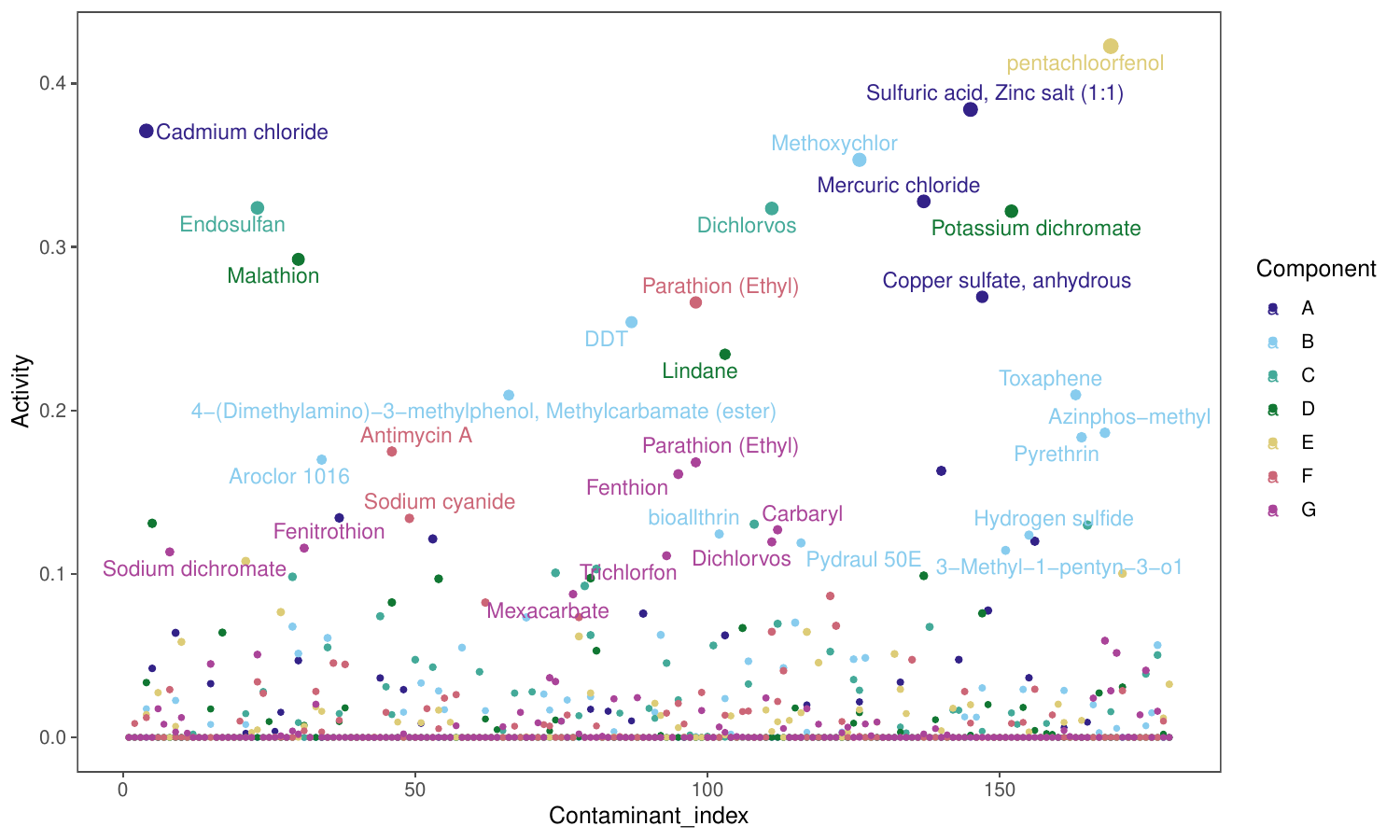}
    \caption{Weight of each contaminant in the various components. The size of the point is proportional to the intensity of the activity of the contaminants.}
    \label{fig:comp_cont}
\end{figure}

We used this methodology on a part of the dataset described in Section \ref{sec:data} (see Figure~\ref{fig:data} for a description of the restricted dataset).
To lower the proportion of missing data, \update{we arbitrarily considered only species for which at least 13 contaminants were tested; with this restriction, 95\% of the data is missing. The tensor on which the factorization is performed has dimensions $179\times71\times71$, so we have $179$ co-clustering matrices of $71$ species, one matrix for each contaminant.}
Using the cross-validation approach, the rank $41$ was selected (see Figure~\ref{CrossV}). We obtained $41$ components of the contaminants and the species. We present seven components among these, selecting those with the highest contrast using Figure~\ref{Kmeans} and Figure~\ref{Kmeans_spec}, deciding to only select the well-separated components in the sense of K-means clustering.
These seven components are analyzed in Figure~\ref{fig:comp_species} and Figure~\ref{fig:comp_cont}.

Figure \ref{fig:comp_species} presents the quasi-taxonomic composition of the seven components in count (left) and in relative proportion (right). Figure~\ref{fig:maj} and Figure~\ref{fig:phyl} present the same result for other taxonomic ranks.
We cannot observe a clear difference in composition among the different components.
This suggests that quasi-taxonomy does not appear to be the main driver for species to be co-clustered, or in other words, quasi-taxonomy does not appear to strongly determine species sensitivity.
This observation should be modulated by the fact that, as we see in Figure \ref{fig:data}, fish species are over-represented in the dataset, so they form a substantial part of every cluster.

Figure \ref{fig:comp_cont} presents the weights of each contaminant in the various components.
We can observe that only one contaminant (Pentachlorophenol) has a significant weight in component E.
In components C, D, and G, the most weighted contaminants are mostly pesticides. 
More precisely, component C is composed of two insecticides. 
Component D contains two neurotoxic insecticides and potassium dichromate, which is an oxidizing agent used in various reactions in laboratories and industry.
Component G contains seven insecticides, including five organophosphate insecticides and two carbamate insecticides, along with sodium dichromate, which is mainly used as an intermediate in the production of other chromium compounds.
In component A, the four most weighted contaminants are inorganic compounds: two chlorinated compounds and two sulfated compounds. 
Finally, components B and F are composed of contaminants with different properties or usages. 
Component B includes mostly pesticides, with six insecticides and one ectoparasiticide, but also four compounds with various uses, such as a flame retardant or a reactant in chemical synthesis.
Component F is composed of two pesticides, including one insecticide and one piscicide, and sodium cyanide, which is mainly used in gold extraction but was also historically applied as an insecticide.
Together, these observations suggest that components are associated to contaminants of a similar type, or, in other words, that species which respond similarly to one type of contaminant could tend to respond similarly to another contaminant of the same type.
This could be seen as providing support to the idea that species sensitivity across contaminants may be correlated, which has been used for instance in \cite{Awkerman2008}.
\section{Discussion and future research\label{sec:discussion}}

We presented a novel approach to \gls{ssd} based on a Bayesian nonparametric mixture model. We performed an extensive comparison to the current \gls{ssd} models both on simulated and on real datasets, to demonstrate the added value of the proposed approach. 
The \gls{bnp}-\gls{ssd} performs particularly well when the dataset deviates from a log-normal distribution, {which allows to leverage its great flexibility in describing the data. 
At the same time,} the proposed approach turns out to be relatively robust and does not seem prone to over-fitting. 
The \gls{bnp}-\gls{ssd} can be thought of as an intermediate model between the single component log-normal-\gls{ssd}, and the \gls{kde} with as many components as there are species. 
Indeed, in all practical cases of the \gls{rivm} dataset, the number of clusters necessary to describe the data was no greater than 3. 

The \gls{bnp}-\gls{ssd} provides several benefits for risk assessment: it is an effective and robust standard model that adapts to many datasets. As such, the \gls{bnp}-\gls{ssd} represents a safe tool to remove one of the arbitrary parametric assumptions of \gls{ssd} \citep{Forbes2002}. 
Moreover, as a Bayesian method, it readily provides credible intervals.

%We could have chosen to base the \gls{bnp}-\gls{ssd} on P\'olya tree processes or on mixtures of Dirichlet processes. 
%We chose to base the \gls{bnp}-\gls{ssd} on \gls{nrmi} rather than on the more common Dirichlet process, because it is more robust in case of misspecification in the number of clusters \citep{barrios2013modeling}. \tm{the previous two sentences are a bit redundant here, maybe i would cut or anticipate them to the 'methods' section where we discuss the BNP prior}
The traditional approach to \gls{ssd} is to consider contaminants independently. 
In a context of data scarcity, only exacerbated by the drive to reduce animal testing, it would be desirable to leverage the long history of ecotoxicity testing to borrow information from experiments regarding different contaminants. 
Large databases are already available. 
We can hope that the ongoing discussion about transparency on the regulation of chemicals will even push towards greater public availability of data. 
Therefore, it is a timely effort to develop models that harness all the information already available about species' sensitivity to other contaminants.
This is, in essence, the proposal brought forward by \cite{Awkerman2008} and \cite{PeterCraig2013} which use taxonomic information to predict the sensitivity of unknown species to a contaminant. 
An important by-product of the \gls{bnp}-\gls{ssd} approach is that it provides interesting opportunities for subsequent cluster analysis.
We presented such an example of a post-processing analysis using non-negative tensor factorization, summarizing insights over 179 contaminants, which suggested some regularities in species sensitivity based on contaminant type. 
However, we did not observe a strong relation between taxonomy and species sensitivity across contaminants. 
As such, we may hypothesize that there could be structures other than taxonomic which would be relevant to species sensitivity, such as ecological niche.

This idea motivates a natural extension of the present work: the BNP-SSD approach could be expanded to model inter-contaminant variation by leveraging a dependent BNP model. This is particularly relevant because some contaminants can exhibit very similar toxicity, often due to belonging to the same chemical class. There exists a rich variety of dependent BNP models that employ different mechanisms to induce dependence, including hierarchical \citep{teh2006hierarchical, camerlenghiDistributionTheoryHierarchical2019}, additive \citep{mueller2004,lijoi2014}, nested \citep{rodriguez2008,camerlenghi2019}, or combinations thereof. \update{These have recently been unified within a general framework in \cite{flpr2025}. Future work will focus on an in-depth investigation to determine which dependence structure, and corresponding sharing of information, is most appropriate for the SSD context.}

%This idea motivates a natural extension of the present work: the \gls{bnp}-\gls{ssd} could be augmented to model inter-contaminant variation in a hierarchical model. This is particularly relevant as some contaminants can have very similar toxicity because they belong to the same class of chemicals. A variety of Bayesian nonparametric hierarchical models already exist \citep{teh2006hierarchical, camerlenghi2018distribution}, and have demonstrated their usefulness and applicability in various contexts, so the tools necessary for this extension {would only need some tailoring}. 

Moreover, as discussed when studying the species clustering, the groups could be specified either based on the known chemical nature of the contaminants or learned in a flexible manner using a Bayesian nonparametric approach. This would open the door to a principled investigation of similarities among potentially very different contaminants. 
Additionally, one would also like to move forward from summarizing the sensitivity of species by a single value. Ecotoxicological tests are usually analyzed by fitting a dose-response model which describes several aspects of the species' response to a contaminant, such as the time between exposure and effect. Using all parameters from the dose-response model would imply performing the clustering in a higher dimensional space, increasing the discriminating power between groups and ultimately resulting in more meaningful clusters.

% \textcolor{red}{Extend with inter-contaminant modelling in a hierarchical model. Definitely makes sense because contaminants are related.}

% Future work to support the \gls{bnp}-\gls{ssd} will include a comparison of methods on simulated data, an extension to the case of censored data and an emphasis on the potential benefits of the approach from a biological point of view.

%\ju{Mention also the idea of working with additional covariates, with the extra Binomial layer, again with BNP infinite mixtures.}

% \begin{itemize}
%  \item will also work for censored data
%  \item clustering of species ? \begin{itemize}
%  \item No sure there is consistency but probably some clusters are meaningful
%  \item Probably more relevant for large datasets
% \end{itemize}
%  \item confidence intervals
%  \item good standard model, will work for most distributions and easy to use once wrapped in a tool.
% \end{itemize}

% Why choose NRMI over DPM ?
% 
% \begin{itemize}
%  \item More robust to model misspecification in the number of clusters
% \end{itemize}
% 
% Reasonably good when distribution looks normal ?

% Seem generally more protective, but check on more datasets.
% 
% Maybe want to assess fit in the left-tail region ?
% 
% Future work on simulated data to evaluate bias, coverage ?
% 

\section*{Acknowledgements}
L. Alamichel is partially supported by the LabEx PERSYVAL-Lab (ANR-11-LABX-0025-01) funded by the French program Investissement d'avenir. 
J. Arbel is partially supported by ANR-21-JSTM-0001 grant.
I. Pr\"unster is partially supported European Union--NextGenerationEU PRIN-PNRR (project P2022H5WZ9). 

% %\section*{Acknowledgement}
% %J. Arbel and I. Pr\"unster are supported by the European Research Council (ERC) through StG ``N-BNP'' 306406.

% %\bibliographystyle{spbasic}
% \bibliographystyle{apalike}
\bibliographystyle{rss}
%\bibliography{biblio_duplicates_free,biblio_saved}
\bibliography{biblio_duplicates_free,biblio_julyan}

@article{arbel2017moment,
  title   = {A Moment-Matching Ferguson \& Klass Algorithm},
  author  = {Arbel, Julyan and Pr{\"u}nster, Igor},
  year    = {2017},
  journal = {Statistics and Computing},
  volume  = {27},
  number  = {1},
  pages   = {3--17},
  doi     = {10.1007/s11222-016-9676-8}
}

@article{camerlenghiDistributionTheoryHierarchical2019,
  title      = {Distribution {{Theory}} for {{Hierarchical Processes}}},
  author     = {Camerlenghi, Federico and Lijoi, Antonio and Orbanz, Peter and Pr{\"u}nster, Igor},
  year       = {2019},
  journal    = {The Annals of Statistics},
  volume     = {47},
  number     = {1},
  eprint     = {26581840},
  eprinttype = {jstor},
  pages      = {67--92},
  publisher  = {Institute of Mathematical Statistics},
  issn       = {0090-5364},
  urldate    = {2025-03-11}
}

@article{dalgarno2021shinyssdtools,
  title   = {{shinyssdtools: A web application for fitting Species Sensitivity Distributions (SSDs)}},
  author  = {Dalgarno, Seb},
  journal = {Journal of Open Source Software},
  volume  = {6},
  number  = {57},
  pages   = {2848},
  year    = {2021}
}

@article{VanStraalen2002,
  author   = {{Van Straalen}, Nico M.},
  isbn     = {3120444707},
  issn     = {13826689},
  journal  = {Environmental Toxicology and Pharmacology},
  month    = {jul},
  number   = {3-4},
  pages    = {167--172},
  pmid     = {21782599},
  title    = {{Threshold models for species sensitivity distributions applied to aquatic risk assessment for zinc}},
  volume   = {11},
  year     = {2002}
}

@article{Awkerman2008,
  author   = {Awkerman, Jill A. and Raimondo, Sandy and Barron, Mace G},
  isbn     = {0013-936X},
  issn     = {0013936X},
  journal  = {Environmental Science and Technology},
  month    = {may},
  number   = {9},
  pages    = {3447--3452},
  pmid     = {18522132},
  title    = {{Development of species sensitivity distributions for wildlife using interspecies toxicity correlation models}},
  volume   = {42},
  year     = {2008}
}

@article{VanderHoeven2001,
  author   = {{Van Der Hoeven}, N.},
  issn     = {09639292},
  journal  = {Ecotoxicology},
  month    = {feb},
  number   = {1},
  pages    = {25--34},
  pmid     = {11227815},
  title    = {{Estimating the 5-percentile of the species sensitivity distributions without any assumptions about the distribution}},
  volume   = {10},
  year     = {2001}
}

@article{coda,
  author   = {Plummer, Martyn and Best, Nicky and Cowles, Kate and Vines, Karen},
  isbn     = {1609-3631},
  issn     = {1662-4025},
  journal  = {R News},
  number   = {March},
  pages    = {7--11},
  pmid     = {21196786},
  title    = {{CODA: convergence diagnosis and output analysis for MCMC}},
  volume   = {6},
  year     = {2006}
}

@article{He2014,
  author    = {He, Wei and Qin, Ning and Kong, Xiangzhen and Liu, Wenxiu and Wu, Wenjing and He, Qishuang and Yang, Chen and Jiang, Yujiao and Wang, Qingmei and Yang, Bin and Xu, Fuliu},
  issn      = {1470160X},
  journal   = {Ecological Indicators},
  pages     = {209--218},
  publisher = {Elsevier Ltd},
  title     = {{Ecological risk assessment and priority setting for typical toxic pollutants in the water from Beijing-Tianjin-Bohai area using Bayesian matbugs calculator (BMC)}},
  volume    = {45},
  year      = {2014}
}

@incollection{ECHA2008,
  address   = {Helsinki},
  author    = {ECHA},
  booktitle = {Guidance on information requirements and chemical safety assessment},
  chapter   = {R.10},
  number    = {May},
  publisher = {European Chemicals Agency},
  title     = {{Characterisation of dose concentration-response for environment}},
  year      = {2008}
}

@article{Zhao2016,
  author   = {Zhao, Jinsong and Chen, Boyu},
  issn     = {1090-2414},
  journal  = {Ecotoxicology and Environmental Safety},
  month    = {mar},
  pages    = {161--9},
  pmid     = {26701839},
  title    = {{Species sensitivity distribution for chlorpyrifos to aquatic organisms: Model choice and sample size.}},
  volume   = {125},
  year     = {2016}
}

@incollection{canada_guidelines,
  address   = {Winnipeg},
  author    = {CCME},
  booktitle = {Canadian Environmental Quality Guidelines},
  number    = {Ccme 1991},
  publisher = {Canadian Council of Ministers of the Environment},
  title     = {{A protocol for the derivation of water quality guidelines for the protection of aquatic life}},
  year      = {2007}
}

@article{ferguson1972representation,
  author    = {Ferguson, Thomas and Klass, Michael},
  issn      = {0003-4851},
  journal   = {The Annals of Mathematical Statistics},
  number    = {5},
  pages     = {1634--1643},
  publisher = {JSTOR},
  title     = {{A Representation of Independent Increment Processes without Gaussian Components}},
  volume    = {43},
  year      = {1972}
}

@article{SuterIi1999,
  author   = {{Suter II}, Glenn W and Barnthouse, Lawrence W and Efroymson, Rebecca A and Jager, Henriette},
  issn     = {0730-7268},
  journal  = {Environmental Toxicology and Chemistry},
  number   = {4},
  pages    = {589--598},
  title    = {{Ecological Risk Assessment in a Large River–Reservoir: 2. Fish Community}},
  volume   = {18},
  year     = {1999}
}

@article{teh2006hierarchical,
  archiveprefix = {arXiv},
  arxivid       = {arXiv:1210.6738v2},
  author        = {Teh, Y. and Jordan, M. and Beal, Matthew J. and Blei, David M.},
  eprint        = {arXiv:1210.6738v2},
  isbn          = {0162-1459},
  issn          = {0162-1459},
  journal       = {J. Am. Stat. Assoc.},
  number        = {476},
  pages         = {1566--1581},
  pmid          = {242869700023},
  publisher     = {ASA},
  title         = {{Hierarchical Dirichlet processes}},
  volume        = {101},
  year          = {2006}
}

@article{Rastelli2018,
  title    = {Optimal {{Bayesian}} Estimators for Latent Variable Cluster Models},
  author   = {Rastelli, Riccardo and Friel, Nial},
  year     = {2018},
  month    = nov,
  journal  = {Statistics and Computing},
  volume   = {28},
  number   = {6},
  pages    = {1169--1186},
  issn     = {1573-1375},
  doi      = {10.1007/s11222-017-9786-y}
}

@article{Xu2015,
  author    = {Xu, Fu-Liu and Li, Yi-Long and Wang, Yin and He, Wei and Kong, Xiang-Zhen and Qin, Ning and Liu, Wen-Xiu and Wu, Wen-Jing and Jorgensen, Sven Erik},
  issn      = {1470160X},
  journal   = {Ecological Indicators},
  pages     = {227--237},
  publisher = {Elsevier Ltd},
  title     = {{Key issues for the development and application of the species sensitivity distribution (SSD) model for ecological risk assessment}},
  volume    = {54},
  year      = {2015}
}

@article{Gauvin2014,
  archiveprefix = {arXiv},
  arxivid       = {arXiv:1308.0723v3},
  author        = {Gauvin, Laetitia and Panisson, Andr{\'{e}} and Cattuto, Ciro},
  eprint        = {arXiv:1308.0723v3},
  issn          = {19326203},
  journal       = {PLoS ONE},
  number        = {1},
  pmid          = {24497935},
  title         = {{Detecting the community structure and activity patterns of temporal networks: A non-negative tensor factorization approach}},
  volume        = {9},
  year          = {2014}
}

@book{cichocki2009nonnegative,
  author    = {Cichocki, Andrzej and Zdunek, Rafal and Phan, Anh Huy and Amari, Shun-ichi},
  publisher = {John Wiley {\&} Sons},
  title     = {{Nonnegative matrix and tensor factorizations: applications to exploratory multi-way data analysis and blind source separation}},
  year      = {2009}
}

@article{Garnier-laplace2006,
  author   = {Garnier-Laplace, Jacqueline and Della-Vedova, Claire and Gilbin, Rodolphe and Copplestone, David and Hingston, Joanne and Ciffroy, Philippe},
  isbn     = {0013-936X},
  issn     = {0013936X},
  journal  = {Environmental Science and Technology},
  month    = {oct},
  number   = {20},
  pages    = {6498--6505},
  pmid     = {17120586},
  title    = {{First derivation of predicted-no-effect values for freshwater and terrestrial ecosystems exposed to radioactive substances}},
  volume   = {40},
  year     = {2006}
}

@article{gelfand1996model,
  author    = {Gelfand, Alan E.},
  journal   = {Markov chain Monte Carlo in practice},
  pages     = {145--161},
  publisher = {London: Chapman and Hall},
  title     = {{Model determination using sampling-based methods}},
  year      = {1996}
}

@article{Forbes2002,
  author   = {Forbes, Valery E. and Calow, Peter},
  isbn     = {1080-7039},
  issn     = {1080-7039},
  journal  = {Human and Ecological Risk Assessment},
  number   = {3},
  pages    = {473--492},
  title    = {{Species Sensitivity Distributions Revisited: A Critical Appraisal}},
  volume   = {8},
  year     = {2002}
}

@incollection{lijoi2010models,
  author    = {Lijoi, Antonio and Pr{\"{u}}nster, Igor},
  booktitle = {Bayesian nonparametrics},
  editor    = {Hjort, N L and Holmes, C C and M{\"{u}}ller, P and Walker, S G},
  pages     = {80},
  publisher = {Cambridge University Press, Cambridge},
  title     = {{Models beyond the Dirichlet process}},
  volume    = {28},
  year      = {2010}
}

@incollection{Jordan2010hierarchical,
  address       = {New York},
  archiveprefix = {arXiv},
  arxivid       = {arXiv:1312.6184v5},
  author        = {Jordan, Michael I.},
  booktitle     = {Frontiers of statistical decision making and Bayesian analysis: In honor of James O. Berger.},
  eprint        = {arXiv:1312.6184v5},
  isbn          = {9781441969446},
  pages         = {207--218},
  publisher     = {Springer},
  title         = {{Hierarchical Models, Nested Models and Completely Random Measures}},
  year          = {2010}
}

@article{Wang2015,
  author   = {Wang, Ying and Wu, Fengchang and Giesy, John P. and Feng, Chenglian and Liu, Yuedan and Qin, Ning and Zhao, Yujie},
  isbn     = {1135601546028},
  issn     = {16147499},
  journal  = {Environmental Science and Pollution Research},
  number   = {18},
  pages    = {13980--13989},
  title    = {{Non-parametric kernel density estimation of species sensitivity distributions in developing water quality criteria of metals}},
  volume   = {22},
  year     = {2015}
}

@article{Chen2004,
  author   = {Chen, Ling},
  issn     = {03783758},
  journal  = {Journal of Statistical Planning and Inference},
  month    = {jul},
  number   = {2},
  pages    = {243--258},
  title    = {{A conservative, nonparametric estimator for the 5th percentile of the species sensitivity distributions}},
  volume   = {123},
  year     = {2004}
}

@incollection{liu2014setting,
  author    = {Liu, Yuedan and Wu, Fengchang and Mu, Yunsong and Feng, Chenglian and Fang, Yixiang and Chen, Lulu and Giesy, John P.},
  booktitle = {Reviews of Environmental Contamination and Toxicology volume},
  pages     = {35--57},
  publisher = {Springer},
  title     = {{Setting Water Quality Criteria in China: Approaches for Developing Species Sensitivity Distributions for Metals and Metalloids}},
  year      = {2014}
}

@article{Hickey2012,
  author   = {Hickey, Graeme L. and Craig, Peter S. and Luttik, Robert and de Zwart, Dick},
  issn     = {07307268},
  journal  = {Environmental Toxicology and Chemistry},
  month    = {aug},
  number   = {8},
  pages    = {1903--1910},
  pmid     = {22619109},
  title    = {{On the quantification of intertest variability in ecotoxicity data with application to species sensitivity distributions}},
  volume   = {31},
  year     = {2012}
}

@article{Rodriguez2008,
  author   = {Rodr{\'{i}}guez, Abel and Dunson, David B. and Gelfand, Alan E.},
  issn     = {0162-1459},
  journal  = {Journal of the American Statistical Association},
  number   = {483},
  pages    = {1131--1154},
  title    = {{The Nested Dirichlet Process}},
  volume   = {103},
  year     = {2008}
}

@book{silverman1986density,
  author    = {Silverman, Bernard W},
  publisher = {CRC press},
  title     = {{Density estimation for statistics and data analysis}},
  volume    = {26},
  year      = {1986}
}

@article{Zajdlik2009,
  author  = {Zajdlik, BA A and Dixon, DG G and Stephenson, G},
  journal = {Human and Ecological Risk Assessment},
  number  = {3},
  pages   = {554--564},
  title   = {{Estimating Water Quality Guidelines for Environmental Contaminants Using Multimodal Species Sensitivity Distributions: A Case Study with Atrazine}},
  volume  = {15},
  year    = {2009}
}

@article{rlp2003,
  author    = {Regazzini, Eugenio and Lijoi, Antonio and Pr{\"{u}}nster, Igor},
  issn      = {00905364},
  journal   = {Annals of Statistics},
  number    = {2},
  pages     = {560--585},
  publisher = {Institute of Mathematical Statistics},
  title     = {{Distributional results for means of normalized random measures with independent increments}},
  volume    = {31},
  year      = {2003}
}

@article{Roux1996,
  author   = {Roux, D. J. and Jooste, S. H J and MacKay, H. M.},
  isbn     = {0038-2353},
  issn     = {00382353},
  journal  = {South African Journal of Science},
  number   = {4},
  pages    = {198--206},
  title    = {{Substance-specific water quality criteria for the protection of South African freshwater ecosystems: Methods for derivation and initial results for some inorganic toxic substances}},
  volume   = {92},
  year     = {1996}
}

@article{KonKamKing2014,
  author   = {{Kon Kam King}, Guillaume and Veber, Philippe and Charles, Sandrine and Delignette-Muller, Marie Laure},
  issn     = {15528618},
  journal  = {Environmental Toxicology and Chemistry},
  month    = {sep},
  number   = {9},
  pages    = {2133--2139},
  pmid     = {24863265},
  title    = {{MOSAIC{\_}SSD: A new web tool for species sensitivity distribution to include censored data by maximum likelihood}},
  volume   = {33},
  year     = {2014}
}

@article{kingman1967completely,
  author   = {Kingman, J F C},
  issn     = {0030-8730},
  journal  = {Pacific Journal Of Mathematics},
  number   = {1},
  pages    = {59--78},
  title    = {{Completely random measures}},
  volume   = {21},
  year     = {1967}
}

@techreport{armcanz2000australian,
  address     = {Canberra, Australia},
  author      = {ANZECC},
  institution = {Australian and New Zealand Environmental and Conservation Council Agriculture and Resource Management Council of Australia and New Zealand},
  title       = {{Australian and New Zealand guidelines for fresh and marine water quality}},
  volume      = {4},
  year        = {2000}
}

@article{Kefford2012a,
  author   = {Kefford, Ben J. and Hickey, Graeme L. and Gasith, Avital and Ben-David, Elad and Dunlop, Jason E. and Palmer, Carolyn G. and Allan, Kaylene and Choy, Satish C. and Piscart, Christophe},
  isbn     = {1932-6203},
  issn     = {19326203},
  journal  = {PLoS ONE},
  month    = {jan},
  number   = {5},
  pages    = {e35224},
  pmid     = {22567097},
  title    = {{Global scale variation in the salinity sensitivity of riverine macroinvertebrates: Eastern Australia, France, Israel and South Africa}},
  volume   = {7},
  year     = {2012}
}

@article{barrios2013modeling,
  archiveprefix = {arXiv},
  arxivid       = {arXiv:1310.0260v1},
  author        = {Barrios, Ernesto and Lijoi, Antonio and Nieto-Barajas, Luis E. and Pr{\"{u}}nster, Igor},
  eprint        = {arXiv:1310.0260v1},
  issn          = {0883-4237},
  journal       = {Statistical Science},
  number        = {3},
  pages         = {313--334},
  publisher     = {Institute of Mathematical Statistics},
  title         = {{Modeling with Normalized Random Measure Mixture Models}},
  volume        = {28},
  year          = {2013}
}

@article{craig2012species,
  archiveprefix = {arXiv},
  arxivid       = {arXiv:0811.2183v2},
  author        = {Craig, Peter S. and Hickey, Graeme L. and Luttik, Robert and Hart, Andy},
  eprint        = {arXiv:0811.2183v2},
  isbn          = {9783642121425},
  issn          = {09641998},
  journal       = {Journal of the Royal Statistical Society. Series A: Statistics in Society},
  number        = {1},
  pages         = {243--262},
  publisher     = {Wiley Online Library},
  title         = {{Species non-exchangeability in probabilistic ecotoxicological risk assessment}},
  volume        = {175},
  year          = {2012}
}

@article{Xing2014,
  author   = {Xing, Liqun and Liu, Hongling and Zhang, Xiaowei and Hecker, Markus and Giesy, John P. and Yu, Hongxia},
  issn     = {09441344},
  journal  = {Environmental Science and Pollution Research},
  number   = {1},
  pages    = {159--167},
  pmid     = {23314707},
  title    = {{A comparison of statistical methods for deriving freshwater quality criteria for the protection of aquatic organisms}},
  volume   = {21},
  year     = {2014}
}

@article{Jones1999,
  author   = {Jones, D S and Barnthouse, Lawrence W and {Suter II}, Glenn W and Efroymson, Rebecca A and Field, J M and Beauchamp, J J},
  isbn     = {0730-7268},
  journal  = {Environmental Toxicology and Chemistry},
  number   = {4},
  pages    = {599--609},
  title    = {{Ecological risk assessment in a large river-reservoir: 3. Benthic invertebrates}},
  volume   = {18},
  year     = {1999}
}

@article{lo1984class,
  author    = {Lo, Albert Y.},
  issn      = {0090-5364},
  journal   = {The Annals of Statistics},
  number    = {1},
  pages     = {351--357},
  publisher = {JSTOR},
  title     = {{On a Class of Bayesian Nonparametric Estimates: I. Density Estimates}},
  volume    = {12},
  year      = {1984}
}

@incollection{de2001observed,
  author    = {de Zwart, Dick},
  booktitle = {Species sensitivity distributions in ecotoxicology},
  publisher = {CRC Press},
  title     = {{Observed regularities in species sensitivity distributions for aquatic species}},
  year      = {2001}
}

@article{xu2013block,
  author    = {Xu, Yangyang and Yin, Wotao},
  journal   = {SIAM Journal on Imaging Sciences},
  number    = {3},
  pages     = {1758--1789},
  publisher = {SIAM},
  title     = {{A block coordinate descent method for regularized multiconvex optimization with applications to nonnegative tensor factorization and completion}},
  volume    = {6},
  year      = {2013}
}

@article{Kooijman1987a,
  author   = {Kooijman, S.A.L.M.},
  issn     = {00431354},
  journal  = {Water Research},
  number   = {3},
  pages    = {269--276},
  title    = {{A safety factor for LC 50 values allowing for differences in sensitivity among species}},
  volume   = {21},
  year     = {1987}
}

@article{Aldenberg2000a,
  author   = {Aldenberg, Tom and Jaworska, J S},
  isbn     = {0147-6513},
  issn     = {0147-6513},
  journal  = {Ecotoxicology and Environmental Safety},
  month    = {may},
  number   = {1},
  pages    = {1--18},
  pmid     = {10805987},
  title    = {{Uncertainty of the hazardous concentration and fraction affected for normal species sensitivity distributions.}},
  volume   = {46},
  year     = {2000}
}

@incollection{aldenberg2002normal,
  address   = {Boca Raton, FL},
  author    = {Aldenberg, Tom and Jaworska, Joanna S and Traas, Theo P},
  booktitle = {Species sensitivity Distribution in Ecotoxicology},
  editor    = {Posthuma, Leo and Suter, GW II and Traas, Theo P},
  isbn      = {1566705789},
  pages     = {49--102},
  publisher = {Lewis Publishers},
  title     = {{Normal Species Sensitivity Distributions and Probalistic Ecological Risk Assessment}},
  year      = {2002}
}

@article{Wang2008,
  author   = {Wang, Bin and Yu, Gang and Huang, Jun and Hu, Hongying},
  isbn     = {0963-9292},
  issn     = {0963-9292},
  journal  = {Ecotoxicology},
  month    = {nov},
  number   = {8},
  pages    = {716--724},
  pmid     = {18463978},
  title    = {{Development of species sensitivity distributions and estimation of HC(5) of organochlorine pesticides with five statistical approaches.}},
  volume   = {17},
  year     = {2008}
}

@book{Posthuma2010,
  author    = {Posthuma, Leo and {Suter II}, Glenn W and Trass, P Teo},
  booktitle = {Ecotoxicology},
  isbn      = {1566705789},
  pages     = {616},
  pmid      = {158},
  publisher = {CRC press},
  title     = {{Species sensitivity distributions in ecotoxicology}},
  year      = {2002}
}

@article{kingman1975random,
  author    = {Kingman, J},
  issn      = {0035-9246},
  journal   = {Journal of the Royal Statistical Society. Series B},
  number    = {1},
  pages     = {1--22},
  publisher = {JSTOR},
  title     = {{Random discrete distributions}},
  volume    = {37},
  year      = {1975}
}

@article{Bro2003,
  author   = {Bro, Rasmus and Kiers, Henk A L},
  isbn     = {0886-9383},
  issn     = {08869383},
  journal  = {Journal of Chemometrics},
  number   = {5},
  pages    = {274--286},
  title    = {{A new efficient method for determining the number of components in PARAFAC models}},
  volume   = {17},
  year     = {2003}
}

@techreport{PeterCraig2013,
  author = {Craig, Peter S.},
  title  = {{Exploring novel ways of using species sensitivity distributions to establish PNECs for industrial chemicals: Final report to Project Steering Group}},
  year   = {2013}
}

@article{Shao2000,
  author   = {Shao, Quanxi},
  issn     = {11804009},
  journal  = {Environmetrics},
  month    = {sep},
  number   = {5},
  pages    = {583--595},
  title    = {{Estimation for hazardous concentrations based on NOEC toxicity data: An alternative approach}},
  volume   = {11},
  year     = {2000}
}

@techreport{USEPASSD,
  address     = {Washington, DC.},
  author      = {USEPA},
  institution = {US Environmental Protection Agency},
  title       = {{Guidelines for ecological risk assessment.}},
  year        = {1998}
}

@article{Lijoi2007,
  author    = {Lijoi, Antonio and Mena, Rams{\'{e}}s H. and Pr{\"{u}}nster, Igor},
  issn      = {13697412},
  journal   = {Journal of the Royal Statistical Society. Series B: Statistical Methodology},
  number    = {4},
  pages     = {715--740},
  publisher = {Wiley Online Library},
  title     = {{Controlling the reinforcement in Bayesian non-parametric mixture models}},
  volume    = {69},
  year      = {2007}
}

@article{Aldenberg1993,
  author   = {Aldenberg, Tom and Slob, W},
  isbn     = {0147-6513},
  issn     = {0147-6513},
  journal  = {Ecotoxicology and Environmental Safety},
  number   = {1},
  pages    = {48--63},
  pmid     = {7682918},
  title    = {{Confidence limits for hazardous concentrations based on logistically distributed NOEC toxicity data.}},
  volume   = {25},
  year     = {1993}
}

@article{Wagner1991,
  author   = {Wagner, Connie and Lokke, Hans},
  isbn     = {0043-1354},
  issn     = {00431354},
  journal  = {Water Research},
  number   = {10},
  pages    = {1237--1242},
  title    = {{Estimation of ecotoxicological protection levels from NOEC toxicity data}},
  volume   = {25},
  year     = {1991}
}

@article{Grist2009a,
  author   = {Grist, Eric P M and Leung, Kenneth M Y and Wheeler, James R and Crane, Mark},
  isbn     = {1552-8618},
  issn     = {07307268 (ISSN)},
  journal  = {Environmental Toxicology and Chemistry},
  number   = {7},
  pages    = {1515--1524},
  pmid     = {12109754},
  title    = {{Better bootstrap estimation of hazardous concentration thresholds for aquatic assemblages}},
  volume   = {21},
  year     = {2002}
}

@article{Jagoe1997,
  author    = {Jagoe, Rosemary H. and Newman, Michael C.},
  isbn      = {0963-9292},
  issn      = {09639292},
  journal   = {Ecotoxicology},
  language  = {English},
  number    = {5},
  pages     = {293--306},
  publisher = {Kluwer Academic Publishers},
  title     = {{Bootstrap estimation of community NOEC values}},
  volume    = {6},
  year      = {1997}
}

@article{mueller2004,
	author = {M\"uller, Peter and Quintana, Fernando and Rosner, Gary L.},
	title = {A method for combining inference across related nonparametric {B}ayesian models},
	journal = {J. R. Stat. Soc. Ser. B},
	year = {2004},
	pages = {735--749},
	volume = {66}}

@article{camerlenghi2019,
	title = {Latent nested nonparametric priors (with discussion)},
	author = {Camerlenghi, Federico and Dunson, David B. and Lijoi, Antonio and Pr{\"{u}}nster, Igor and Rodr{\'{i}}guez, Abel},
	journal = {Bayesian Anal.},
	year = {2019},
	volume ={14},
	pages = {1303--1356}
}

@article{lijoi2014,
	title={Bayesian inference with dependent normalized completely random measures},
	author={Lijoi, Antonio and Nipoti, Bernardo and Pr{\"u}nster, Igor},
	journal={Bernoulli},
	volume={20},
	pages={1260--1291},
	year={2014},
	publisher={Bernoulli Society for Mathematical Statistics and Probability}
}

@misc{flpr2025,
      title={Multivariate Species Sampling Models}, 
      author={Beatrice Franzolini and Antonio Lijoi and Igor Prünster and Giovanni Rebaudo},
      year={2025},
      eprint={2503.24004},
      archivePrefix={arXiv},
      primaryClass={math.ST},
      url={https://arxiv.org/abs/2503.24004}, 
}

@article{dahl2022search,
	author = {David B. Dahl and Devin J. Johnson and Peter M{\"u}ller},
	journal = {Journal of Computational and Graphical Statistics},
	number = {4},
	pages = {1189-1201},
	publisher = {Taylor & Francis},
	title = {Search Algorithms and Loss Functions for {Bayesian} Clustering},
	volume = {31},
	year = {2022},}

@article{arbel2021BNPdensity,
	author = {Arbel, Julyan and Kon Kam King, Guillaume and Lijoi, Antonio and Nieto-Barajas, Luis E. and Pr\"{u}nster, Igor},
	issue = {3},
	journal = {Australian \& New Zealand Journal of Statistics},
	pages = {542-564},
	title = {{BNPdensity: Bayesian nonparametric mixture modeling in R}},
	volume = {63},
	year = {2021},
	}

@article{wade2015bayesian,
	title = {Bayesian {Cluster} {Analysis}: {Point} {Estimation} and {Credible} {Balls}},
	volume = {13},
	issn = {1936-0975, 1931-6690},
	shorttitle = {Bayesian {Cluster} {Analysis}},
	number = {2},
	urldate = {2022-02-02},
	journal = {{Bayesian Analysis}},
	author = {Wade, Sara and Ghahramani, Zoubin},
	month = jun,
	year = {2018},
	pages = {559--626},
}

@article{dahl2006model,
	Author = {Dahl, David B},
	Journal = {Bayesian inference for gene expression and proteomics},
	Pages = {201--218},
	Publisher = {Citeseer},
	Title = {{Model-based clustering for expression data via a Dirichlet process mixture model}},
	Year = {2006}}

@article{lau2007bayesian,
	Author = {Lau, John W and Green, Peter J},
	Journal = {Journal of Computational and Graphical Statistics},
	Number = {3},
	Pages = {526--558},
	Publisher = {Taylor \& Francis},
	Title = {Bayesian model-based clustering procedures},
	Volume = {16},
	Year = {2007}}

@article{meila2007comparing,
	Author = {Meil{\u{a}}, Marina},
	Journal = {Journal of Multivariate Analysis},
	Number = {5},
	Pages = {873--895},
	Publisher = {Elsevier},
	Title = {Comparing clusterings---an information based distance},
	Volume = {98},
	Year = {2007}}

@article{binder1978bayesian,
	Author = {Binder, David A},
	Journal = {Biometrika},
	Number = {1},
	Pages = {31--38},
	Publisher = {Biometrika Trust},
	Title = {Bayesian cluster analysis},
	Volume = {65},
	Year = {1978}}

@article{leeuw_multiway_2011,
	title = {The {Multiway} {Package}},
	language = {en},
	urldate = {2023-12-22},
	author = {Leeuw, Jan de},
	month = oct,
	year = {2011},
}

@inproceedings{rai_scalable_2014,
	title = {Scalable {Bayesian} {Low}-{Rank} {Decomposition} of {Incomplete} {Multiway} {Tensors}},
	language = {en},
	booktitle = {Proceedings of the 31st {International} {Conference} on {Machine} {Learning}},
	publisher = {PMLR},
	author = {Rai, Piyush and Wang, Yingjian and Guo, Shengbo and Chen, Gary and Dunson, David and Carin, Lawrence},
	month = jun,
	year = {2014},
	pages = {1800--1808},
}

\appendix
\section*{Supplementary material}

This supplementary material is organized as follows: Section~\ref{sec:app_results} provides additional results on real data, Section~\ref{sec:conv_diag} presents convergence diagnostics for the results  on real data, Section~\ref{sec:sens} discusses a sensitivity analysis of the model parameters, Section~\ref{sec:details-ntf} contains details on  the non-negative tensor factorization, and Section~\ref{sec:supp-figures} displays additional figures related to Section~\ref{sec:res_clus} of the main document.
%In this supplementary material, we provide additional results on real data in Section~\ref{sec:app_results},  details on  the non-negative tensor factorization in Section~\ref{sec:details-ntf}, and additional figures for Section~\ref{sec:res_clus} of the main document in Section~\ref{sec:supp-figures}.

\section{Results on real data}\label{sec:app_results}
This section illustrates the results stated in Section~\ref{sec:est_data}.
We present the comparison of the three models (\gls{bnp}, \gls{kde} and normal) on real data.
We consider three categories of contaminants: contaminants with large datasets, consisting of more than 60 values, medium datasets, with around 25 values, and small datasets, with a little over 10 values.
The three models were fitted on each dataset and we studied the estimate of the \gls{hc5} and its credible interval, the LOO error and the shape of the estimated density compared to the histogram.
The censored version of the same datasets was also studied with the \gls{bnp} and normal model.

\subsection{Model comparison on non-censored data}
We present the results for the non-censored version of the datasets.
\subsubsection{Large non-censored datasets}

\def\colorlegend{Red (\red) for the BNP model, blue (\blue) for the normal model, and green (\green) for the KDE model.\xspace}

\begin{minipage}[c]{0.32\textwidth}
    \centering
    \includegraphics[width=\textwidth]{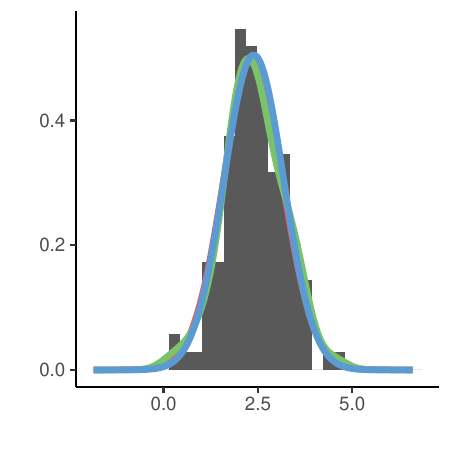}\\
    \includegraphics[width=\textwidth]{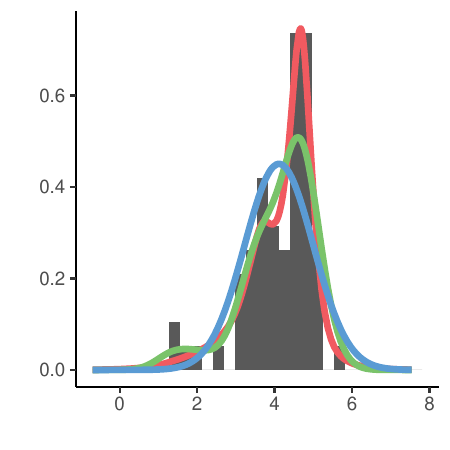}\\
    \includegraphics[width=\textwidth]{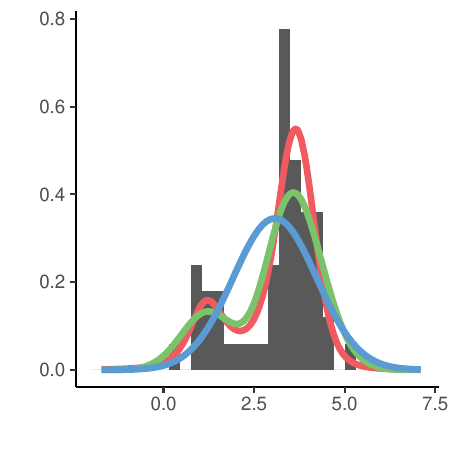}\\
    Density estimates
\end{minipage}
\begin{minipage}[c]{0.32\textwidth}
    \centering
    \includegraphics[width=\textwidth]{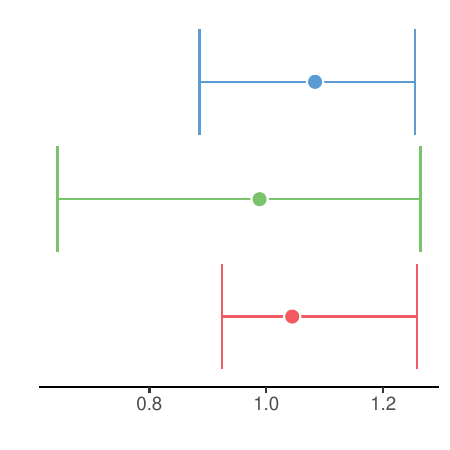}\\
    \includegraphics[width=\textwidth]{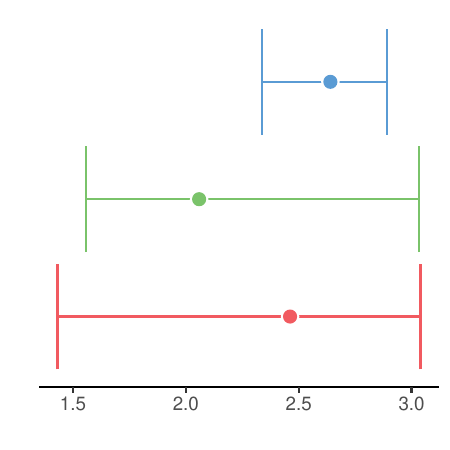}\\
    \includegraphics[width=\textwidth]{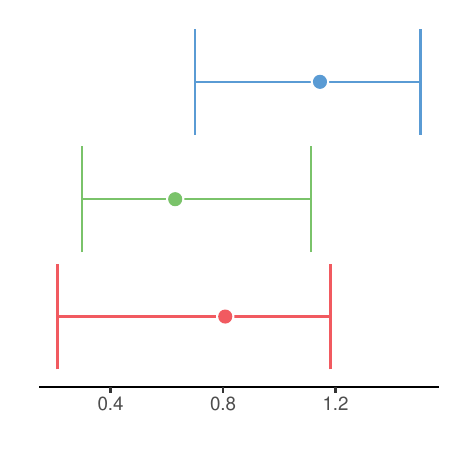} \\
    HC$_5$
\end{minipage}
\begin{minipage}[c]{0.32\textwidth}
    \centering
    \includegraphics[width=\textwidth]{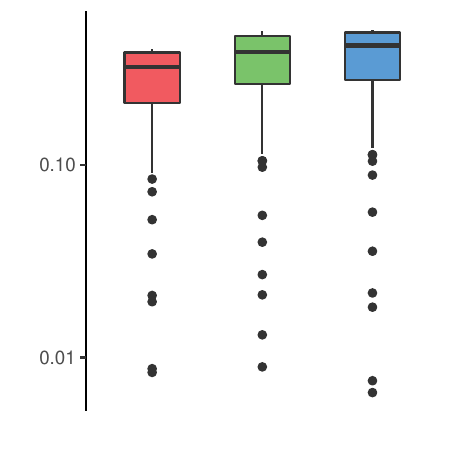}\\
    \includegraphics[width=\textwidth]{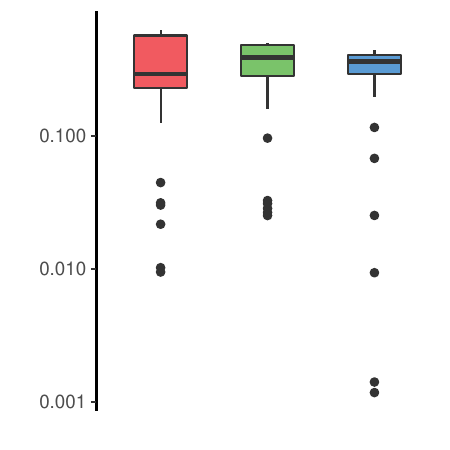}\\
    \includegraphics[width=\textwidth]{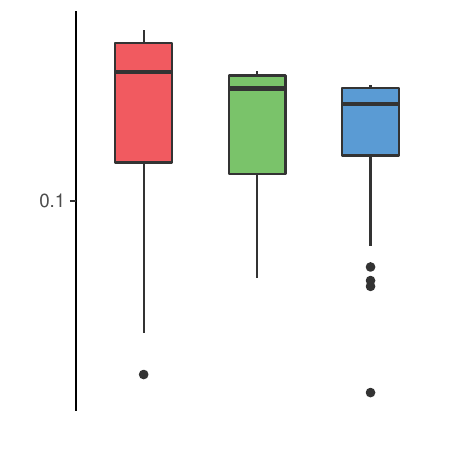}\\
    CPO
\end{minipage}
\captionof{figure}{Density estimates, \gls{hc5} and \gls{cpo} for three large non-censored datasets. \colorlegend From top to bottom: Cadmium chloride, Potassium Dichromate, and Carbaryl.\label{fig:S1}}

\newpage

\subsubsection{Medium non-censored datasets}

\begin{minipage}[c]{0.32\textwidth}
    \centering
    \includegraphics[width=\textwidth]{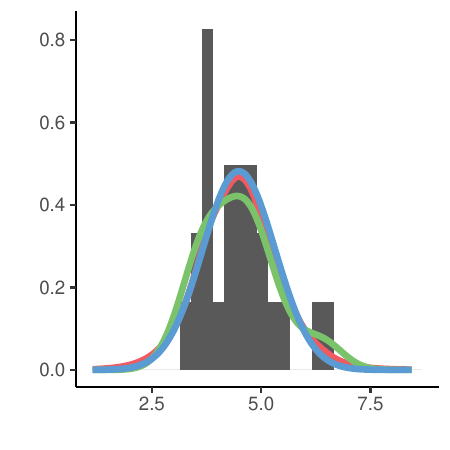}\\
    \includegraphics[width=\textwidth]{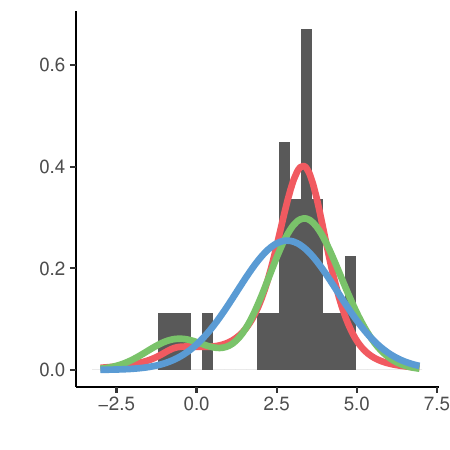}\\
    \includegraphics[width=\textwidth]{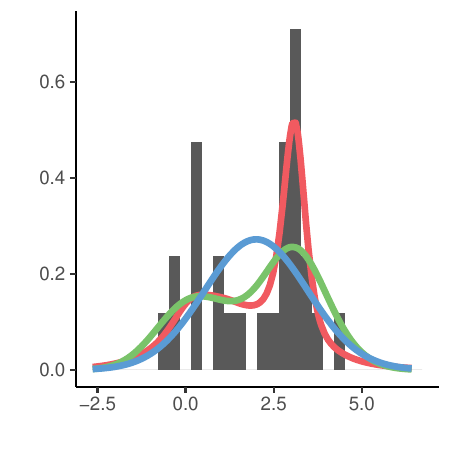}\\
    Density estimates
\end{minipage}
\begin{minipage}[c]{0.32\textwidth}
    \centering
    \includegraphics[width=\textwidth]{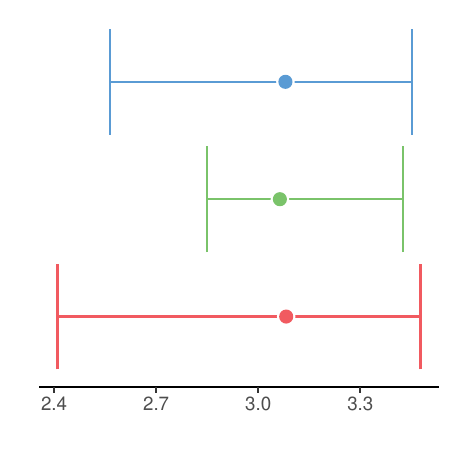}\\
    \includegraphics[width=\textwidth]{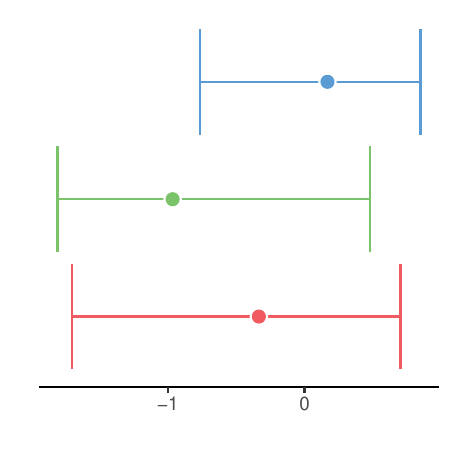} \\
    \includegraphics[width=\textwidth]{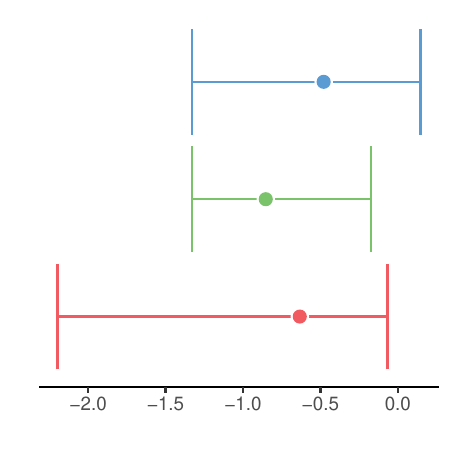} \\
    HC$_5$
\end{minipage}
\begin{minipage}[c]{0.32\textwidth}
    \centering
    \includegraphics[width=\textwidth]{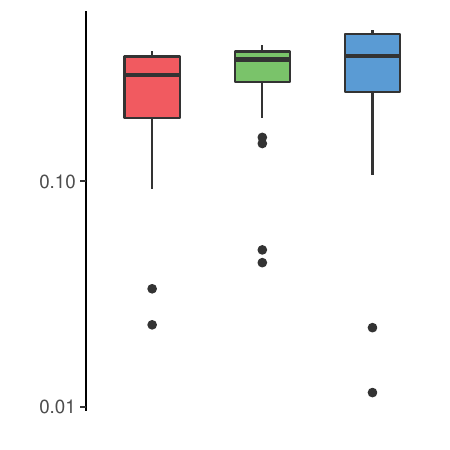}\\
    \includegraphics[width=\textwidth]{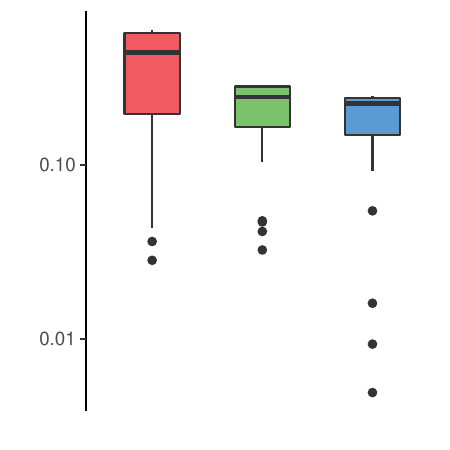}\\
    \includegraphics[width=\textwidth]{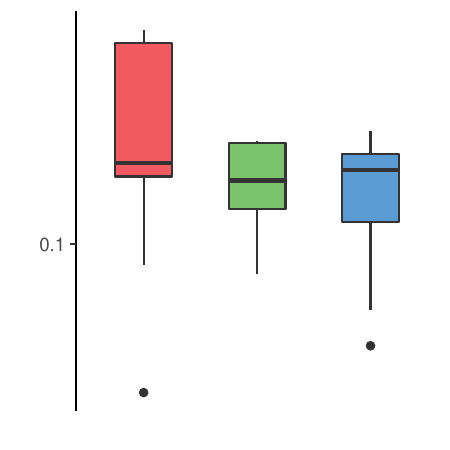}\\
    CPO
\end{minipage}
\captionof{figure}{Density estimates, \gls{hc5} and \gls{cpo} for three medium non-censored datasets. \colorlegend From top to bottom: 2,4-D Acid, Trichlorfon, Parathion.}

\newpage

\subsubsection{Small non-censored datasets}

\begin{minipage}[c]{0.32\textwidth}
    \centering
    \includegraphics[width=\textwidth]{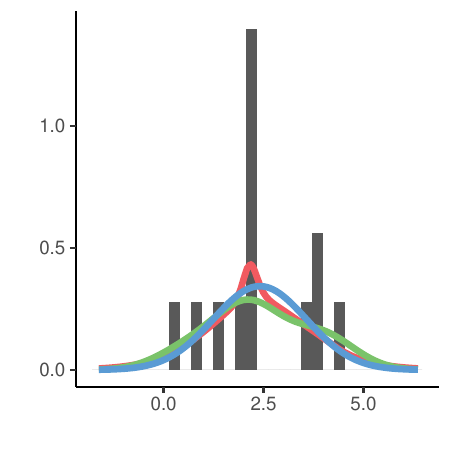}\\
    \includegraphics[width=\textwidth]{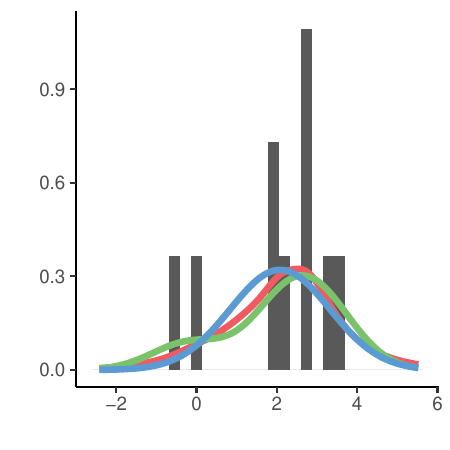}\\
    \includegraphics[width=\textwidth]{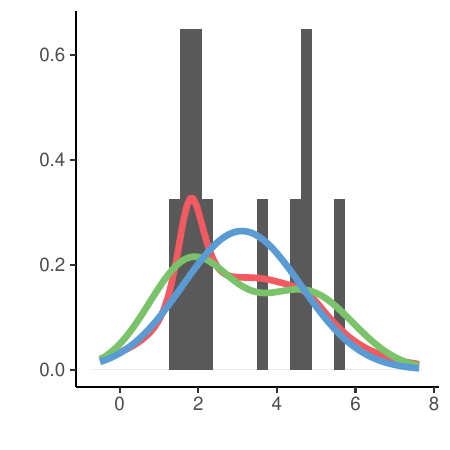}\\
    Density estimates
\end{minipage}
\begin{minipage}[c]{0.32\textwidth}
    \centering
    \includegraphics[width=\textwidth]{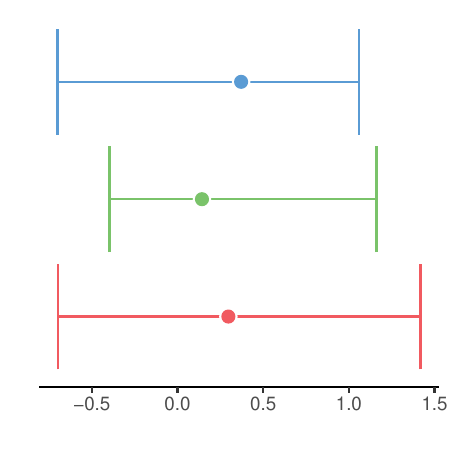}\\
    \includegraphics[width=\textwidth]{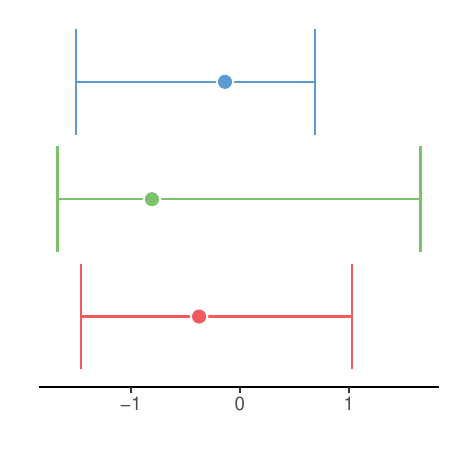}\\
    \includegraphics[width=\textwidth]{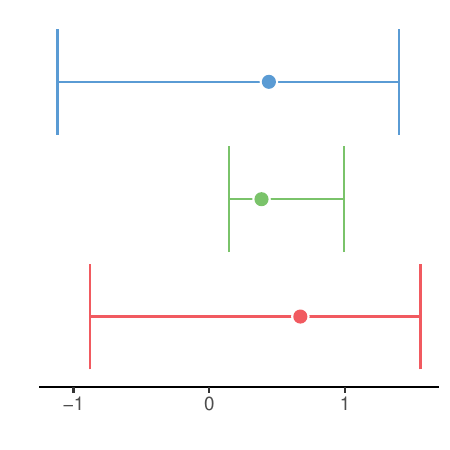}  \\
    HC$_5$
\end{minipage}
\begin{minipage}[c]{0.32\textwidth}
    \centering
    \includegraphics[width=\textwidth]{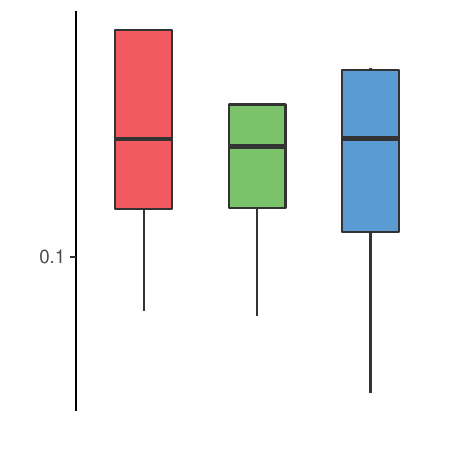}\\
    \includegraphics[width=\textwidth]{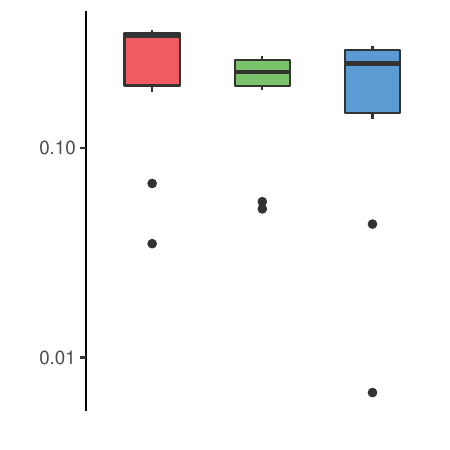}\\
    \includegraphics[width=\textwidth]{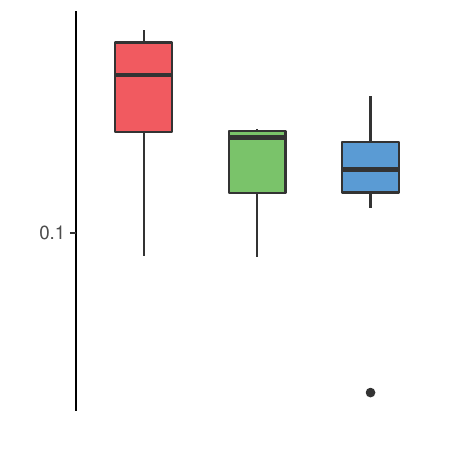}  \\
    CPO
\end{minipage}
\captionof{figure}{Density estimates, \gls{hc5} and \gls{cpo} for three small non-censored datasets. \colorlegend From top to bottom: Phosmet, Naled, Sodium dichromate.}

\newpage

\subsection{Model comparison on censored datasets}
Here we present the results for the censored version of the datasets.
As extensions for kernel density estimators with censored data are not available (see Section~\ref{sec:censored-data}), we only compare the normal and \gls{bnp} models.
% % 
\subsubsection{Large censored datasets}
\def\colorlegendnKDE{Red (\red) for the BNP model, blue (\blue) for the normal model \xspace}
\begin{minipage}[c]{0.32\textwidth}
    \centering
    \includegraphics[width=\textwidth]{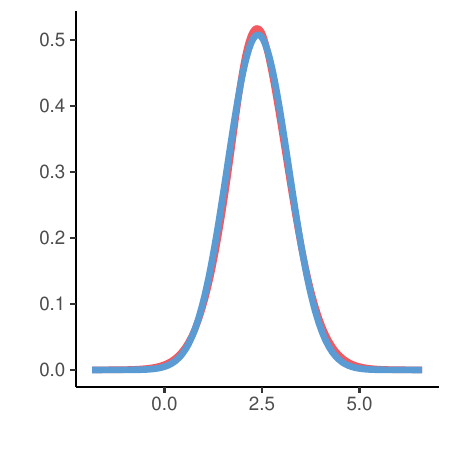}\\
    \includegraphics[width=\textwidth]{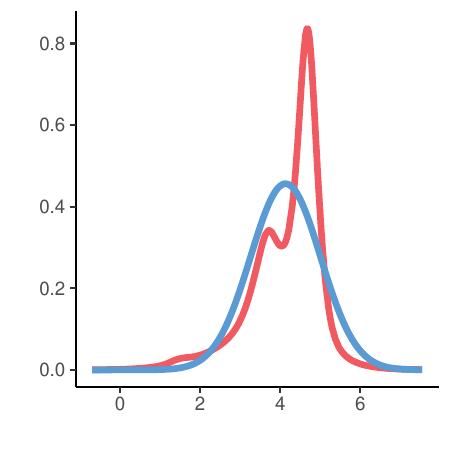}\\
    \includegraphics[width=\textwidth]{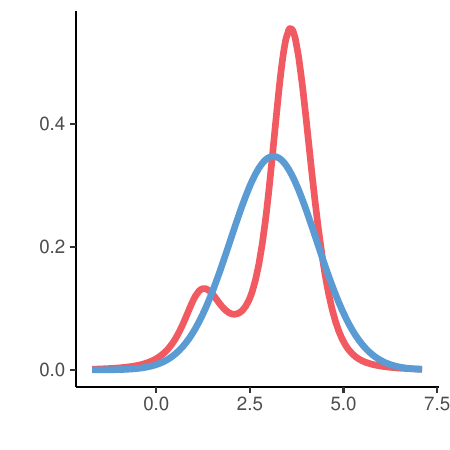}\\
    Density estimates
\end{minipage}
\begin{minipage}[c]{0.32\textwidth}
    \centering
    \includegraphics[width=\textwidth]{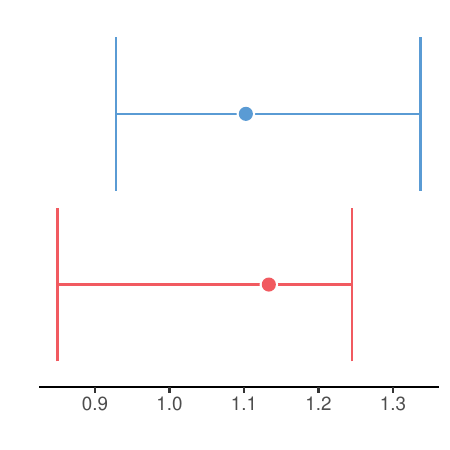}\\
    \includegraphics[width=\textwidth]{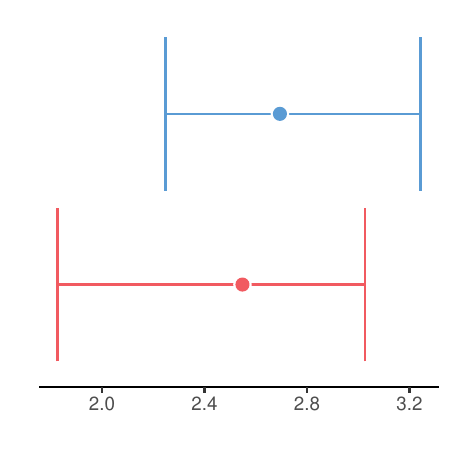}\\
    \includegraphics[width=\textwidth]{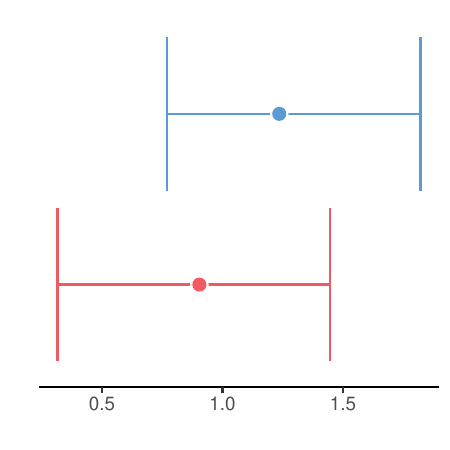}  \\
    HC$_5$
\end{minipage}
\begin{minipage}[c]{0.32\textwidth}
    \centering
    \includegraphics[width=\textwidth]{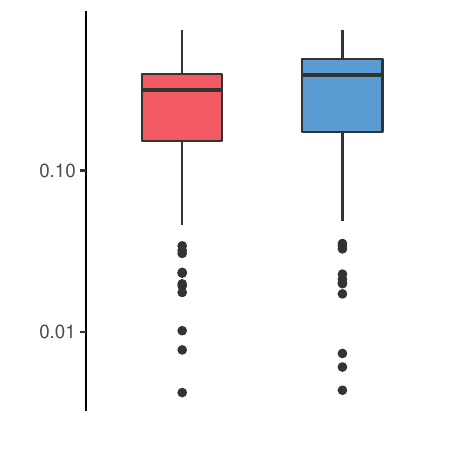}\\
    \includegraphics[width=\textwidth]{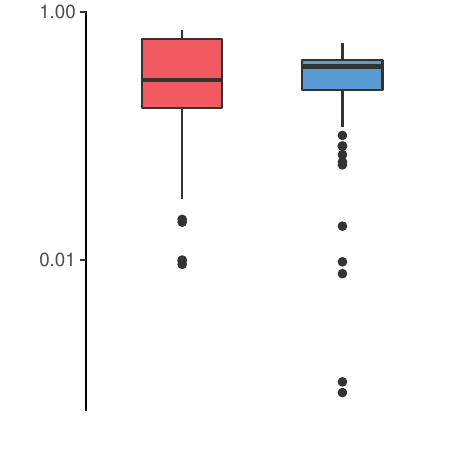}\\
    \includegraphics[width=\textwidth]{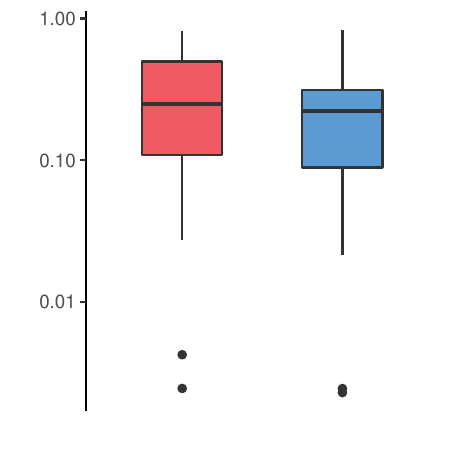}\\
    CPO
\end{minipage}
\captionof{figure}{Density estimates, \gls{hc5} and \gls{cpo} for three large censored datasets. \colorlegendnKDE  (KDE not implemented for censored data). From top to bottom:
    Cadmium chloride, Potassium Dichromate, and Carbaryl.}

\newpage

\subsubsection{Medium censored datasets}

\begin{minipage}[c]{0.32\textwidth}
    \centering
    \includegraphics[width=\textwidth]{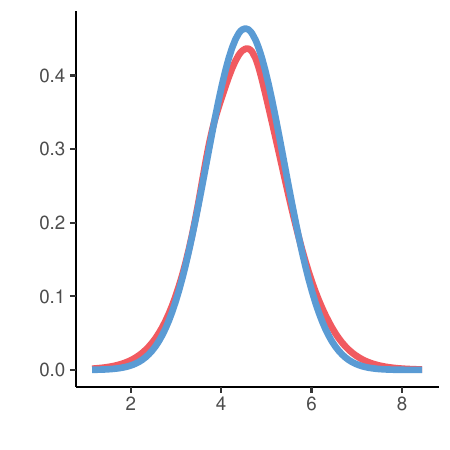}\\
    \includegraphics[width=\textwidth]{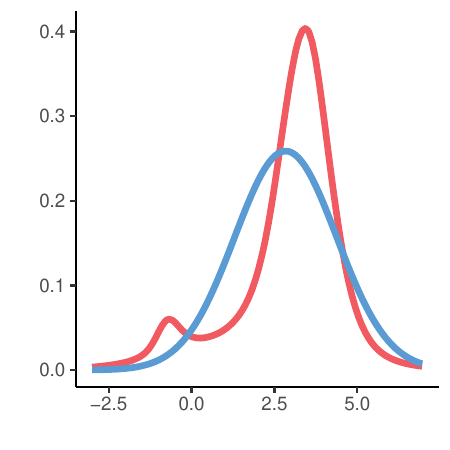}\\
    \includegraphics[width=\textwidth]{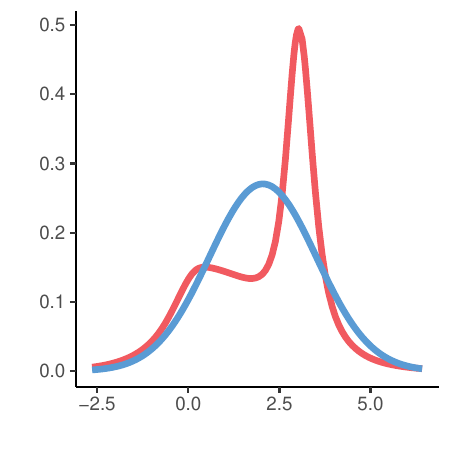}\\
    Density estimates
\end{minipage}
\begin{minipage}[c]{0.32\textwidth}
    \centering
    \includegraphics[width=\textwidth]{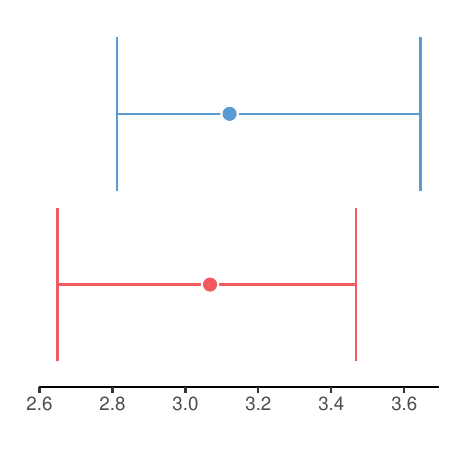}\\
    \includegraphics[width=\textwidth]{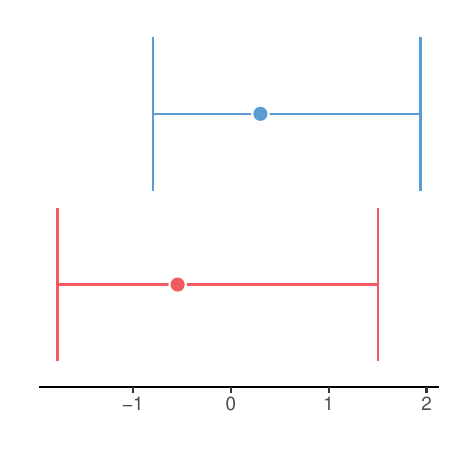}\\
    \includegraphics[width=\textwidth]{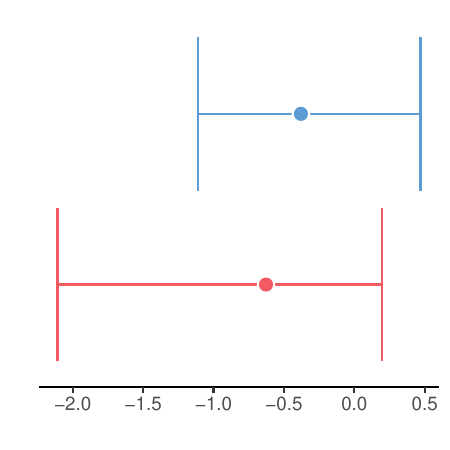} \\
    HC$_5$
\end{minipage}
\begin{minipage}[c]{0.32\textwidth}
    \centering
    \includegraphics[width=\textwidth]{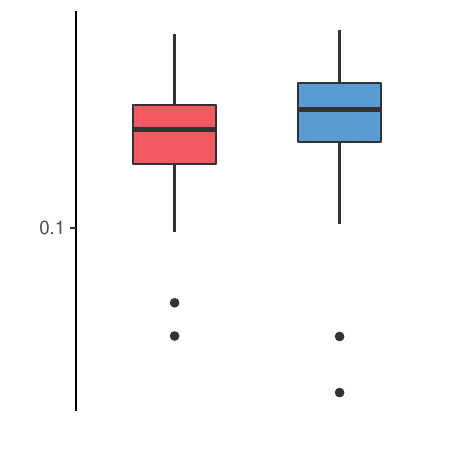}\\
    \includegraphics[width=\textwidth]{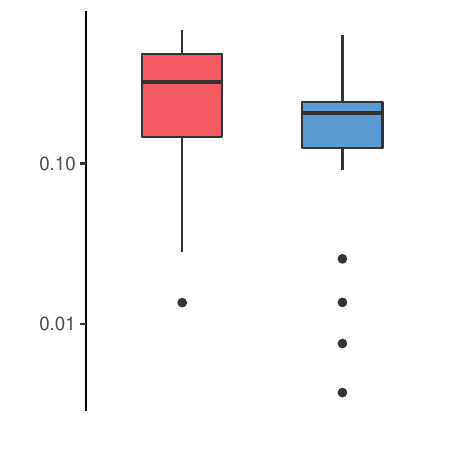}\\
    \includegraphics[width=\textwidth]{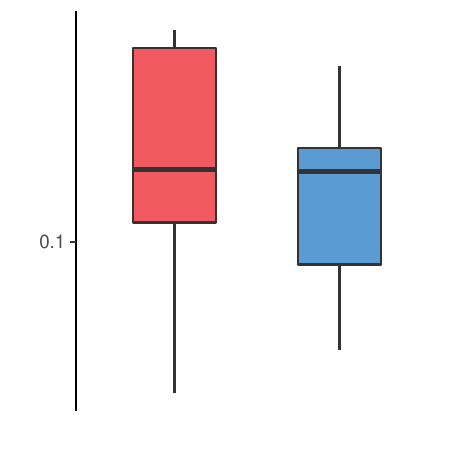} \\
    CPO
\end{minipage}
\captionof{figure}{Density estimates, \gls{hc5} and \gls{cpo} for three medium censored datasets. \colorlegendnKDE (KDE not implemented for censored data). From top to bottom: 2,4-D Acid, Trichlorfon, Parathion.}

\newpage

\subsubsection{Small censored datasets}

\begin{minipage}[c]{0.32\textwidth}
    \centering
    \includegraphics[width=\textwidth]{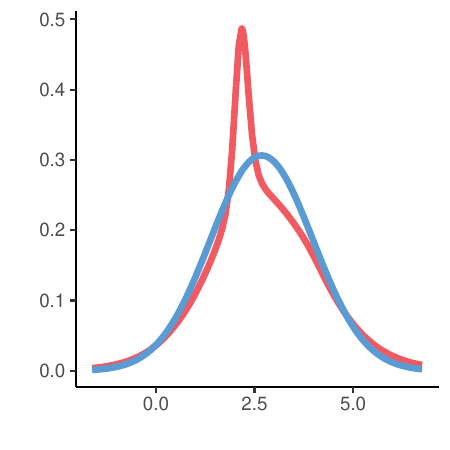}\\
    \includegraphics[width=\textwidth]{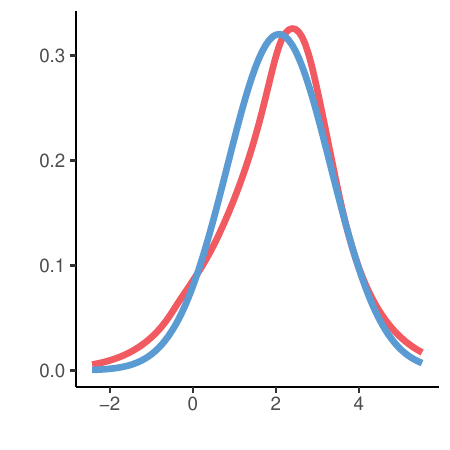}\\
    \includegraphics[width=\textwidth]{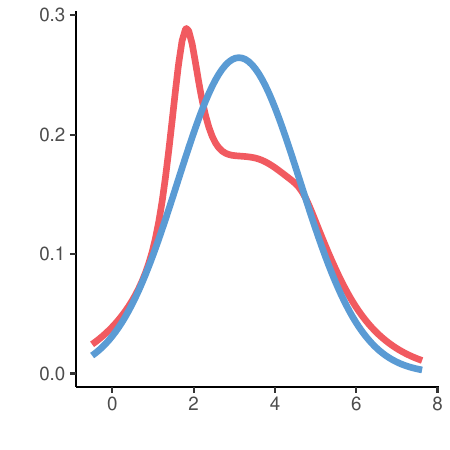}\\
    Density estimates
\end{minipage}
\begin{minipage}[c]{0.32\textwidth}
    \centering
    \includegraphics[width=\textwidth]{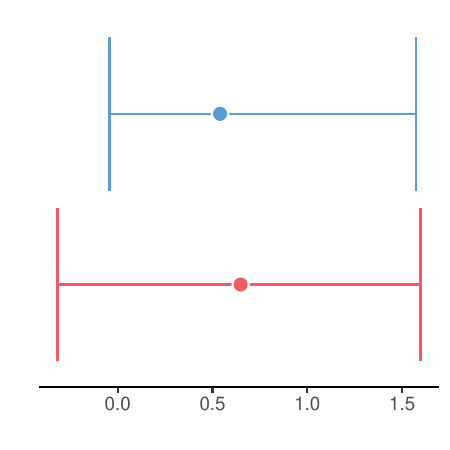}\\
    \includegraphics[width=\textwidth]{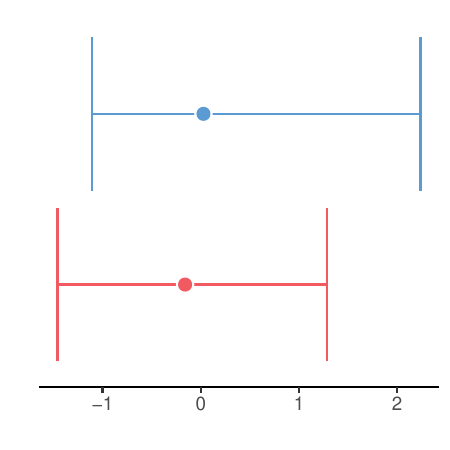}\\
    \includegraphics[width=\textwidth]{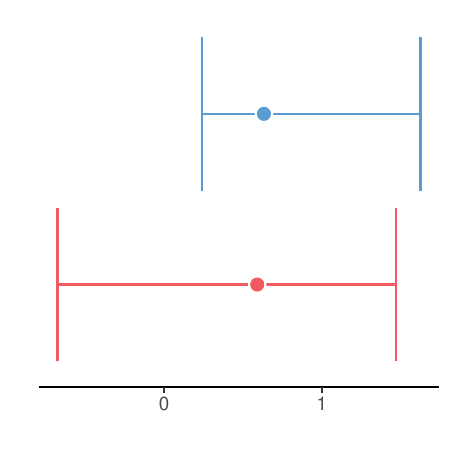}  \\
    HC$_5$
\end{minipage}
\begin{minipage}[c]{0.32\textwidth}
    \centering
    \includegraphics[width=\textwidth]{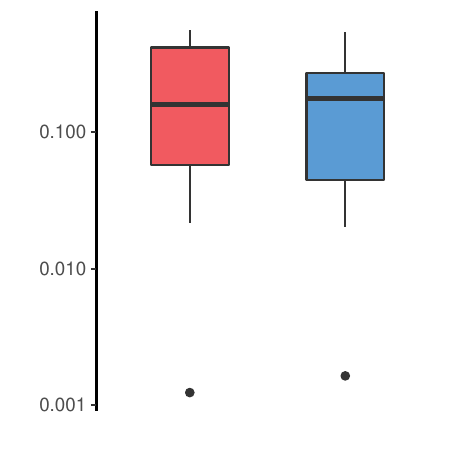}\\
    \includegraphics[width=\textwidth]{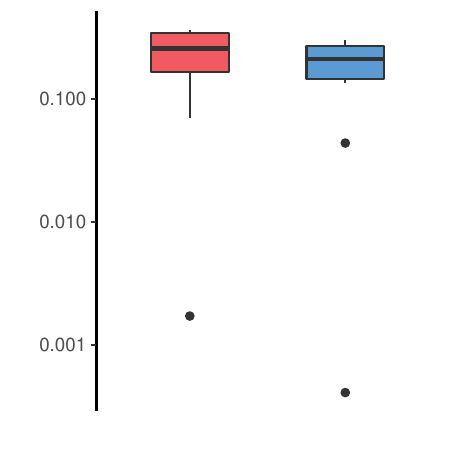}\\
    \includegraphics[width=\textwidth]{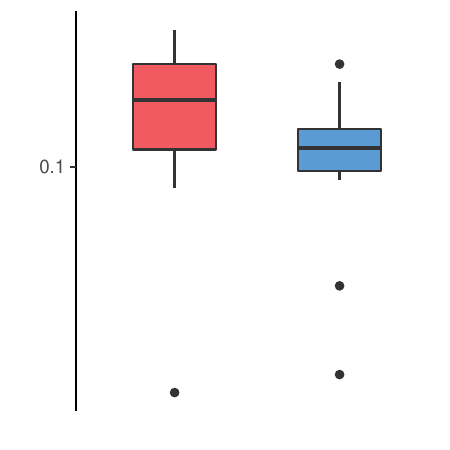} \\
    CPO
\end{minipage}
\captionof{figure}{Density estimates, \gls{hc5} and \gls{cpo} for three small censored datasets. \colorlegendnKDE (KDE not implemented for censored data). From top to bottom: Phosmet, Naled, Sodium dichromate.\label{fig:S6}}

% \section{Contaminant-wise clustering illustration with credible bands on the density estimates}\label{sec:contaminant-wise-clustering}

% \includegraphics[width=\textwidth]{example_clustering_post_analysis_credible_bands.pdf}
% \captionof{figure}{\update{\gls{ssd} for Carbaryl (CAS: 63-25-2) with the quasi-taxonomic group of each species overlaid on the curve. Left: Species coloured by quasi-taxonomic group. Right: Species coloured by cluster membership in the \gls{bnp} model. The solid line denotes the cumulative distribution function estimate, while the gray bands around it denote a pointwise 95\% credible interval.
%     \label{fig:Illustration_clustering_with_credible_bands}}}

\clearpage

\section{Convergence diagnostics for real data analysis in \ref{sec:app_results}}
\label{sec:conv_diag}
\update{The results presented below have been performed on the non-censored datasets. The traceplots are similar for the censored version of the datasets.}

\subsection{Large datasets}
\begin{figure}[H]
    \centering
    \begin{tabular}{ccc}
       \includegraphics[page=1,width=.32\textwidth, trim={1.15cm 0.6cm 1.05cm 0.7cm}, clip]{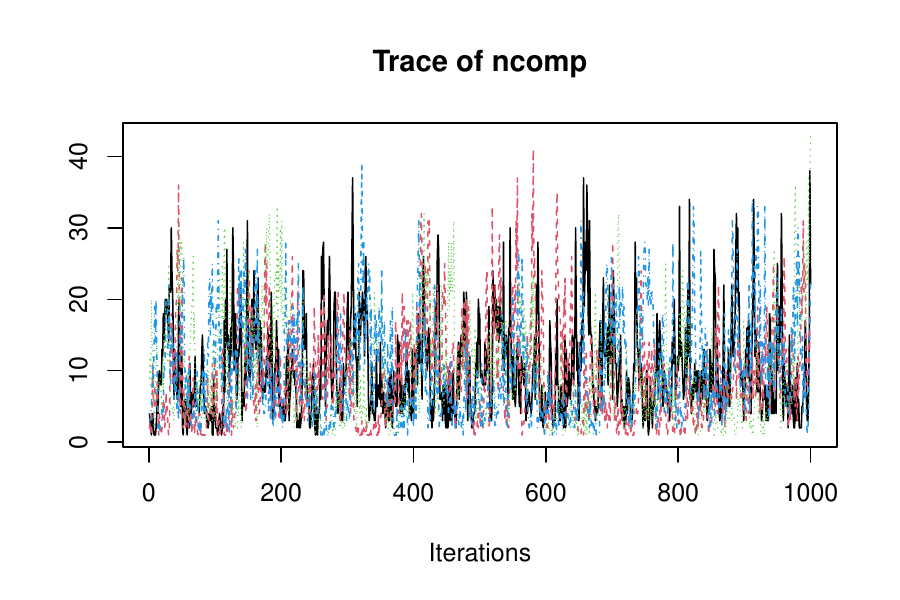} & 
       \includegraphics[page=2,width=.32\textwidth, trim={1.15cm 0.6cm 1.05cm 0.7cm}, clip]{figures/conv_diag/cadmium.pdf} &
       \includegraphics[page=3,width=.32\textwidth, trim={1.15cm 0.6cm 1.05cm 0.7cm}, clip]{figures/conv_diag/cadmium.pdf}    
       \end{tabular} 
    \caption{\update{Traceplots obtained with \texttt{BNPdensity} and \texttt{coda} packages for Cadmium chloride.}}
\end{figure}

\begin{figure}[H]
    \centering
    \begin{tabular}{ccc}
       \includegraphics[page=1,width=.32\textwidth, trim={1.15cm 0.6cm 1.05cm 0.7cm}, clip]{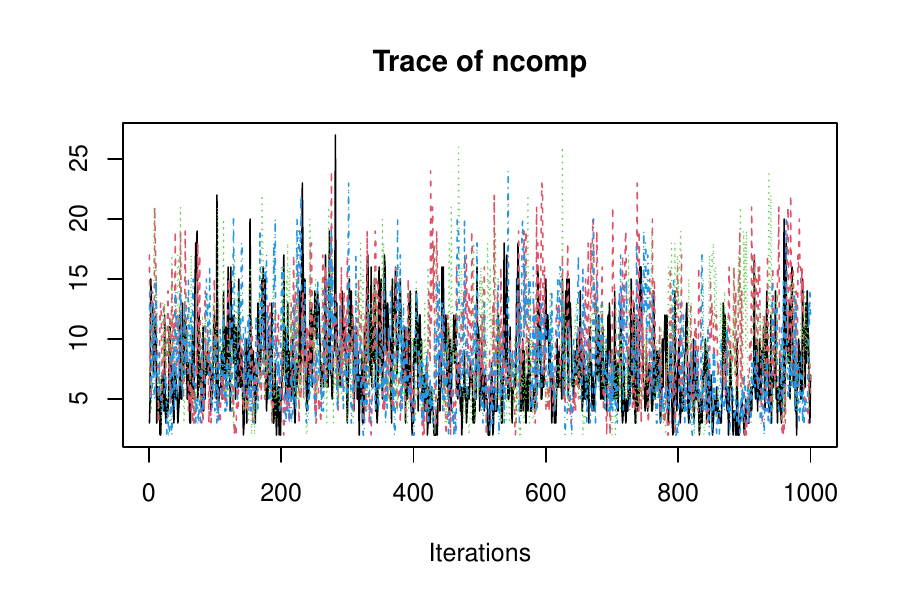} & 
       \includegraphics[page=2,width=.32\textwidth, trim={1.15cm 0.6cm 1.05cm 0.7cm}, clip]{figures/conv_diag/potassium.pdf} &
       \includegraphics[page=3,width=.32\textwidth, trim={1.15cm 0.6cm 1.05cm 0.7cm}, clip]{figures/conv_diag/potassium.pdf}    
       \end{tabular} 
    \caption{\update{Traceplots for Potassium Dichromate.}}
\end{figure}

\begin{figure}[H]
    \centering
    \begin{tabular}{ccc}
       \includegraphics[page=1,width=.32\textwidth, trim={1.15cm 0.6cm 1.05cm 0.7cm}, clip]{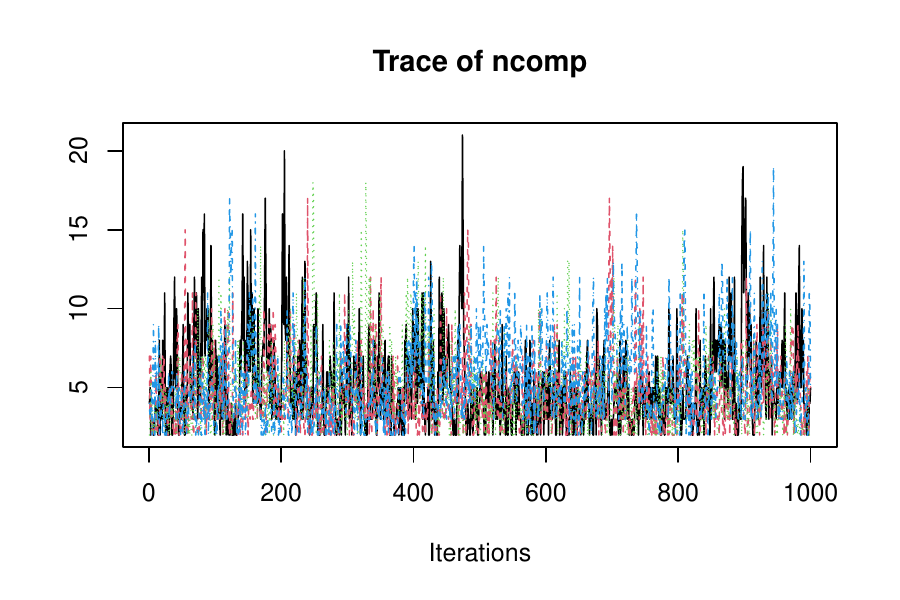} & 
       \includegraphics[page=2,width=.32\textwidth, trim={1.15cm 0.6cm 1.05cm 0.7cm}, clip]{figures/conv_diag/carbaryl.pdf} &
       \includegraphics[page=3,width=.32\textwidth, trim={1.15cm 0.6cm 1.05cm 0.7cm}, clip]{figures/conv_diag/carbaryl.pdf}    
       \end{tabular} 
    \caption{\update{Traceplots for Carbaryl.}}
\end{figure}

\begin{table}[h]
    \centering
    \update{
    \begin{tabular}{|c|c|c|c|}
        \hline
         & ESS (ncomp) & ESS (Latent\_variable) & ESS (log\_likelihood) \\ \hline
        Cadmium chloride & 171 & 341 & 43 \\
        Potassium Dichromate & 71 & 121 & 59\\
        Carbaryl & 242 & 400 & 37 \\
        \hline
    \end{tabular}}
    \caption{\update{Effective sample size of the parameters for the different datasets.}}
\end{table}

\subsection{Medium datasets}

\begin{figure}[H]
    \centering
    \begin{tabular}{ccc}
       \includegraphics[page=1,width=.32\textwidth, trim={1.15cm 0.6cm 1.05cm 0.7cm}, clip]{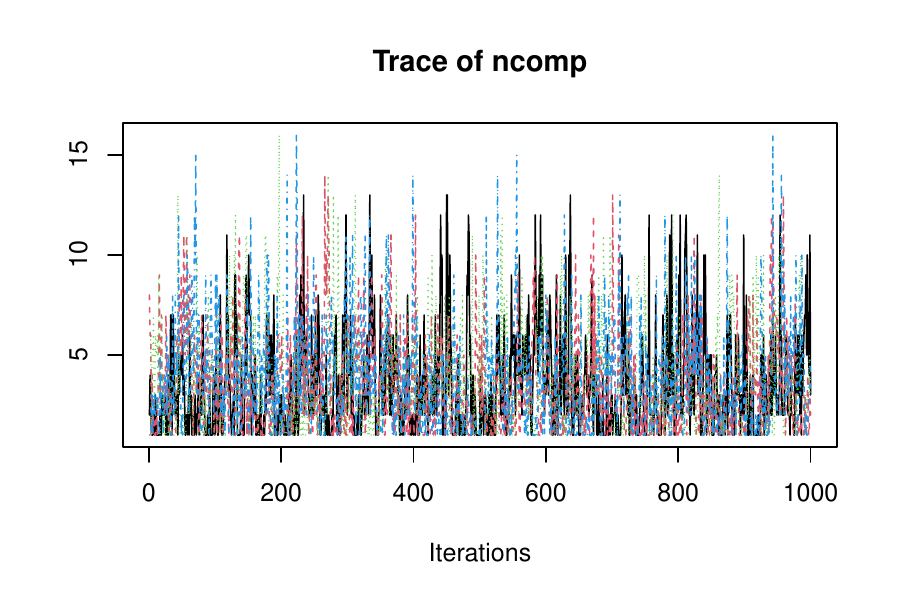} & 
       \includegraphics[page=2,width=.32\textwidth, trim={1.15cm 0.6cm 1.05cm 0.7cm}, clip]{figures/conv_diag/acid.pdf} &
       \includegraphics[page=3,width=.32\textwidth, trim={1.15cm 0.6cm 1.05cm 0.7cm}, clip]{figures/conv_diag/acid.pdf}    
       \end{tabular} 
    \caption{\update{Traceplots for 2,4-D Acid.}}
\end{figure}

\begin{figure}[H]
    \centering
    \begin{tabular}{ccc}
       \includegraphics[page=1,width=.32\textwidth, trim={1.15cm 0.6cm 1.05cm 0.7cm}, clip]{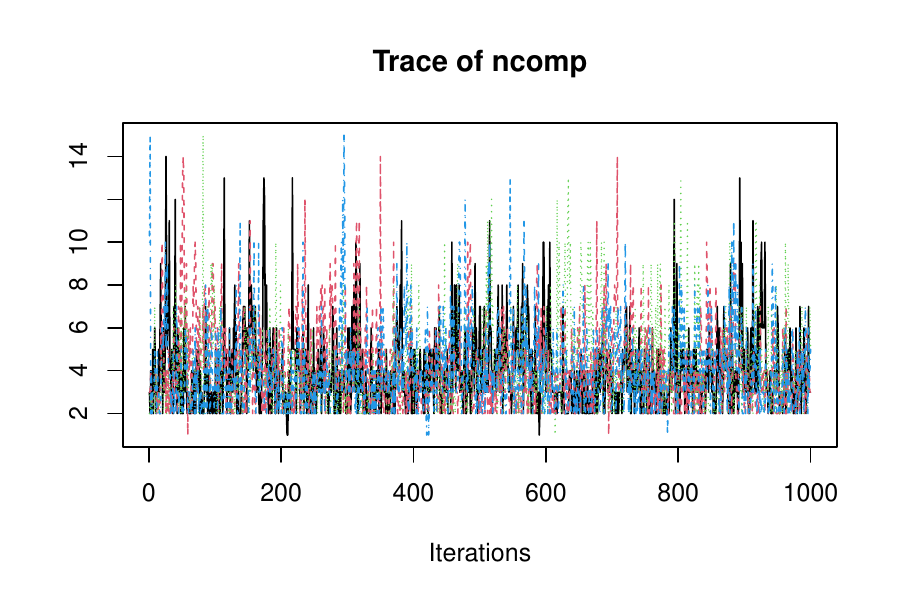} & 
       \includegraphics[page=2,width=.32\textwidth, trim={1.15cm 0.6cm 1.05cm 0.7cm}, clip]{figures/conv_diag/Trichlorfon.pdf} &
       \includegraphics[page=3,width=.32\textwidth, trim={1.15cm 0.6cm 1.05cm 0.7cm}, clip]{figures/conv_diag/Trichlorfon.pdf}    
       \end{tabular} 
    \caption{\update{Traceplots for Trichlorfon.}}
\end{figure}

\begin{figure}[H]
    \centering
    \begin{tabular}{ccc}
       \includegraphics[page=1,width=.32\textwidth, trim={1.15cm 0.6cm 1.05cm 0.7cm}, clip]{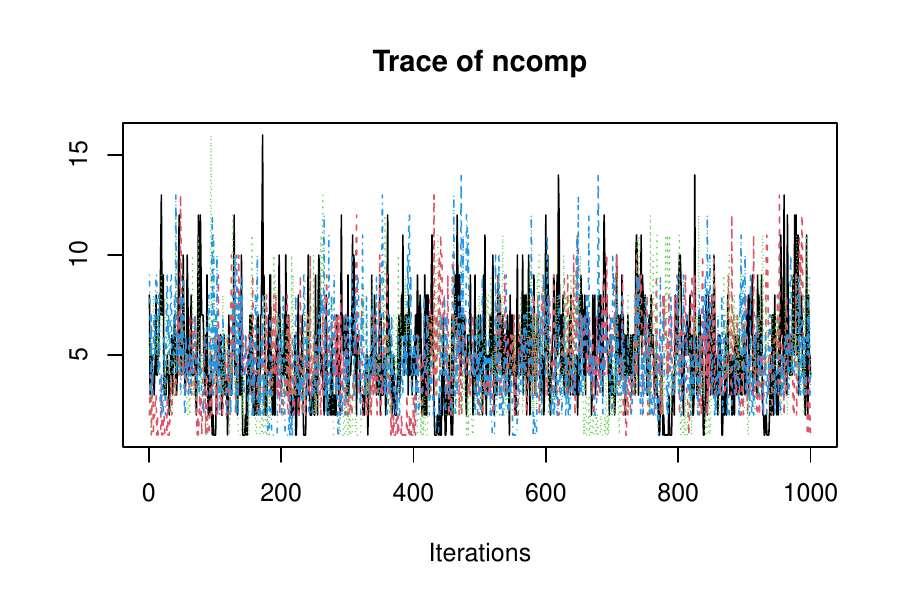} & 
       \includegraphics[page=2,width=.32\textwidth, trim={1.15cm 0.6cm 1.05cm 0.7cm}, clip]{figures/conv_diag/parathion.pdf} &
       \includegraphics[page=3,width=.32\textwidth, trim={1.15cm 0.6cm 1.05cm 0.7cm}, clip]{figures/conv_diag/parathion.pdf}    
       \end{tabular} 
    \caption{\update{Traceplots for Parathion (Ethyl).}}
\end{figure}

\begin{table}[h]
    \centering
    \update{
    \begin{tabular}{|c|c|c|c|}
        \hline
         & ESS (ncomp) & ESS (Latent\_variable) & ESS (log\_likelihood) \\ \hline
        2,4-D Acid & 201 & 316 & 213 \\
        Trichlorfon & 276 & 404 & 37\\
        Parathion (Ethyl) & 190 & 501 & 41 \\
        \hline
    \end{tabular}}
    \caption{\update{Effective sample size of the parameters for the different datasets.}}
\end{table}

\subsection{Small datasets}
\begin{figure}[H]
    \centering
    \begin{tabular}{ccc}
       \includegraphics[page=1,width=.32\textwidth, trim={1.15cm 0.6cm 1.05cm 0.7cm}, clip]{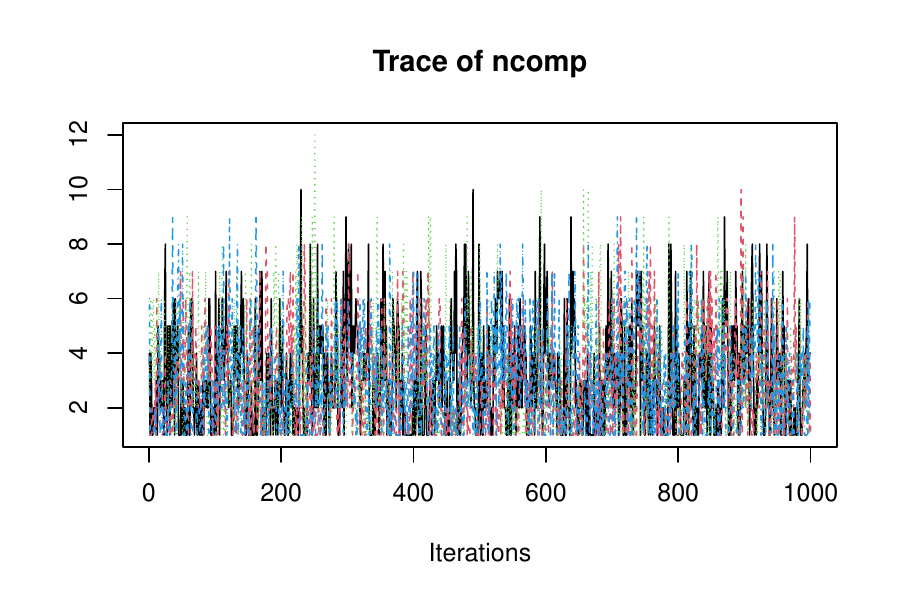} & 
       \includegraphics[page=2,width=.32\textwidth, trim={1.15cm 0.6cm 1.05cm 0.7cm}, clip]{figures/conv_diag/phosmet.pdf} &
       \includegraphics[page=3,width=.32\textwidth, trim={1.15cm 0.6cm 1.05cm 0.7cm}, clip]{figures/conv_diag/phosmet.pdf}    
       \end{tabular} 
    \caption{\update{Traceplots for Phosmet.}}
\end{figure}

\begin{figure}[H]
    \centering
    \begin{tabular}{ccc}
       \includegraphics[page=1,width=.32\textwidth, trim={1.15cm 0.6cm 1.05cm 0.7cm}, clip]{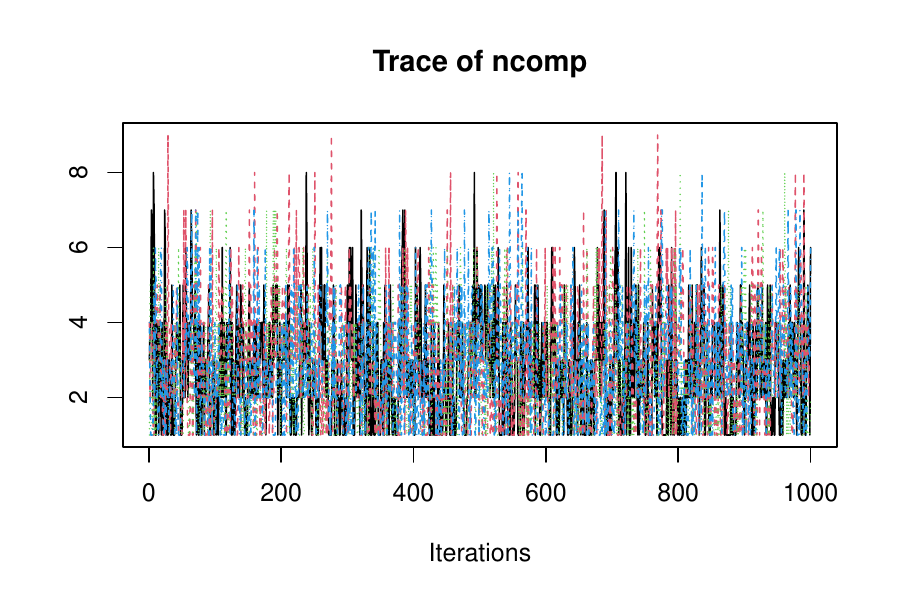} & 
       \includegraphics[page=2,width=.32\textwidth, trim={1.15cm 0.6cm 1.05cm 0.7cm}, clip]{figures/conv_diag/Naled.pdf} &
       \includegraphics[page=3,width=.32\textwidth, trim={1.15cm 0.6cm 1.05cm 0.7cm}, clip]{figures/conv_diag/Naled.pdf}    
       \end{tabular} 
    \caption{\update{Traceplots for Naled.}}
\end{figure}

\begin{figure}[H]
    \centering
    \begin{tabular}{ccc}
       \includegraphics[page=1,width=.32\textwidth, trim={1.15cm 0.6cm 1.05cm 0.7cm}, clip]{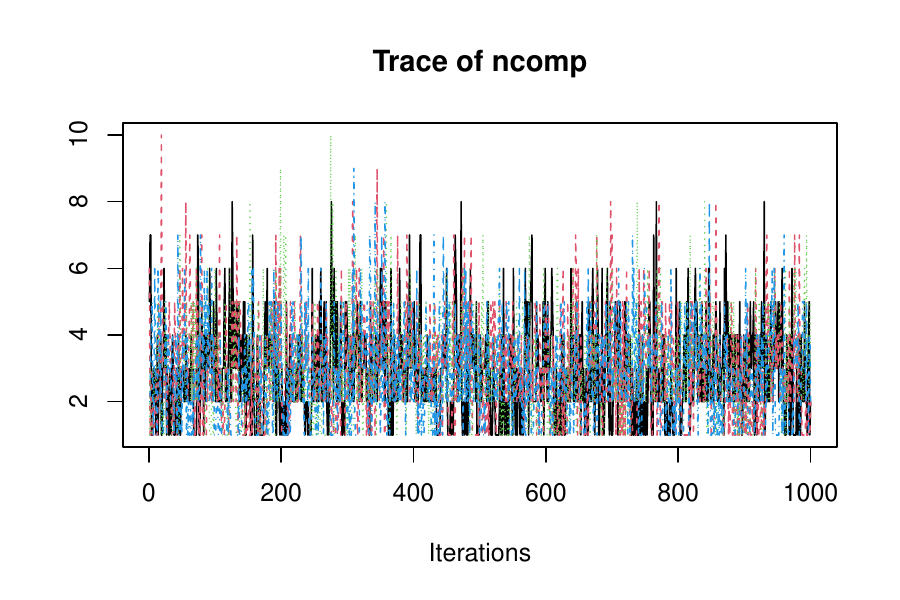} & 
       \includegraphics[page=2,width=.32\textwidth, trim={1.15cm 0.6cm 1.05cm 0.7cm}, clip]{figures/conv_diag/sodium.pdf} &
       \includegraphics[page=3,width=.32\textwidth, trim={1.15cm 0.6cm 1.05cm 0.7cm}, clip]{figures/conv_diag/sodium.pdf}    
       \end{tabular} 
    \caption{\update{Traceplots for Sodium dichromate.}}
\end{figure}

\begin{table}[h]
    \centering
    \update{
    \begin{tabular}{|c|c|c|c|}
        \hline
         & ESS (ncomp) & ESS (Latent\_variable) & ESS (log\_likelihood) \\ \hline
        Phosmet & 286 & 585 & 153 \\
        Naled & 433 & 624 & 318 \\
        Sodium dichromate & 350 & 590 & 77 \\
        \hline
    \end{tabular}}
    \caption{\update{Effective sample size of the parameters for the different datasets.}}
\end{table}

\section{Sensitivity analysis of the model}
\label{sec:sens}
\subsection{Parameter $\gamma$}
\begin{figure}[h]
  \centering
  \includegraphics[width=0.6\linewidth]{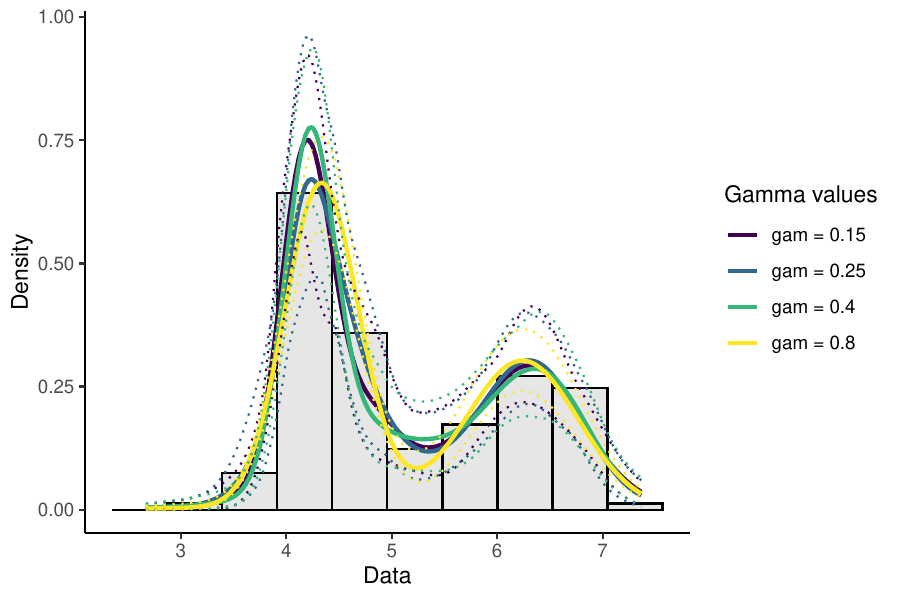}
  \caption{\update{Study of the sensitivity of the model to the parameter $\gamma$. A toy dataset (acidity from \texttt{BNPdensity}) has been considered here. The values of $\gamma$ are chosen in order to have different prior behaviors.}}
  \label{fig:sens_gam}
\end{figure}

\update{
An experiment was conducted to test the robustness of the model described in Section \ref{sec:methods} with respect to the choice of parameter $\gamma$. 
We used our model to estimate the density of a toy dataset. We took the toy dataset named acidity from the \texttt{BNPdensity} package. 
We tested different values of the parameter $\gamma \in (0,1)$. These values of $\gamma$ induce different a priori behaviors.  
Our results are summarized in Figure \ref{fig:sens_gam}.}
\update{In this figure, we can see that the different densities estimated for models with different values of $\gamma$ are very similar.
% A slight difference for $\gamma$ with a value of $0.15$ can be noticed, but the density remains very close to the others. 
This results supports the choice of a fixed value for $\gamma$. In this paper we consider $\gamma=0.4$.}

\subsection{Parameter $\sigma$}
\begin{figure}[h]
  \centering
  \includegraphics[width=0.6\linewidth]{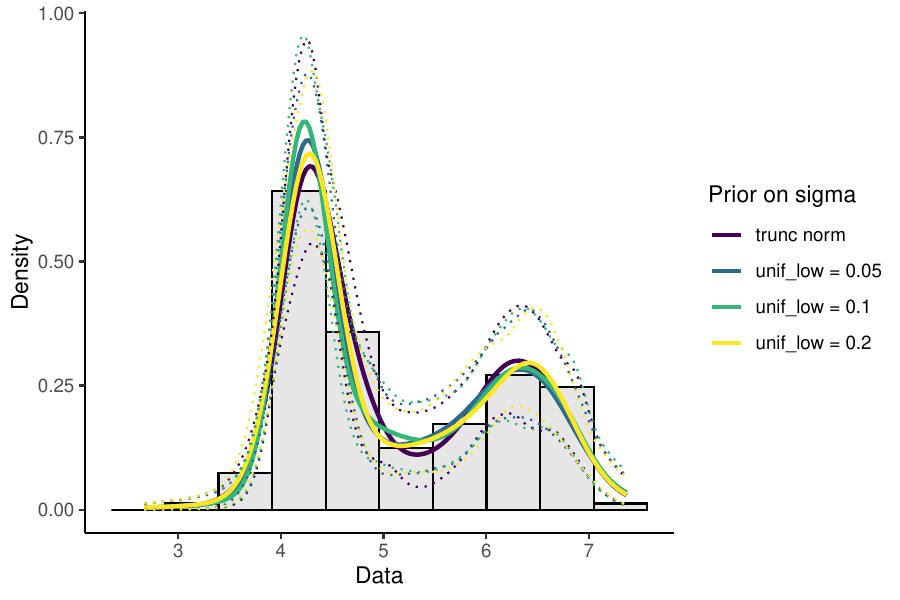}
  \caption{\update{Study of the sensitivity of the model to the choice of the prior for the parameter $\sigma$. A toy dataset (acidity from \texttt{BNPdensity}) has been considered here. Different priors for $\sigma$ have been considered such as an uniform with different bounds and truncated normal.}}
  \label{fig:sens_sig}
\end{figure}

\update{We conducted a similar analysis concerning the choice of the prior distribution for the parameter $\sigma$. 
Various prior distributions can be considered, in particular a uniform distribution and a truncated normal distribution.
The results presented in the paper are obtained considering a uniform prior distribution on the set $[0,1,1,5]$. 
We studied the sensitivity of the model with respect to this prior specification by varying its extreme points, as well as with respect to a left-truncated normal prior distribution with mean $0.5$, variance $1$, and lower bound $0.1$.
Our results are presented in Figure \ref{fig:sens_sig}.}
\update{For the uniform prior distribution, the results presented here only concern the case where the lower bound is modified, but we also took the upper bound into account and obtained similar results.
In light of these results, we concluded that the model is robust with regard to the choice of prior distribution on $\sigma$.
Note that we could consider other prior distributions, but we want $\sigma$ to have certain properties, as described in section \ref{sec:methods}, which explains the choice to consider only uniform and truncated normal distributions.}

%\newpage

\section{Details on the non-negative tensor factorization}\label{sec:details-ntf}
%%%%%%%% Change Louise %%%%%%%%%
\label{app:ntf}
In this paper, a tensor is a high-dimensional form of matrices.
We consider only three-order tensors, it means any tensor with three dimensions: $\mathbf{Y}\in\mathbb{R}^{I\times J\times K}$ where $I,J,K\in \mathbb{N}$.
We introduce one useful product for tensor factorization or decomposition, the outer product. The outer product of two vectors $a\in\mathbb{R}^I$ and $b\in\mathbb{R}^J$, denoted by $\circ$, yields a matrix $A\in\mathbb{R}^{I\times J}$,
\[ A = a \circ b = ab^T. \]
The outer product of three vectors $a\in\mathbb{R}^I$, $b\in\mathbb{R}^J$ and $c\in\mathbb{R}^K$ yields a third-order tensor $\mathbf{Y}\in\mathbb{R}^{I\times J\times K}$,
\[ \mathbf{Y} = a \circ b \circ c, \qquad \text{with } y_{ijk} = a_i b_j c_k. \]
A three-order tensor defined as the outer product of three vectors is called a rank-one tensor. The rank of a tensor $\mathbf{Y}$ if defined as the minimal number $R$ of rank-one tensors $\mathbf{Y}_1,\ldots,\mathbf{Y}_R$ such that $\mathbf{Y} = \sum_{r=1}^R\mathbf{Y}_r$.
% We also define diagonal tensors as special tensors. A tensor $\mathbf{Y} \in \mathbb{R}^{I\times J\times K}$ is a diagonal tensor if $y_{ijk} \neq 0$ only if $i=j=k$, i.e. if all elements outside the superdiagonal are zero.

The tensor factorization generalizes the matrix factorization techniques.
The different matrix factorizations are useful notably for feature selection or dimensionality reduction.
The tensor or multi-way array factorization allows to consider applications where the data contains high-order structures.
One of the most popular models for the factorization of high-order tensors is the PARAFAC model.
In the following, we will describe the decomposition for a third-order tensor, but the model can be extended to decompose a higher-order tensor.

% \subsection{Tucker decomposition}
% The Tucker decomposition \citep{tucker1966some} is a form of higher-order Principal Components Analysis (PCA).  
% The idea of the Tucker decomposition is to decompose a tensor $\mathbf{Y} \in \mathbb{R}^{I\times J\times K}$ into a core tensor denoted $\mathbf{G}\in \mathbb{R}^{P\times Q\times R}$ multiplied by three matrices, $A\in \mathbb{R}^{I\times P}$, $B\in \mathbb{R}^{J\times Q}$ and $C\in \mathbb{R}^{K\times R}$. 
% With previous notations, we denote by $[\![\mathbf{G}; A,B,C]\!]$ the Tucker decomposition, defined by
% \[ \mathbf{Y} = \sum_{p=1}^P \sum_{q=1}^Q \sum_{r=1}^R g_{pqr} (a_p\circ b_q\circ c_r) + \mathbf{E} = [\![\mathbf{G}; A,B,C]\!]+ \mathbf{E} , \]
% where $\mathbf{E}$ is a tensor of approximation error.
% The tensor $\mathbf{G}$, known as the core tensor, expresses the interaction between the different components. $\mathbf{G}$ can be seen as a compressed version of the initial tensor $\mathbf{Y}$.
% The matrices $A$, $B$ and $C$ are known as the factor matrices. $P$, $Q$ and $R$ are the number of components in the factor matrices.
% They are usually orthogonal, in this case, the Tucker decomposition is also called the three-way PCA.

\subsection{PARAFAC factorization}
We recall the Parallel Factors Analysis (PARAFAC) described in Section \ref{sec:res_clus}.
Given a tensor $\mathbf{Y}\in \mathbb{R}^{I\times J\times K}$, the PARAFAC factorization is denoted by $\mathbf{Y} = [\![A,B,C]\!]$ where $A=[a_1,\ldots,a_R] \in \mathbb{R}^{I\times R}$, $B=[b_1,\ldots,b_R] \in \mathbb{R}^{J\times R}$ and $C=[c_1,\ldots,c_R] \in \mathbb{R}^{K\times R}$ are three components matrices. More formally,
\[ \mathbf{Y} = \sum_{r=1}^R a_j \circ b_j \circ c_j + \mathbf{E} = [\![A,B,C]\!] + \mathbf{E}, \]
where the tensor $\mathbf{E}\in \mathbb{R}^{I\times J\times K}$ represents the approximation error.

% The PARAFAC factorization can be stated in the form of a restricted Tucker decomposition \citep{Bro2003},
% \begin{equation}
%     \label{eq:CP_Tucker}
%     \mathbf{Y} = [\![\mathbf{T}; A,B,C]\!],
% \end{equation} 
% where the core tensor $\mathbf{T}\in \mathbb{R}^{R\times R\times R}$ is a diagonal tensor with ones on the superdiagonal, which means that $t_{ijk} = 1$ if $i=j=k$ and $t_{ijk} = 0$ else, this tensor is also called the cubical identity tensor.

%\subsubsection{Selection of the rank decomposition}
A difficult problem for the PARAFAC model is to choose the appropriate number of components $R$.
This problem is equivalent to determining the rank of a tensor, in the sense that the rank of a tensor is the smallest number of $R$ components in an exact PARAFAC decomposition, i.e. with a null approximation error tensor $\mathbf{E}$.
Determining the rank of a given tensor is known to be a NP-hard problem.

In the ideal case, there is no noise in the data, so we can fit the PARAFAC model for different values of $R$ until we have an exact decomposition.
This assumes a perfect procedure for fitting the PARAFAC model, which is not the case in practice.
Furthermore, in a more realistic case, the data is noisy and the procedure described is no longer applicable.
% \cite{Bro2003} propose a heuristic method called core consistency diagnostic or CORCONDIA. 
% This method is based on the fact that the PARAFAC model can be expressed as a special case of the Tucker model as in \eqref{eq:CP_Tucker}. 
% The idea is to quantify the similarity between the Tucker decomposition core tensor, denoted by $\mathbf{G}$ previously, and the core tensor $\mathbf{T}$ introduced in \eqref{eq:CP_Tucker} where both models are applied to the same tensor $\mathbf{Y}$.
% As the core tensor $\mathbf{T}$ is a super diagonal tensor of ones, the way to assess the similarity between $\mathbf{T}$ and $\mathbf{G}$ is to compare the element of the superdiagonal of $\mathbf{G}$ with $1$ and the off-superdiagonal elements with $0$.
% More formally, the similarity between $\mathbf{G}$ and $\mathbf{T}$ is quantified as
% \[ \text{CORCONDIA} =1-\frac{\sum_{i=1}^R\sum_{j=1}^R\sum_{k=1}^R (g_{ijk}-t_{ijk})^2}{R}. \]
% A core consistency above $0.9$ indicates an appropriate model. In this paper, we take the smallest number of components yielding a CONCORDIA index above $0.9$.

There are different proposed methods to solve this problem, such as core consistency diagnostic, residual analysis, visual appearance of loadings (also called factors represented by the components matrices), or cross-validation.
The method we used in this paper is the cross-validation as described in Section \ref{sec:res_clus}.

\subsection{Non-negative tensor factorization}
On the PARAFAC model presented previously, it is possible to impose some non-negativity constraints.
The non-negativity constraints allow to give physical meaning to the different components found.

In practice, adding constraints on the PARAFAC model leads to solving an optimization problem under constraints.
Indeed, finding the PARAFAC factorization means solving the following optimization problem,
\[ \min_{A,B,C} \left\|\mathbf{Y}-[\![A,B,C]\!]\right\|_F^2, \]
where $\|\cdot\|_F$ denotes the tensor Frobenius norm defined by $\|\mathbf{Y}\|_F^2=\sum_{i=1}^I\sum_{j=1}^J\sum_{k=1}^K y_{ijk}^2$.
The non-negative PARAFAC factorization is a PARAFAC factorization with the following non-negative constraints
\[ \min_{A,B,C \text{ s.t. } a_{ir},\,b_{jr},\,c_{kr} \in \mathbb{R}_+} \left\|\mathbf{Y}-[\![A,B,C]\!]\right\|_F^2 . \]
%The idea is the same for the Tucker model with some non-negativity constraints.

\section{Additional figures for Section~\ref{sec:res_clus}}\label{sec:supp-figures}
We now present some additional figures illustrating the results presented in Section~\ref{sec:res_clus}.

\begin{figure}[h]
    \centering
    \includegraphics[width=\textwidth]{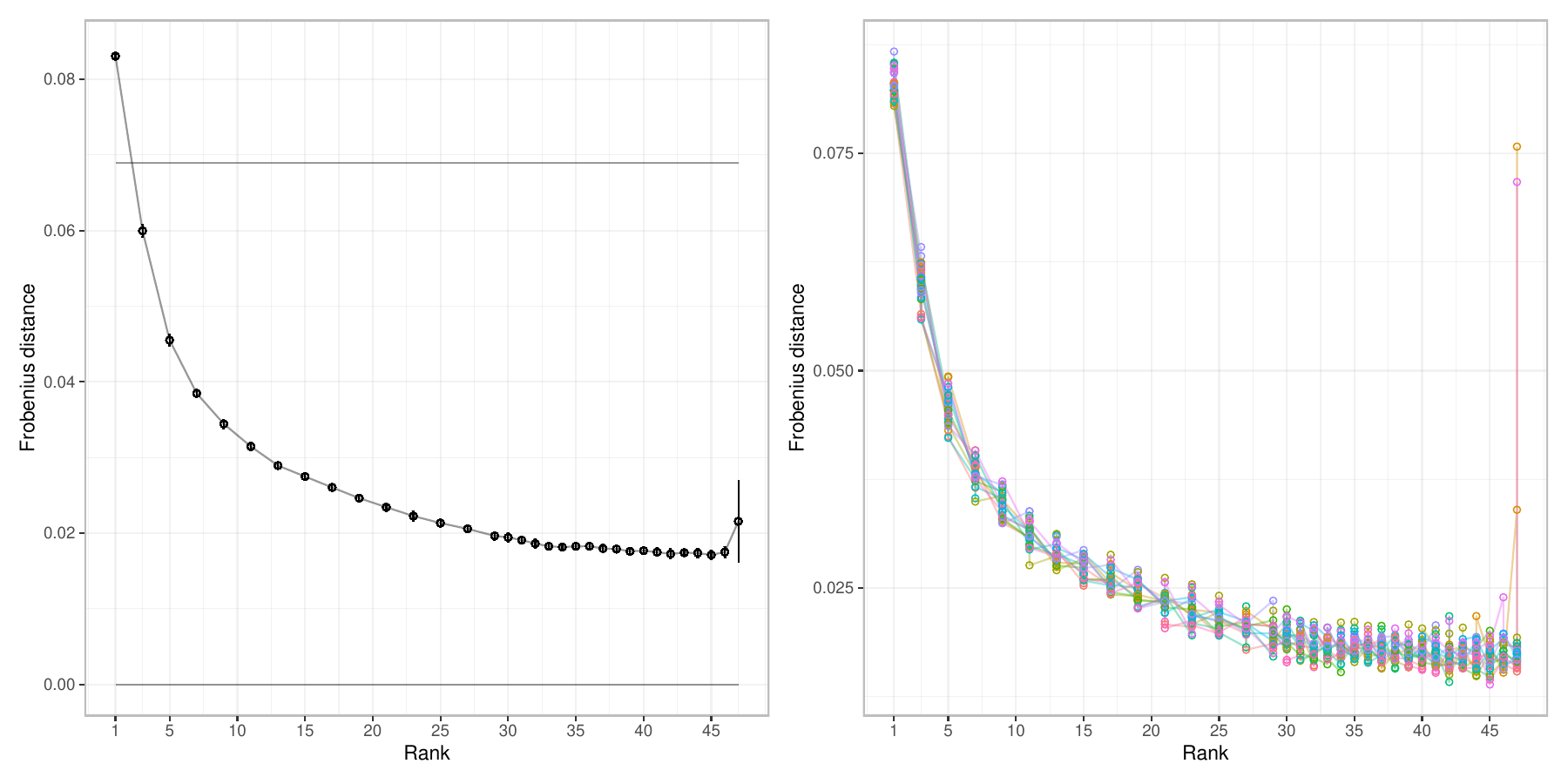}
    \caption{Cross-validation results for the tensor decomposition. Left: Frobenius distance mean of the 30-fold cross-validation with a confidence interval at $95\%$ around each point with the rank varying between 1 and 47. Right: Frobenius distance for each of the 30 folds. The rank of the decomposition is chosen by taking the lowest rank (here 39) contained in the confidence interval from the rank minimizing the error (here 45).}
    \label{CrossV}
\end{figure}

Figure~\ref{CrossV} illustrates the choice of the rank decomposition using the cross-validation method.

\begin{figure}[ht]
    \centering
    \includegraphics[width=0.75\textwidth]{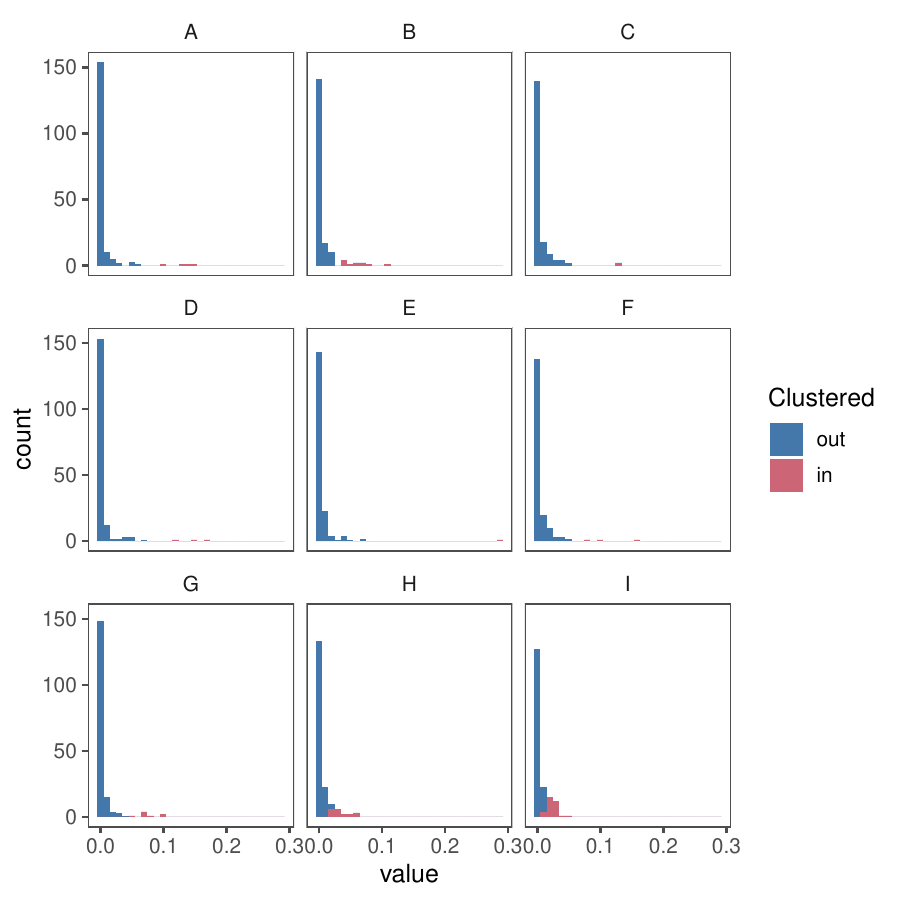}
    \caption{Distribution of the contaminant weight in each component. The presence of two well-separated groups in the distribution suggests separating the contaminants that belong to the component and those that do not. The clustering is performed using the K-means method. In Section~\ref{sec:res_clus} only the components where there is no overlap between the red part and the blue part, components A, B, C, D, E, F, and G are considered while components H and I are not.}
    \label{Kmeans}
\end{figure}

\begin{figure}[hb]
    \centering
    \includegraphics[width=0.75\textwidth]{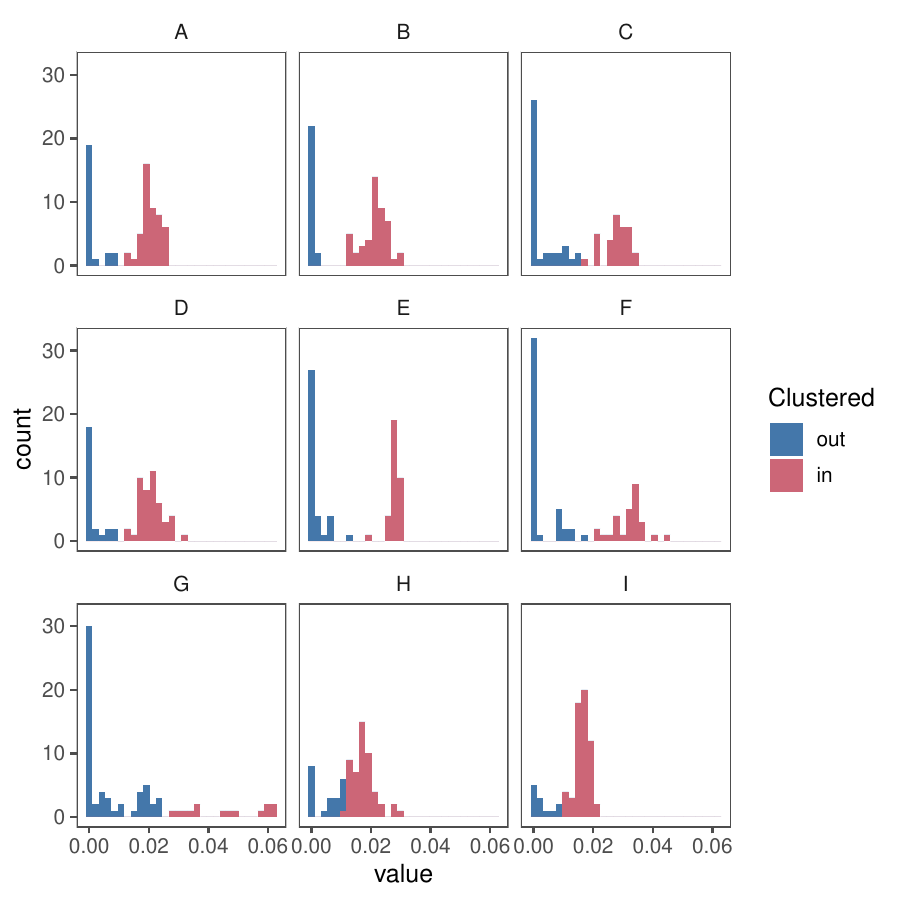}
    \caption{Distribution of the species weight in each component. The presence of two well-separated groups in the distribution suggests separating the contaminants that belong to the component and those that do not. The clustering is performed using the K-means method.}
    \label{Kmeans_spec}
\end{figure}

Figure~\ref{Kmeans} and Figure~\ref{Kmeans_spec} illustrate the K-means clustering used on each components resulting from the tensor decomposition.
The 7 components considered in Section~\ref{sec:res_clus}, components A, B, C, D, E, F and G, are chosen using the K-means clustering and are the only ones well-separated in the K-means clustering considering both the contaminants and species.
These chosen components are the 2nd, 4th, 12th, 14th, 20th, 30th and 34th components in the decomposition.

\begin{figure}[h]
    \centering
    \includegraphics[page = 2, height = 0.95\textheight]{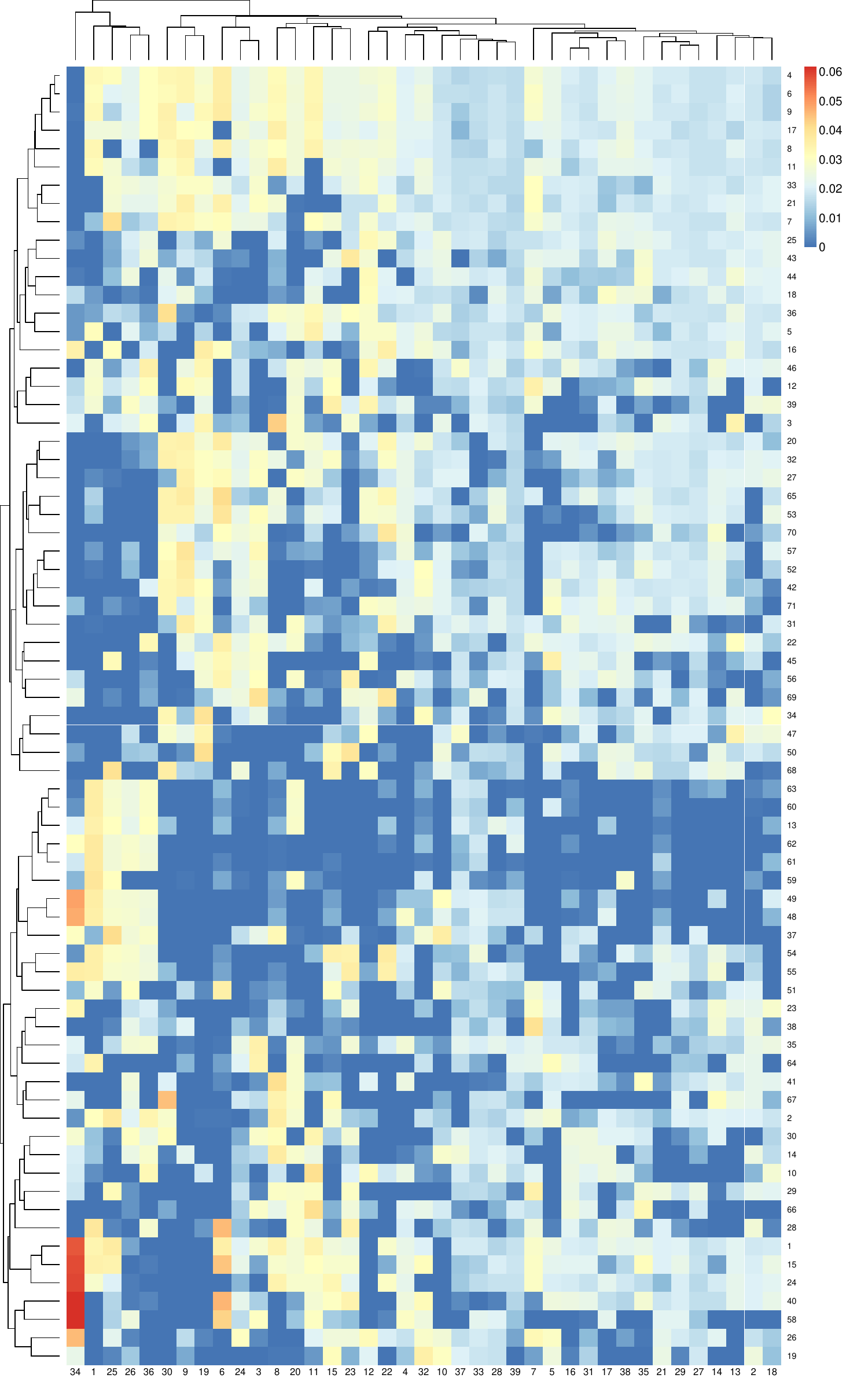}
    \caption{Heatmap of all the species and components.}
    \label{fig:Heat_spe}
\end{figure}

\begin{figure}[h]
    \centering
    \includegraphics[page = 2, height = 0.95\textheight]{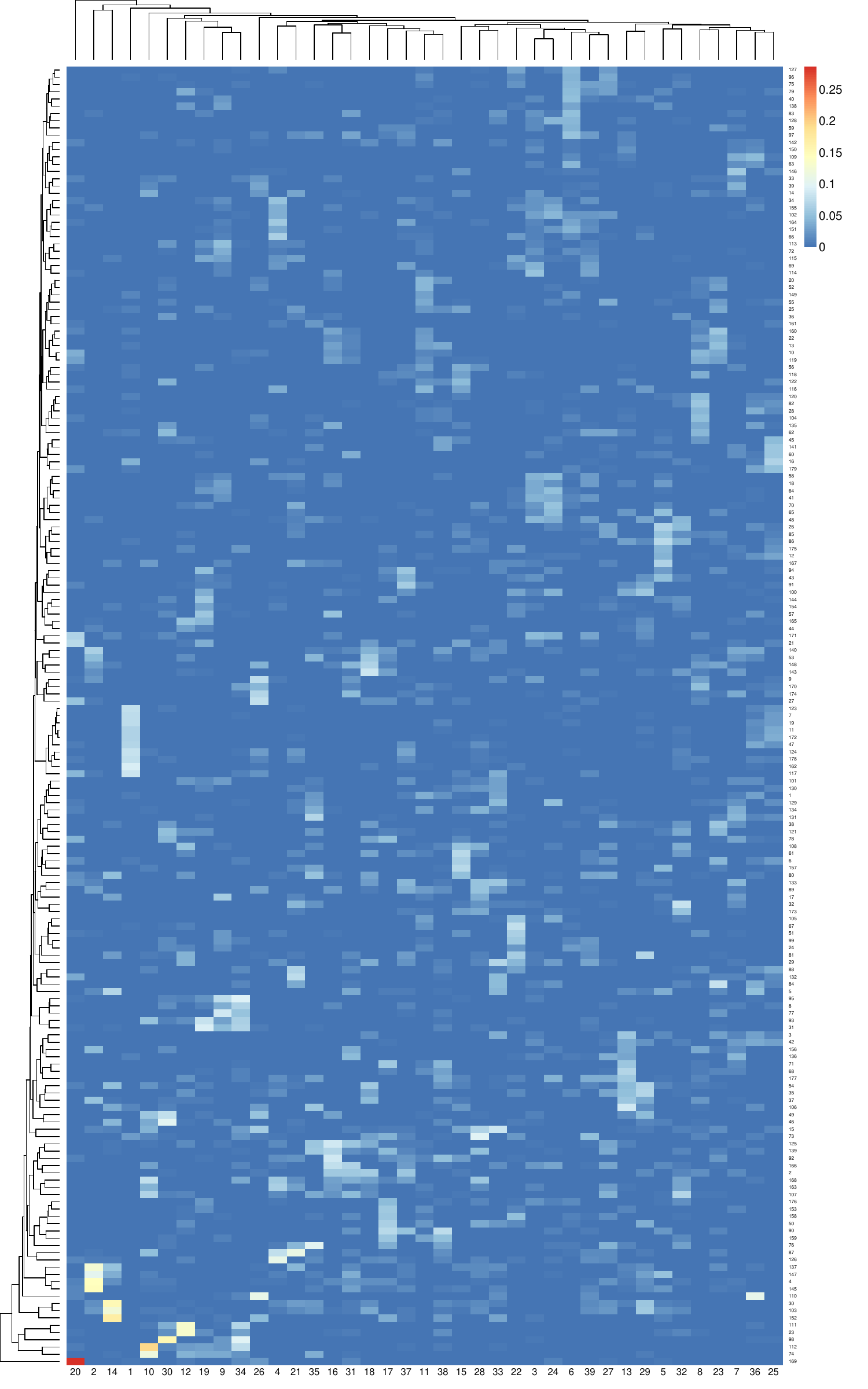}
    \caption{Heatmap of all the contaminants and components.}
    \label{fig:Heat_cont}
\end{figure}

Figure~\ref{fig:Heat_spe} presents the heatmap of the components and the species, and Figure~\ref{fig:Heat_cont} presents the heatmap of the components and the contaminants.
The heatmaps indicate which contaminants or species have more weights in a component.
This is another way to illustrate the composition of the components.

\begin{figure}[h]
    \centering
    \includegraphics[width=\textwidth]{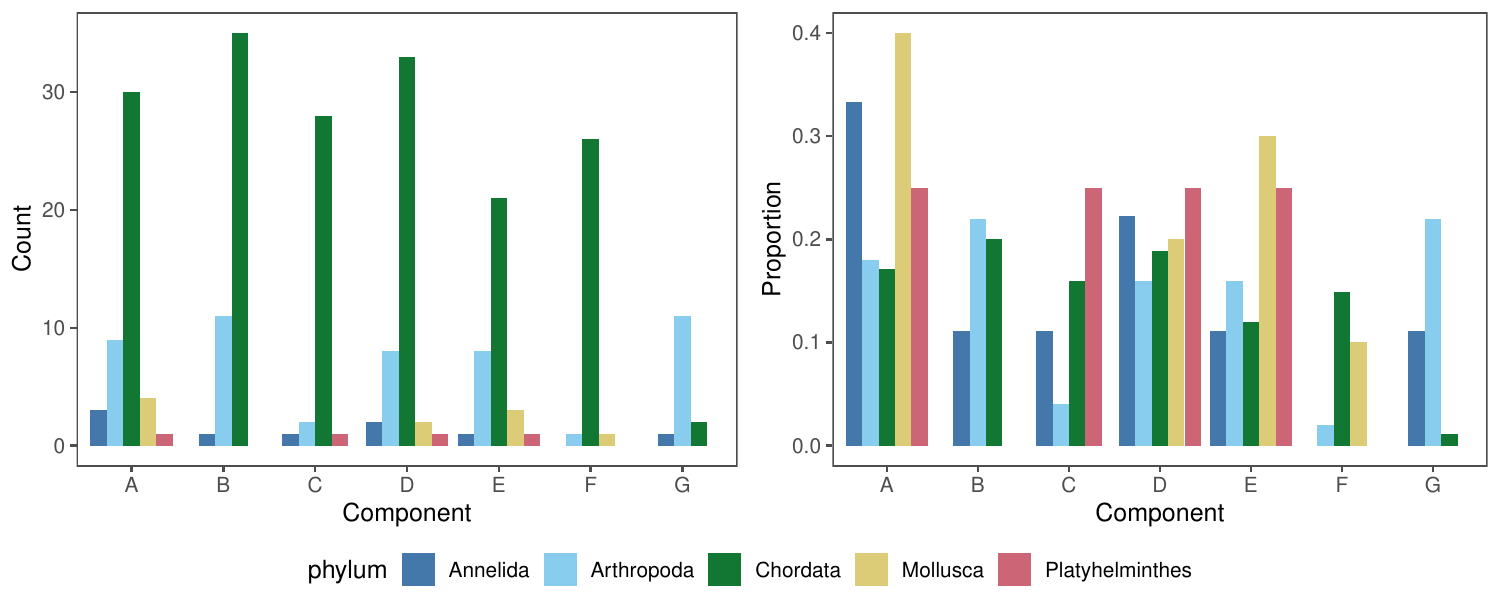}
    \caption{Taxonomic composition of the selected components at the Phylum level.}
    \label{fig:phyl}
\end{figure}

\begin{figure}[h]
    \centering
    \includegraphics[width=\textwidth]{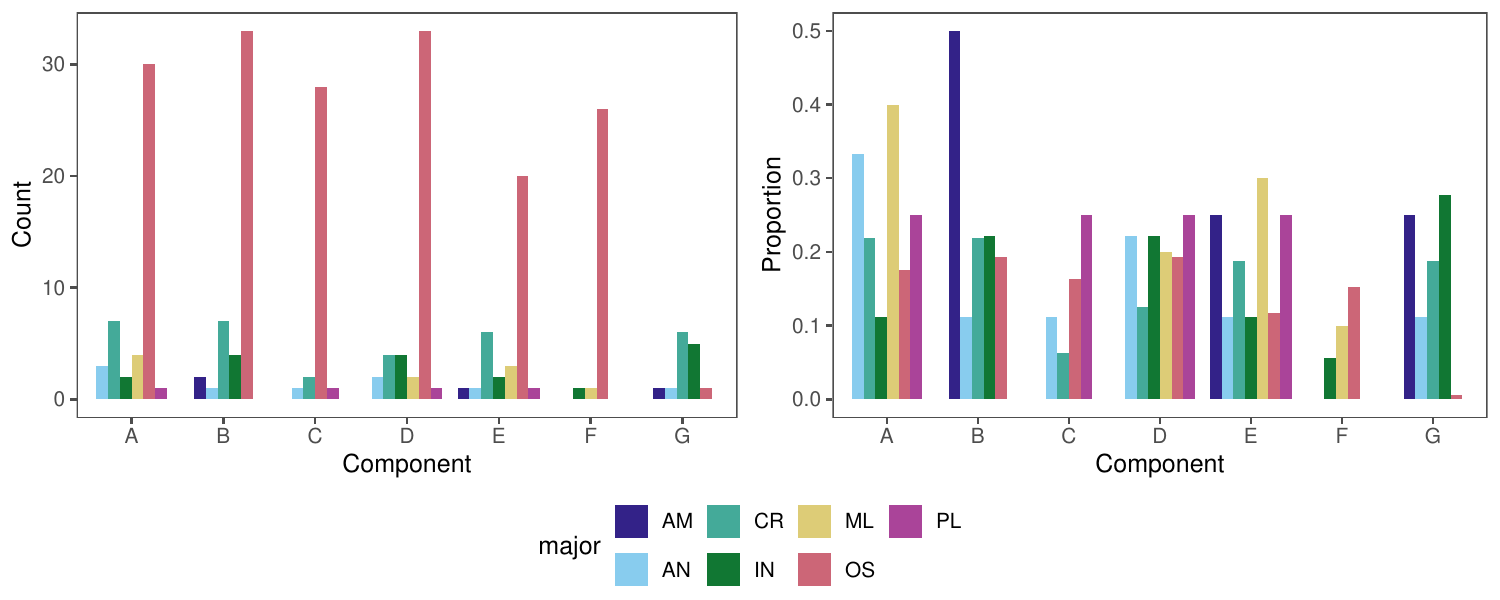}
    \caption{Taxonomic composition of the selected components at the major level.}
    \label{fig:maj}
\end{figure}

Figure \ref{fig:maj} and \ref{fig:phyl} present the taxonomic composition of the seven components previously selected at two different taxonomic ranks.
These two figures support the idea that the taxonomy does not seem to strongly determine species sensitivity.

% \includegraphics[width = \textwidth]{core_consistency}
% \includegraphics[width = \textwidth]{corcondia.png}
% \captionof{figure}{Core consistency diagnostic for the tensor decomposition. \label{CoreConsistency}} 

\end{document}